\documentclass[a4paper,11pt]{article}
\pdfoutput=1
\usepackage{jheppub}
\usepackage[T1]{fontenc}
\usepackage[english]{babel}
\usepackage{amsthm,amsfonts}
\usepackage{empheq}
\usepackage{mathdots}
\usepackage{dsfont}
\usepackage{pifont}

\usepackage[textsize=small,shadow]{todonotes}
\usepackage{ulem}
\hyphenchar\font=\string"7F
\usepackage{colortbl}
\definecolor{fillcolor}{HTML}{C4C4C4}
\usepackage[boxsize=1em,aligntableaux=center]{ytableau}
\usepackage{array}

\usepackage{setspace,cancel,url,verbatim,color}
\usepackage{booktabs}
\usepackage{caption} 
\usepackage{stmaryrd} 
\usepackage{breqn}
\usepackage{nicefrac}

\usepackage{tikz}
\usetikzlibrary{arrows,matrix,calc,shapes.multipart}
\tikzset{>=stealth',every on chain/.append style={join},
         every join/.style={->}}
\tikzstyle{shiftup}=[transform canvas={yshift=0.25em}]
\tikzstyle{shiftdn}=[transform canvas={yshift=-0.25em}]
\makeatletter
\renewcommand{\@dotsep}{10000}
\makeatother

\DeclareMathOperator{\diag}{diag}
\DeclareMathOperator{\ord}{ord}
\newcommand{\DFTwzw}{DFT${}_\mathrm{WZW}$}
\allowdisplaybreaks
\newcommand{\HIDDEN}[1]{}

\title{\boldmath Generalized Parallelizable Spaces from\\Exceptional Field Theory}
\preprint{LMU-ASC 32/17\\MPP-2017-105}

\author[a,b]{Pascal du Bosque,}
\emailAdd{dubosque@mpp.mpg.de}
\author[c]{Falk Hassler,}
\emailAdd{fhassler@unc.edu}
\author[a,b]{Dieter L\" ust,}
\emailAdd{dieter.luest@lmu.de}

\affiliation[a]{Max-Planck-Institut f\"ur Physik\\
F\"ohringer Ring 6, 80805 M\"unchen, Germany}
\affiliation[b]{Arnold-Sommerfeld-Center f\"ur Theoretische Physik\\
Fakult\"at f\"ur Physik, Ludwig-Maximilians-Universit\"at M\"unchen\\
Theresienstra\ss e 37, 80333 M\"unchen, Germany}
\affiliation[c]{University of North Carolina\\Department of Physics and Astronomy\\
Phillips Hall, CB \#3255, 120 E. Cameron Ave., Chapel Hill, NC 27599-3255, USA}

\abstract{Generalized parallelizable spaces allow a unified treatment of consistent maximally supersymmetric truncations of ten- and eleven-dimensional supergravity in generalized geometry. Known examples are spheres, twisted tori and hyperboloides. They admit a generalized frame field over the coset space $M$=$G/H$ which reproduces the Lie algebra $\mathfrak{g}$ of $G$ under the generalized Lie derivative. An open problem is a systematic construction of these spaces and especially their generalized frames fields. We present a technique which applies to $\dim M$=4 for SL(5) exceptional field theory. In this paper the group manifold $G$ is identified with the extended space of the exceptional field theory. Subsequently, the section condition is solved to remove unphysical directions from the extended space. Finally, a SL(5) generalized frame field is constructed from parts of the left-invariant Maurer-Cartan form on $G$. All these steps impose conditions on $G$ and $H$.}

\begin{document}
\maketitle

\section{Introduction and Summary}
Dualities are a key element of string theory and play an essential role in our understanding of it. This explains the recent interest in Exception Field Theory (EFT) \cite{Berman:2010is,Berman:2011jh,Berman:2012vc,Hohm:2013vpa,Hohm:2013uia,Hohm:2014fxa,Abzalov:2015ega,Musaev:2015ces}. It aims at making the U-duality \cite{Hull:1994ys} groups in $d$ dimensions a manifest symmetry in the low-energy effective description of string/M-theory. By doing so, it is an excellent tool to study maximal supergravities in lower dimensions which for example arise from compactifications of eleven-dimensional supergravity on a $d$-torus. Intriguingly, their global symmetry is captured by the corresponding duality group (see e.g. \cite{Cremmer:1979up,Cremmer:1997ct}). In order to appreciate the power of the EFT formalism, we try to understand these distinguished global symmetries from the eleven-dimensional perspective. All of them admit a GL($d$) subgroup which originates from diffeomorphisms on the torus. If we study the generators of the global symmetry group, we observe that $d^2$ of them generate this subgroup. Furthermore, there exist additional generators from internal gauge transformations of $p$-form fields. However, they are not sufficient to enhance GL($d$) to the full duality groups listed in table~\ref{tab:Udualitygroups}. In addition, there have to be hidden symmetries without obvious explanation from an eleven-dimensional point of view \cite{Samtleben:2008pe}.
\begin{table}
  \centering\caption{U-duality groups \cite{Hull:1994ys} which have the T-duality groups O($d$-1,$d$-1) as subgroup. Moreover, the coordinate and section condition irreps \cite{Berman:2011jh,Berman:2012vc} of the corresponding EFTs \cite{Berman:2015rcc,Hohm:2015xna,Musaev:2015ces,Abzalov:2015ega,Hohm:2013vpa,Hohm:2013uia,Hohm:2014fxa} and the embedding tensor irreps \cite{deWit:2002vt,deWit:2008ta} after the linear constraint are given.}\label{tab:Udualitygroups}
  \begin{tabular}{llllllll}\toprule
    d & 2 & 3 & 4 & 5 & 6 & 7 & 8 \\
    \hline
    U-d. group & SL(2)$\times\mathbb{R}^+$ & SL(3)$\times$SL(2) & SL(5) & Spin(5,5) & E${}_{6(6)}$ & E${}_{7(7)}$ & E${}_{8(8)}$ \\
    coord. irrep & $\mathbf{2}_1$ + $\mathbf{1}_{-1}$ & $(\mathbf{3},\mathbf{2})$ & $\mathbf{10}$ & $\mathbf{16}$ & $\mathbf{27}$ & $\mathbf{56}$ & $\mathbf{248}$ \\
    SC irrep & & $(\overline{\mathbf{3}},\mathbf{1})$ & $\overline{\mathbf{5}}$ & $\mathbf{10}$ & $\overline{\mathbf{27}}$ & $\mathbf{133}$ & $\mathbf{3875}+\mathbf{1}$ \\
    emb. tensor & & & $\mathbf{15} + \overline{\mathbf{40}}$ & $\mathbf{144}$ & $\mathbf{351}$ & $\mathbf{912}$ & $\mathbf{3875}$ \\
    \bottomrule
  \end{tabular}
\end{table}

EFT succeeds in making the full duality group manifest by considering an extended spacetime. There are no hidden symmetries. It is very important to keep in mind that the additional directions added to the eleven-dimensional spacetime of M-theory are not physical\footnote{In the $d$-torus compactification outlined above, these additional coordinates allow for the interpretation of being conjugate to certain brane wrapping modes. But taking the section condition into account, EFT is defined for more general backgrounds. Take for instance a $d$-dimensional sphere instead of a torus. For this case there do not exist non-contractible cycles and thus no winding modes.}. Nevertheless, they are a powerful book keeping device. But at the end of the day, one has to get rid of them by imposing the section condition (SC). An example where this extra book keeping pays off are generalized Scherk-Schwarz reductions\footnote{For results in Double Field Theory see for example \cite{Aldazabal:2011nj, Geissbuhler:2011mx,Berman:2013cli}.}\cite{Berman:2012uy,Musaev:2013rq,Aldazabal:2013mya,Lee:2014mla,Hohm:2014qga,Baguet:2015sma} which result in maximal gauged supergravities. In conventional Scherk-Schwarz reductions \cite{Scherk:1978ta,Scherk:1979zr}, the compactification ansatz for the metric is given by the left- or right-invariant Maurer-Cartan form on a group manifold $G$. More specifically, the Maurer-Cartan form gives rise to a vielbein $e_a$ from which the metric follows. Here, $a$ marks an index in the adjoint representation of the Lie algebra $\mathfrak{g}$ of $G$ and we suppress vector indices. Considering the Lie derivative
\begin{equation}\label{eqn:lieonMC}
  L_{e_a} e_b = f_{ab}{}^c e_c\,,
\end{equation}
we observe that the vielbein implements $\mathfrak{g}$ with the structure constants $f_{ab}{}^c$ on every point of $G$. As a result, the lower dimensional theory after the compactification has $G$ as a gauge symmetry. In EFT one chooses a generalized vielbein $\mathcal{E}_A$ where $A$ is an index in the coordinate irrep of the corresponding duality group (see table~\ref{tab:Udualitygroups}). Again, we suppress the vector index of the extended space. Under the generalized Lie derivative $\widehat{\mathcal{L}}$, which mediates infinitesimal gauge transformations in EFT, an analogous relation to \eqref{eqn:lieonMC},
\begin{equation}\label{eqn:genlieongenviel}
  \widehat{\mathcal{L}}_{\mathcal{E}_A} \mathcal{E}_B = X_{AB}{}^C \mathcal{E}_C\,,
\end{equation}
holds. The compactified theory is a maximal gauged supergravity. Its gauge algebra $\mathfrak{g}$ is restricted by the embedding tensor formalism \cite{deWit:2002vt,deWit:2008ta,Samtleben:2008pe,Samtleben:2005bp}. We denote the EFT analogue to the structure constants $f_{ab}{}^c$ as $X_{AB}{}^C$. Even if \eqref{eqn:lieonMC} and \eqref{eqn:genlieongenviel} look very similar, there are three important differences. First, the generalized Lie derivative is not the conventional Lie derivative $L$ on the extended space. Moreover, the generalized frame field is not the left- or right invariant Maurer-Cartan form on a group manifold $G$ associated to the Lie algebra $\mathfrak{g}$. Finally, the generalized frame is constrained by the SC\footnote{There are exceptions which give rise to non-geometric backgrounds. But here, we are only concerned with the simplest case where the SC has to hold.}. Thus, $\widehat{\mathcal{L}}$ in EFT reduces to the generalized Lie derivative of exceptional Generalized Geometry (GG) \cite{Hull:2007zu,Pacheco:2008ps,Aldazabal:2010ef,Baraglia:2011dg,Coimbra:2011ky} on the physical space $M$. If one is able to find a generalized frame field $\mathcal{E}_A$ which fulfills \eqref{eqn:genlieongenviel}, $M$ is called a generalized parallelizable space \cite{Grana:2008yw,Lee:2014mla}. There exists no algorithm yet to construct those spaces. Still there are some known examples such as spheres \cite{Lee:2014mla,Hohm:2014qga}, twisted tori, and hyperboloides \cite{Hohm:2014qga}. In combination with the generalized Scherk-Schwarz ansatz, they are crucial to show that dimensional reductions on certain coset spaces are consistent \cite{Cvetic:2003jy,Lee:2014mla,Hohm:2014qga,Baguet:2015sma,Baguet:2015iou,Cassani:2016ncu}. Hence, presenting a methodical way to construct the generalized frame field $\mathcal{E}_A$ in \eqref{eqn:genlieongenviel} is the objective of this work.

In this paper we follow a different approach to EFT which makes the group $G$ manifest. It is based on geometric Exceptional Field Theory (gEFT) introduced in \cite{Bosque:2016fpi}. This theory treats the extended space as a conventional manifold. Compared to the conventional formulation, it has a modified SC and an additional linear constraint. Following \cite{Bosque:2016fpi}, we mainly consider manifolds with GL(5)$^+$-structure. More specifically, we study group manifolds $G$ which arise as a solution of the SL(5) embedding tensor \cite{Samtleben:2005bp}. They represent an explicit example for an extended manifold whose structure group is a subgroup of GL(5)$^+$. In this setup, the background vielbein of gEFT is the left-invariant Maurer-Cartan form on $G$. The resulting theory is closely related to \DFTwzw{} \cite{Blumenhagen:2014gva,Blumenhagen:2015zma}, a version of Double Field Theory (DFT) \cite{Siegel:1993th,Hull:2009mi,Hull:2009zb,Hohm:2010pp,Aldazabal:2013sca,Hohm:2013bwa}, derived from the worldsheet theory of a Wess-Zumino-Witten model. Perhaps most remarkably, it allows us to give a direct construction of a large class of generalized parallelizable spaces.

Our main results can be summarized as follows. We start by following the ideas of \cite{Cederwall:2014kxa,Cederwall:2014opa,Bosque:2016fpi} to implement generalized diffeomorphisms that are compatible with standard diffeomorphisms (mediated by the Lie derivative $L$). To this end, we introduce a covariant derivative $\nabla$ on $G$ and use it to write the generalized Lie derivative
\begin{equation}\label{eqn:genLieintro}
  \mathcal{L}_\xi V^A = \xi^B \nabla_B V^A - V^B \nabla_B \xi^A + Y^{AB}{}_{CD} \nabla_B \xi^C V^D\,.
\end{equation}
In this equation we use flat indices $A, B, C$ for the group manifold and $Y^{AB}{}_{CD}$ denotes the $Y$-tensor measuring the deviance from Riemann geometry for the corresponding EFT \cite{Berman:2012vc}. The left-invariant Maurer-Cartan form $E_A{}^I$ connects flat indices with tangent space indices on $G$. Additionally, we need to impose the modified SC
\begin{equation}\label{eqn:SCintro}
  Y^{CD}{}_{AB} \, D_C \cdot D_D \, \cdot = 0
\end{equation}
for the algebra of infinitesimal generalized diffeomorphisms to close \cite{Blumenhagen:2014gva,Bosque:2016fpi}. It involves flat (curvature free but torsionful) derivatives $D$ which are connected to $\nabla$ by $\nabla_A V^B = D_A V^B + \Gamma_{AC}{}^B V^C$. By itself, the SC is not sufficient for \eqref{eqn:genLieintro} to close. Furthermore, we have to impose two linear and a quadratic constraint. At this point, the results are quite general and do not require to explicitly fix the T-/U-duality group. However, solving the linear constraints heavily depends on the representation theory of the chosen duality group. A linear constraint is known in the context of the embedding tensor formalism as well. It reduces the irreps resulting from the tensor product coordinate irrep $\times$ adjoint to the embedding tensor irreps given in table~\ref{tab:Udualitygroups}. Considering SL(5) as a specific duality group, the linear constraints we find result in the same restriction. On top of that, they come with an additional subtlety. For gaugings in the $\overline{\mathbf{40}}$ the dimension of the resulting group manifold is between nine and six. Thus, we are not able to identify the coordinates on $G$ with the irrep $\mathbf{10}$ stated in table~\ref{tab:Udualitygroups}. In order to still obtain well-defined generalized diffeomorphisms on these group manifolds, we break SL(5) into smaller U-duality subgroups whose irreps can be chosen such that they add up to $\dim G$. This situation is special to gEFT, it does not occur for the T-duality subgroup O(3,3) for which we reproduce the gauge algebra of \DFTwzw{} \cite{Blumenhagen:2014gva,Blumenhagen:2015zma}. Finally, the quadratic constraint is equivalent to the Jacobi identity on the Lie algebra of $G$ and therefore automatically fulfilled in our setup.

Additionally, the SC in \eqref{eqn:SCintro} has to be solved. It acts on fluctuations (denoted by $\cdot$) around the background group manifold $G$. A trivial solution is given by constant fluctuations. They are sufficient to capture the lightest modes in the generalized Scherk-Schwarz reduction. If we want to incorporate heavier modes, we have to find the most general SC solutions. They depend on all $d$ physical coordinates of the extended space. We apply a technique introduced in \cite{Hassler:2016srl} for \DFTwzw{} to construct them. It interprets the group manifold as a $H$-principal bundle over the physical manifold $M=G/H$. For DFT, $H$ is a maximally isotropic subgroup of $G$ and the embedding of $H$ in $G$ is parameterized by the irrep $\mathbf{2}^{d-2}$ of the T-duality group O($d$-1,$d$-1). We show that for gEFT the subgroup $H$ is fixed by the SC irrep in table~\ref{tab:Udualitygroups}. Following \cite{Hassler:2016srl}, the data selecting $d$ physical directions out of the $\dim G$ coordinates on $G$ are encoded in a connection one-form on the principle bundle. By pulling this one-form back to the physical manifold $M$, one obtains a gauge connection. If it vanishes, which implies that the corresponding field strength is zero, the SC is solved. In this case, we show how the generalized Lie derivative \eqref{eqn:genLieintro} on $G$ is connected to GG on $M$. From the data of the $H$-principal bundle and its connection a generalized frame field $\hat E_A{}^{\hat I}$ on the generalized tangent bundle $T M \oplus \Lambda^2 T^* M$ (indices $\hat I$) is constructed. It allows us to rewrite \eqref{eqn:genLieintro}, if we restrict it to $M\subset G$, as
\begin{equation}\label{eqn:genLietwisted}
  \mathcal{L}_\xi V^{\hat{I}} = \widehat{\mathcal{L}}_\xi V^{\hat I} + \mathcal{F}_{\hat J\hat K}{}^{\hat I} \xi^{\hat J} V^{\hat K}\,,
\end{equation}
the $\mathcal{F}$-twisted generalized Lie derivative of GG. Moreover, the generalized frame field fulfills the relation
\begin{equation}\label{eqn:genLiegEFT}
  \mathcal{L}_{\hat E_A} \hat E_B = X_{AB}{}^C \hat E_C
\end{equation}
similar to \eqref{eqn:genlieongenviel}. However, we are still left with a twist term in \eqref{eqn:genLietwisted} and hence $\hat E_A$ is not equivalent to the generalized frame $\mathcal{E}_A$ in \eqref{eqn:genlieongenviel}. But it is now possible to construct $\mathcal{E}_A$. Subsequently, we use the splitting of group elements $G\ni g = m h$ ($h\in H$) induced by the $H$-principal bundle construction. Especially, each point of $M$ is in one-to-one correspondence with a $m\in G$ which gives rise to the SL(5) transformation
\begin{equation}
  M_A{}^B t_B = m^{-1} t_A m 
    \quad \text{for} \quad
  t_A \in\mathfrak{g}\,.
\end{equation}
If the structure constants $X_{AB}{}^C$ satisfy an additional linear constraint which is required for the embedding tensor solution to describe a geometric background, the generalized frame field
\begin{equation}\label{eqn:genparaframeintro}
  \mathcal{E}_A{}^{\hat I} = - M_A{}^B \begin{pmatrix}
    E_\beta{}^i & E_\beta{}^k \mathcal{C}_{kij} \\
    0 & \eta_{\delta\epsilon,\tilde\beta} E^\delta{}_i E^\epsilon{}_j
  \end{pmatrix}{}_B{}^{\hat I}
    \quad \text{with} \quad
  \mathcal{C} = \frac1{3!} \mathcal{C}_{ijk}\, d x^i \wedge d x^j \wedge d x^k = \lambda\, \mathrm{vol}
\end{equation}
represents a generalized parallelization \eqref{eqn:genLieintro} of $M$. The constant factor $\lambda$ follows directly from the embedding tensor and $\mathrm{vol}$ is the volume form on $M$ induced by the frame $E_{\alpha}{}^i$. It is the inverse transpose of
\begin{equation}
  t_{\alpha} E^\alpha{}_i = m^{-1} \partial_i m\,,
\end{equation}
where the splitting of the $\mathfrak{g}$=$\mathfrak{m}\oplus\mathfrak{h}$ generators $t_A$=$(t_\alpha,\,t_{\tilde\alpha})$ into the subalgebra $t_{\tilde\alpha}\in\mathfrak{h}$ and a complement coset part $t_\alpha\in\mathfrak{m}$ is used. For more details on the $\eta$-tensor see \eqref{eqn:etatensor} in section~\ref{sec:linkvomega}.

In general, the choice of $H$ for a given $G$ is not unique. To show that different choices result in backgrounds related by a duality transformation, we take a closer look at the duality chain for the four-torus with four-form $G$-flux in M-theory. Its extended space corresponds to the ten-dimensional group manifold CSO(1,0,4) resulting from the $\mathbf{15}$ of the embedding tensor. It is a priori not compact and requires the modding out of the discrete subgroup CSO(1,0,4,$\mathbb{Z}$) from the left. There are two choices of subgroups $H$ which reproduce the duality chain four-torus with $G$- $\leftrightarrow$ $Q$-flux \cite{Blair:2014zba}. Another T-duality transformation results in a type IIB background with $f$- $\leftrightarrow$ $R$-flux. This chain is captured by an embedding tensor solution in the $\overline{\mathbf{40}}$. It gives rise to a nine-dimensional group manifold with an unique subgroup $H$. This subgroup realizes the background with geometric flux only. We do not find a SC solution for the $R$-flux background which is in agreement with the fact that there exists no GG description for the locally non-geometric flux $R$-flux \cite{Grana:2008yw,Hull:2009sg}. As an example for a physical manifold $M$ without any non-contractible cycles, we discuss the four-sphere with $G$-flux. For all these backgrounds we construct the generalized frame $\mathcal{E}_A$.

The rest of this paper is organized into three main parts. First, we show in section~\ref{sec:gendiffonG} how to implement generalized diffeomorphisms on group manifolds. We approach this question from a slightly different perspective than \cite{Bosque:2016fpi} by keeping the discussion as general as possible and only fix specific duality groups if needed. Additionally, we emphasis the similarities and differences to \DFTwzw{}. At the same time, the notation is set and a short review of relevant results in DFT and EFT is given. Section~\ref{sec:genLie} presents the derivation of the two linear and the quadratic constraints from demanding closure of the generalized Lie derivative \eqref{eqn:genLieintro} under the SC \eqref{eqn:SCintro}. In order to solve the linear constraints, we consider the U-duality group SL(5) in section~\ref{sec:linconstsl5}. Moreover, a detailed picture of the SL(5) breaking for group manifolds with $\dim G$<$10$, originating from embedding tensor solutions in the $\overline{\mathbf{40}}$, is given. The second part is covered by section~\ref{sec:solSC}. It shows how the techniques to solve the SC in \DFTwzw{} \cite{Hassler:2016srl} carry over to gEFT. The presented SC solutions admit a description in terms of GG which is discussed in section~\ref{sec:gg}. Based on these results, section~\ref{sec:genframe} explains the construction of the generalized frame field $\mathcal{E}_A$ and the additional linear constraint it requires. Finally, the four-torus with $G$-flux, the backgrounds contained in its duality chain and the four-sphere with $G$-flux are worked out as explicit examples in section~\ref{sec:examples}. Section~\ref{sec:conclusion} concludes this work.
  
\section{Generalized Diffeomorphisms on Group Manifolds}\label{sec:gendiffonG}
Covariance with respect to diffeomorphisms plays an essential role in general relativity. In EFT, diffeomorphisms are replaced by generalized diffeomorphisms. They combine the former with gauge transformations on the physical subspace, which emerge after solving the SC. It is important to distinguish between these generalized and standard diffeomorphisms on the extended space. They are not identical, but we show in this section that they can be modified to become compatible with each other on a group manifold $G$. By compatible, we mean that the generalized Lie derivative transforms covariantly on $G$ in the sense known from general relativity. A similar approach, which works for arbitrary Riemannian manifolds, has been suggested by Cederwall for DFT \cite{Cederwall:2014kxa,Cederwall:2014opa}. Cederwall introduces a torsion free, covariant derivative with curvature to obtain a closing algebra of infinitesimal generalized diffeomorphisms. Here, we use a different approach. Our covariant derivative has both torsion and curvature. It is motivated by \DFTwzw{} \cite{Blumenhagen:2014gva,Blumenhagen:2015zma} whose gauge transformations we review in subsection~\ref{sec:dftwzwmotiv} (for a complete review see \cite{Hassler:2015pea,Blumenhagen:2017noc}). In the following subsections~\ref{sec:sectioncond} and \ref{sec:genLie}, we extend the structure from \DFTwzw{} to EFT. Doing so, we see that closure requires two linear and a quadratic constraint in addition to the SC. For the U-duality group SL(5), we present the solution to the linear constraints in subsection~\ref{sec:linconstsl5} and discuss the quadratic one in subsection~\ref{sec:quadrconstr}. For this particular U-duality group, the constraints found agree with the ones in \cite{Bosque:2016fpi}. 

\subsection{From Double Field Theory to Exceptional Field Theory}\label{sec:dftwzwmotiv}
First, we want to review the most important features of generalized and standard diffeomorphisms which we want to combine. In DFT, the infinitesimal version of the former is mediated by the generalized Lie derivative \cite{Hohm:2010pp}
\begin{equation}\label{eqn:genLieDFT}
  \mathcal{L}_\xi V^I = \xi^J \partial_J V^I + ( \partial^I \xi_J - \partial_J \xi^I ) V^J\,. 
\end{equation}
It closes according to
\begin{equation}\label{eqn:genLieDFTclosure}
  [ \mathcal{L}_{\xi_1}, \mathcal{L}_{\xi_2} ] = \mathcal{L}_{[\xi_1, \xi_2]_{\mathrm C}} \,,
\end{equation}
if the SC
\begin{equation}\label{eqn:SCDFT}
  \partial_I \cdot \partial^I \cdot = 0
\end{equation}
is fulfilled \cite{Hull:2009zb}. This constraint applies to arbitrary combinations of fields $V^I$ and parameters of gauge transformations $\xi^I$, represented by the placeholder $\cdot$\,. Furthermore, we make use of the C-bracket
\begin{equation}\label{eqn:CbracketDFT}
  [\xi_1, \xi_2]_{\mathrm C} = \frac{1}{2} ( \mathcal{L}_{\xi_1} \xi_2 - \mathcal{L}_{\xi_2} \xi_1 )
\end{equation}
in \eqref{eqn:genLieDFTclosure}. For the canonical solution of the SC where all $\cdot$ only depend on the momentum coordinates $x^i$ it reduces to the Courant bracket of GG \cite{Hull:2009zb}. As can be easily checked, the generalized Lie derivative
\begin{equation}\label{eqn:etainvgenLie}
  \mathcal{L}_\xi \, \eta_{IJ} = 0
\end{equation}
leaves the coordinate independent O($d$-1,$d$-1) metric $\eta_{IJ}$ invariant. This metric is used to raise and lower indices $I,J,K, \ldots\,$ running from $1, \ldots, 2D$. This completes the short review of the relevant DFT structures. On the other hand, infinitesimal diffeomorphisms are mediated by the Lie derivative
\begin{equation}\label{eqn:Lie}
  L_\xi V^I = \xi^J \partial_J V^I - V^J \partial_J \xi^I
\end{equation}
which closes according to
\begin{equation}
  [ L_{\xi_1}, L_{\xi_2} ] = L_{[\xi_1, \xi_2]}\,.
\end{equation}
The Lie bracket
\begin{equation}
  [\xi_1, \xi_2] = L_{\xi_1} \xi_2 = \frac{1}{2} (L_{\xi_1} \xi_2 - L_{\xi_2} \xi_1 )
\end{equation}
is defined analogous to the C-bracket in \eqref{eqn:CbracketDFT}. In contrast to generalized diffeomorphisms, neither does its closure require an additional constraint, nor is $\eta_{IJ}$ invariant under the Lie derivative.

In order to make \eqref{eqn:genLieDFT} and \eqref{eqn:Lie} compatible with each other, we require that the generalized Lie derivative transforms covariantly under standard diffeomorphisms. In this case
\begin{equation}\label{eqn:genLiecov}
  L_{\xi_1} \mathcal{L}_{\xi_2} = \mathcal{L}_{L_{\xi_1} \xi_2} + \mathcal{L}_{\xi_2} L_{\xi_1}
    \quad \text{or equivalently} \quad
  [L_{\xi_1}, \mathcal{L}_{\xi_2} ] = \mathcal{L}_{[\xi_1, \xi_2]}
\end{equation}
holds. Assume that $V^I$ and $\xi^I$ in the definition of the generalized Lie derivative transform covariantly, namely
\begin{equation}
  \delta_\lambda V^I = L_\lambda V^I
    \quad \text{and} \quad
  \delta_\lambda \xi^I = L_\lambda \xi^I\,. 
\end{equation}
Then, the partial derivative in \eqref{eqn:genLieDFT} spoil \eqref{eqn:genLiecov}. We fix this problem by replacing all partial derivatives with covariant derivatives
\begin{equation}
  \nabla_I V^J = \partial_I V^J+ \Gamma_{IL}{}^J V^L\,.
\end{equation}
In this case, we obtain the generalized Lie derivative
\begin{equation}\label{eqn:genLieDFTnabla}
  \mathcal{L}_\xi V^I = \xi^J \nabla_J V^I + ( \nabla^I \xi_J - \nabla_J \xi^I ) V^J\,.
\end{equation}
Before we study it in more detail we have to choose the connection $\Gamma$. In order to fix it, we impose some additional constraints. First of all, the covariant derivative has to be compatible with the metric $\eta_{IJ}$. Thus, it has to fulfill
\begin{equation}\label{eqn:metriccomptDFTwzw}
  \nabla_I \, \eta_{JK} = 0\,.
\end{equation}
Otherwise \eqref{eqn:etainvgenLie} would be violated. With the SC \eqref{eqn:SCDFT} expressed in terms of a covariant derivative as well, the new generalized Lie derivative still has to close. As shown in \cite{Cederwall:2014kxa}, these two constraints are sufficient to identify $\Gamma$ with the torsion-free Levi-Civita connection. From a purely symmetry oriented point of view, this is the most straightforward approach to merge generalized and standard diffeomorphisms. However, string theory on group manifolds leads to an interesting subtlety in this construction providing additional structure.

In order to explain the additional ingredient for the construction, consider a group manifold $G$ and identify $\eta_{IJ}$ with a bi-invariant metric of split signature on it. Then, $G$ is parallelizable and comes with the torsion-free (flat) derivative
\begin{equation}
  D_A = E_A{}^I \partial_I\,,
\end{equation}
where $E_A{}^I$ (generalized background vielbein) denotes the left-invariant Maurer-Cartan form on $G$. $D_A$ carries a flat index, like $A, B, C, \ldots\,$, running from one to $2D$ and is compatible with the flat metric
\begin{equation}
  \eta_{AB} = E_A{}^I \eta_{IJ} E_B{}^J \,.
\end{equation}
Its torsion
\begin{equation}\label{eqn:defFABCDFTWZW}
  [D_A, D_B] = F_{AB}{}^C D_C
\end{equation}
is given by the structure constants of the Lie algebra $\mathfrak{g}$ associated to $G$. Hence, on a group manifold it appears more natural to use the flat derivative $D_A$ instead of $\nabla_I$ with a torsion-free Levi-Civita connection. Indeed, Closed String Field Theory (CSFT) calculations for bosonic strings on $G$ suggest that we write the SC \cite{Blumenhagen:2014gva}
\begin{equation}\label{eqn:scDFTWZW}
  D_A \cdot D^A \cdot = 0
\end{equation}
with flat derivatives. In CSFT, $D_A$ has a very clear interpretation as a zero mode in the Ka\v{c}-Moody current algebra on the world sheet and the SC arises as a direct consequence of level matching. On the other hand, the generalized Lie derivative \eqref{eqn:genLieDFTnabla} does not close with only flat derivatives $D_A$ anymore. The only way out is to accept the presence of two different covariant derivatives. The flat derivatives needed for the SC and the covariant derivative $\nabla_A$ for everything else. This approach completely fixes
\begin{equation}
  \nabla_A V^B = D_A V^B - \frac{1}{3} F_{CA}{}^B V^C
\end{equation}
and reproduces exactly the results arising from CSFT \cite{Blumenhagen:2014gva}. Vectors with flat indices are formed by contracting vectors with curved indices with the generalized background vielbein, e.g. $V^A = V^I E^A{}_I$. The Christoffel symbols are obtained from the compatibility condition for the vielbein
\begin{equation}
  \nabla_A \, E_B{}^I = 0\,.
\end{equation}
In order to generalize this structure to EFT, we have to
\begin{itemize}
  \item fix the Lie algebra $\mathfrak{g}$ of the group manifold $G$ by specifying the torsion of the flat derivative
  \item fix the connection of $\nabla$ to obtain a closing generalized Lie derivative
\end{itemize}
This outlines the steps we perform in the following two subsections. Another guiding principle is that the U-duality groups in table~\ref{tab:Udualitygroups} include the T-duality groups O($d$-1,$d$-1) as subgroup. Thus, we are always able to consider the \DFTwzw{} limit to check our results.

\subsection{Section Condition}\label{sec:sectioncond}
While in DFT all indices live in the fundamental representation of the Lie algebra $\mathfrak{o}(d$-1,$d$-1), the situation in EFT is more involved. Here, we use different indices in different representations of the U-duality group. Let us start with the coordinate irrep denoted by capital letters $I, J, \dots\,$. Our main example in this paper is SL(5) EFT for which this irrep is the two index anti-symmetric $\mathbf{10}$ of $\mathfrak{sl}(5)$. Moreover, a convenient way to express the SC in a uniform manner \cite{Berman:2012vc}
\begin{equation}\label{eqn:sectioncond}
  Y^{MN}{}_{LK} \partial_M \partial_N \cdot = 0
\end{equation}
uses the invariant $Y$-tensor. It is a projector from the symmetric part of the tensor product of two coordinate irreps to the SC irrep. Both irreps are given in table~\ref{tab:Udualitygroups}. For SL(5) the SC irrep is the fundamental $\mathbf{5}$ and denoted by small lettered indices $a, b, \ldots\,$. In this particular case, the $Y$-tensor reads \cite{Musaev:2015ces}
\begin{equation}\label{eqn:SL5Y}
  Y^{MN}{}_{LK} = \frac{1}{4} \epsilon^{MNa}\epsilon_{LKa}
    \quad\text{with the normalization}\quad
  Y^{MN}{}_{MN} = 30
\end{equation}
where $\epsilon$ is the totally anti-symmetric tensor with five fundamental indices. For the SC itself the normalization can be neglected. However, it is essential if we express the generalized Lie derivative \eqref{eqn:genLieintro} in terms of the $Y$-tensor. With the flat indices defined in analogy to the curved ones, the flat derivative
\begin{equation}
  D_A = E_A{}^I \partial_I
\end{equation}
has the same form as in \DFTwzw. The generalized background vielbein $E_A{}^I$ describes a non-degenerate frame field on the group manifold and is valued in GL($n$) with $n = \dim G$. At this point it is natural to ask: What happens when the dimension of $G$ is not the same as the dimension of the coordinate irrep. We postpone the answer to section~\ref{sec:linconstsl5}. For the moment let us assume that the dimensions match. Now $n$-dimensional standard diffeomorphisms act through the Lie derivative on curved indices and the SC reads
\begin{equation}\label{eqn:flatSC}
  Y^{CD}{}_{AB} \, D_C \cdot D_D \, \cdot = 0\,.
\end{equation}
Thus, we have a situation very similar to \DFTwzw{} discussed in the last subsection.

Finally, the torsion of the flat derivative $F_{AB}{}^C$ \eqref{eqn:defFABCDFTWZW} lives in the tensor product
\begin{equation}
  \overline{\mathbf 45} \times \overline{\mathbf 10} = \mathbf{10} + \mathbf{14} + \mathbf{40} + \mathbf{175} + \mathbf{210}\,,
\end{equation}
where the $\overline{\mathbf 45}$ is the anti-symmetric part of $\mathbf{10}\times\mathbf{10}$. At the moment, this is all we know about it. Discussing the closure of the generalized Lie derivative in the next section will refine this statement.

\subsection{Generalized Lie Derivative}\label{sec:genLie}
In analogy to the SC \eqref{eqn:sectioncond}, the generalized Lie derivative of different EFTs in table~\ref{tab:Udualitygroups} can be written in the canonical form 
\begin{equation}\label{eqn:genLieYtensor}
  \mathcal{L}_\xi V^M = L_\xi V^M + Y^{MN}{}_{LK}\partial_N \xi^L V^K
\end{equation}
by using the $Y$-tensor and the standard Lie derivative on the extended space. If the SC holds, the infinitesimal generalized diffeomorphisms mediated by it close to form the algebra \cite{Berman:2012vc}
\begin{equation}
  [\mathcal{L}_{\xi_1}, \mathcal{L}_{\xi_2}] V^M= \mathcal{L}_{[\xi_1, \xi_2]_{\mathrm{E}}} V^M
    \quad \text{with} \quad
  [\xi_1, \xi_2]_{\mathrm{E}} = \frac{1}{2} \big(\mathcal{L}_{\xi_1} \xi_2 - \mathcal{L}_{\xi_1} \xi_2 \big)\,.
\end{equation}
It should be noted that this formulation includes the DFT results for $Y^{MN}{}_{LK} = \eta^{MN}\eta_{LK}$ and therefore extends naturally to EFT, e.g. for the SL(5) $Y$-tensor in \eqref{eqn:SL5Y}. Hence, it is the natural starting point for our discussion. It is instructive to keep the rough structure of the closure calculation in mind, because we have to repeat it after replacing partial derivatives with covariant ones. Evaluating
\begin{equation}\label{eqn:closureterm}
  [\mathcal{L}_{\xi_1}, \mathcal{L}_{\xi_2}] V^M - \mathcal{L}_{[\xi_1, \xi_2]_{\mathrm{E}}} V^M\,,
\end{equation}
one is left with sixteen different terms. All of them contain two partial derivatives. But only in four terms the partial derivatives act on the same variable. Because the $Y$-tensor has the properties
\begin{align}
  & \delta_F^{(B} Y^{AC)}{}_{DE} - Y^{(AC}{}_{FG} Y^{B)G}{}_{DE} = 0 \quad \text{and} \nonumber \\
  & \delta_{(F}^{B} Y^{AC}{}_{DE)} - Y^{AC}{}_{G(F} Y^{BG}{}_{DE)} = 0 \label{eqn:Yzero}
\end{align}
for $d\le 6$ only terms which are annihilated by the SC remain. For the other U-duality groups with $d>6$ the closure calculation gets more involved \cite{Berman:2012vc}. Here, we are interested in a proof of concept. Thus, we focus on the simplest cases and postpone the rest to future work. For scalars the generalized and standard Lie derivative coincide
\begin{equation}
  \mathcal{L}_\xi s = L_\xi s\,.
\end{equation}
Applying the Leibniz rule we obtain the action of generalized diffeomorphisms on one-forms
\begin{equation}
  \mathcal{L}_\xi V_M = L_\xi V_M - Y^{PQ}{}_{NM} \partial_Q \xi^N V_P\,,
\end{equation}
the objects dual to the vector representation. Finally, we remember that $Y^{MN}{}_{PQ}$ has to be an invariant tensor
\begin{equation}
  \mathcal{L}_\xi Y^{MN}{}_{PQ} = 0
\end{equation}
in the same fashion as $\eta_{IJ}$ is in DFT. This completes the list of properties for the EFT generalized Lie derivative we require to make it compatible with standard diffeomorphisms.

As a first step into this direction, we switch to flat indices and replace all partial derivatives in \eqref{eqn:genLieYtensor} by covariant ones to find
\begin{equation}\label{eqn:genLieCov}
  \mathcal{L}_\xi V^A = \xi^B \nabla_B V^A - V^B \nabla_B \xi^A + Y^{AB}{}_{CD} \nabla_B \xi^C V^D\,.
\end{equation}
This expression can be rewritten in terms of flat derivatives
\begin{equation}
\label{eqn:covderiv}
  \nabla_A V^B = D_A V^B + \Gamma_{AC}{}^B V^C \quad \text{and} \quad
  \nabla_A V_B = D_A V_B - \Gamma_{AB}{}^C V_C
\end{equation}
by introducing the spin connection $\Gamma_{AB}{}^C$. Expanding the generalized Lie derivative yields
\begin{align}
  \mathcal{L}_\xi V^A &= \xi^B D_B V^A - V^B D_B \xi^A + Y^{AB}{}_{CD} D_B \xi^C V^D + X_{BC}{}^A \xi^B V^C \quad \text{and} \nonumber \\
  \mathcal{L}_\xi V_A &= \xi^B D_B V_A + V_B D_A \xi^B - Y^{CD}{}_{BA} D_D \xi^B V_C - X_{BA}{}^C \xi^B V_C \label{eqn:genLieVA}
\end{align}
with
\begin{equation}\label{eqn:XfromGamma}
  X_{AB}{}^C = 2 \Gamma_{[AB]}{}^C + Y^{CD}{}_{BE} \, \Gamma_{DA}{}^E
\end{equation}
collecting all terms depending on the spin connection. We will see later that $X_{AB}{}^C$ is closely related to the embedding tensor of gauged supergravities. Under the modified generalized Lie derivative the $Y$-tensor should still be invariant, which translates to the first linear constraint
\begin{equation*}\label{eqn:linconst1}\tag{C1}
  \nabla_C Y^{AB}{}_{DE} := C_1^{AB}{}_{CDE} = 2 Y^{F(A}{}_{DE} \Gamma_{CF}{}^{B)} - 2 Y^{AB}{}_{(D|F} \Gamma_{C|E)}{}^F = 0
\end{equation*}
on the spin connection $\Gamma$ after imposing $D_A Y^{BC}{}_{DE} = 0$. It is a direct generalization of the metric compatibility \eqref{eqn:metriccomptDFTwzw} in \DFTwzw{}.

Now, we demand closure of the modified generalized Lie derivative. Equivalently, all terms \eqref{eqn:closureterm} which spoil the closure have to vanish. Let us start with the ones containing no flat derivatives. They only vanish if the quadratic constraint
\begin{equation}\label{eqn:quadconstr}
  X_{BE}{}^A X_{CD}{}^E - X_{BD}{}^E X_{CE}{}^A + X_{[CB]}{}^E X_{ED}{}^A = 0
\end{equation}
is fulfilled. In order to analyze it, we decompose $X_{AB}{}^C$ into a symmetric part $Z^C{}_{AB}$ and an anti-symmetric one
\begin{equation}
  X_{AB}{}^C = Z^C{}_{AB} + X_{[AB]}{}^C\,.
\end{equation}
Moreover, all terms with only one flat derivative acting on $V^A$ in \eqref{eqn:closureterm} vanish, if we identify the torsion of the flat derivative with
\begin{equation}
\label{eqn:commutator}
  [D_A, D_B] = X_{[AB]}{}^C D_C\,.
\end{equation}
Note that we have used $D_A \,X_{BC}{}^D = 0$ and $Y^{AB}{}_{BC} = Y^{(AB)}{}_{(BC)}$, which holds only for $d\le 6$, in all calculations. For a consistent theory, it is essential that the Bianchi identify
\begin{equation}\label{eqn:BianchiFlatD}
  [D_A, [D_B, D_C]] + [D_C, [D_A, D_B]] + [D_B, [D_C, D_A]] = 0
\end{equation}
is fulfilled. Evaluating the commutator above, we find that this constraint is equivalent to the Jacobi identity
\begin{equation}\label{eqn:jacobiident}
  \Big(X_{[AB]}{}^E X_{[CE]}{}^D + X_{[CA]}{}^E X_{[BE]}{}^D + X_{[BC]}{}^E X_{[AE]}{}^D\Big) D_D = 0\,.
\end{equation}
But after antisymmetrizing \eqref{eqn:quadconstr} with respect to $B,C,D$, we obtain
\begin{equation}
  X_{[BC]}{}^E X_{[CE]}{}^A + X_{[DB]}{}^E X_{[CE]}{}^D + X_{[CD]}{}^E X_{[BE]}{}^D = - Z^A{}_{E[B} X_{CD]}{}^E
\end{equation}
instead of zero. Hence, we are left with
\begin{equation}\label{eqn:betterzero}
  Z^A{}_{E[B} \, X_{CD]}{}^E D_A = 0
\end{equation}
which in general does not vanish. For \DFTwzw\, $Z^A{}_{BC}$ vanishes and this problem does not occur. Thus, it is special to gEFT. As we show in section~\ref{sec:linconstsl5}, it is solved by reducing the dimension of the group manifold representing the extended space.

An important property of the generalized Lie derivative is that the Jacobiator of its E-bracket only vanishes up to trivial gauge transformations. Let us take a closer look at these transformations
\begin{equation}\label{eqn:trivialgaugetr}
  \xi^A = Y^{AB}{}_{CD} D_B \chi^{CD}
\end{equation}
in the context of our modified generalized Lie derivative. Ultimately, this will help us to better organize terms in the closure calculation with one flat derivative acting either on $\xi_1$ or $\xi_2$. Inserting \eqref{eqn:trivialgaugetr} into the generalized Lie derivative \eqref{eqn:genLieVA}, we obtain
\begin{equation}
  \mathcal{L}_{\xi} V^A = C_{2a}{}^{AB}{}_{CDE} D_B \chi^{CD} V^E + \dots = 0
\end{equation}
where $\dots$ denotes terms which vanish under the SC and due to the properties of the $Y$-tensor \eqref{eqn:Yzero}. The tensor 
\begin{equation*}\label{eqn:linconst2a}\tag{C2a}
  C_{2a}^{AB}{}_{CDE} = Y^{BF}{}_{CD} X_{FE}{}^A + \frac{1}{2} Y^{AF}{}_{CD} X_{[FE]}{}^B + \frac{1}{2} Y^{AF}{}_{EH} Y^{GH}{}_{CD} X_{[FG]}{}^B
\end{equation*}
has to vanish if trivial gauge transformations have the form \eqref{eqn:trivialgaugetr}.

Terms with two derivatives in \eqref{eqn:closureterm} vanish under the SC and due to \eqref{eqn:Yzero}. All we are left with are terms with one flat derivative acting on the gauge parameters $\xi_1$ or $\xi_2$. Because \eqref{eqn:closureterm} is anti-symmetric with respect to the gauge parameters, it is sufficient to check whether all $D_A \xi_1^B$ contributions vanish. We write them in terms of the tensor
\begin{align*}
  C_{2b}^{AB}{}_{CDE} &= Z^A{}_{DC} \delta_E^B - Z^B{}_{DE} \delta_C^A - Y^{BF}{}_{EC} Z^A{}_{DF} + Y^{AB}{}_{CF}{} Z^F{}_{DE} \nonumber \\
  &+ Y^{AB}{}_{EF} X_{[DC]}{}^F + Y^{AB}{}_{CF} X_{[DE]}{}^F - 2 Y^{F(A}{}_{EC} X_{[DF]}{}^{B)} = 0 \tag{C2b}
\end{align*}
as
\begin{equation}
  \Big(-\frac{1}{2} C_{2a}^{AB}{}_{CDE} + C_{2b}^{AB}{}_{CDE} \Big) \, D_B \xi_1^C \xi_2^D V^E = 0\,,
\end{equation}
which has a contribution from trivial gauge transformations \eqref{eqn:linconst2a} as well. This is reasonable since the E-bracket closes up to trivial gauge transformations. However, in general there is no reason why the two contributions have to vanish independently. Only the second linear constraint
\begin{equation*}\label{eqn:linconst2}\tag{C2}
  -\frac{1}{2} C_{2a}^{AB}{}_{CDE} + C_{2b}^{AB}{}_{CDE} = 0
\end{equation*}
has to hold in conjunction with the first linear constraint \eqref{eqn:linconst1} and the quadratic constraint \eqref{eqn:quadconstr} for closure of generalized diffeomorphisms under the SC. Thus, one has to restrict the connection $\Gamma_{AB}{}^C$ in such a way that all these three constraints are fulfilled. This is exactly what we do in the next two subsections.

Without too much effort we can already perform a first check of our results at this point. To this end, consider the O($d$-1,$d$-1) T-duality group with
\begin{equation}
  Y^{AB}{}_{CD} = \eta^{AB} \, \eta_{CD}\,, \quad
    \Gamma_{AB}{}^C = \frac{1}{3} F_{AB}{}^C
    \quad \text{and} \quad
    X_{AB}{}^C = F_{AB}{}^C\,.
\end{equation}
In this case, we have
\begin{align}
  C_1^{AB}{}_{CDE}&= \frac23 \eta_DE F_C{}^{(AB)} - \frac23 \eta^{AB} F_{C(DE)} = 0 \\
  C_{2a}^{AB}{}_{CDE}&= \eta_{CD} ( F^B{}E{}^A + F^A{}_E{}^B ) = 0 \\
  C_{2b}^{AB}{}_{CDE}&= \eta^{AB} ( F_{DCE} + F_{DEC} ) - 2 \eta_{EC} F_D{}^{(AB)} = 0
\end{align}
due to the total antisymmetry of the structure constants $F_{ABC}$. Hence, this short calculation is in agreement with the closure of the gauge algebra of \DFTwzw{} presented in \cite{Blumenhagen:2014gva}.

\subsection{Linear Constraints}\label{sec:linconstsl5}
Solving the linear constraints for gEFT is more involved than for \DFTwzw{}, which we presented as a simple example at the end of the last subsection. It requires more sophisticated tools from representation theory. Especially, we need to obtain projectors which filter out certain irreps from tensor products of the coordinate irrep in table~\ref{tab:Udualitygroups}. Here, our initial choice of SL(5) as the duality group pays off. Irreps (or more precisely projectors onto them) of SL($n$) and their tensor products can be conveniently organized in terms of young tableaux making their representation theory very traceable. All required techniques are reviewed in appendix~\ref{app:SLnrepresentations}. As an explicit example, the T-duality subgroup SL(4) is discussed there. Its Lie algebra $\mathfrak{sl}(4)$ is isomorphic to $\mathfrak{so}(3,3)$. Hence, we already know the solutions to the linear constraints, which allows us to check the machinery developed in the appendix.

We start our discussion with the spin connection $\Gamma_{AB}{}^C$. The indices are in the $\mathbf{10}$ and $\overline{\mathbf{10}}$ of $\mathfrak{sl}(5)$. We express said indices through the fundamental $\mathbf{5}$ indices and lower the raised indices with the totally anti-symmetric tensor to obtain
\begin{equation}
  \Gamma_{a_1 a_2, b_1 b_2, c_1 c_2 c_3} = \Gamma_{a_1 a_2, b_1 b_2}{}^{d_1 d_2} \epsilon_{d_1 d_2 c_1 c_2 c_3}\,.
\end{equation}
In this form, it is manifest that the 1000 independent components of the connection are embedded in the tensor product
\begin{equation}\label{eqn:decomp101010b}
  \mathbf{10} \times \mathbf{10} \times \overline{\mathbf 10} = 3 (\mathbf{10}) + \mathbf{15} + 2 (\mathbf{40}) + 2 (\mathbf{175}) + \mathbf{210} + \mathbf{315}
\end{equation}
which we translate into corresponding Young diagrams
\begin{equation}
  \ydiagram{1,1}\times\left(\,\ydiagram{1,1}\times\ydiagram{1,1,1}\,\right) = 3\, \ydiagram{2,2,1,1,1} + \ydiagram{3,1,1,1,1} + 2\,\ydiagram{2,2,2,1} + 2 \,\ydiagram{3,2,1,1} + \ydiagram{3,2,2} + \ydiagram{3,3,1}\,.
\end{equation}
This decomposition looks similar to \eqref{eqn:P666} in appendix~\ref{app:SLnrepresentations}. However, the $\mathbf{10}$ of $\mathfrak{sl}(5)$ is not self dual as the $\mathbf{6}$ of $\mathfrak{sl}(4)$. Thus, we pick up an additional box in the last irrep on the left hand side. Each of these diagrams is associated with a projector. Because some irreps appear more than once in the decomposition of the tensor product \eqref{eqn:decomp101010b}, we label them as
\begin{equation}
  \mathbf{10} \times ( \mathbf{10}\times \overline{\mathbf 10} ) = \mathbf{10} \times (\mathbf{1} + \mathbf{24} + \mathbf{75}) = \left\{ \begin{array}{ll}
    \mathbf{10} \times \mathbf{1} &= \mathbf{10}a \\
    \mathbf{10} \times \mathbf{24} &= \mathbf{10}b + \mathbf{15} + \mathbf{40}a + \mathbf{175}a \\
    \mathbf{10} \times \mathbf{75} &= \mathbf{10}c + \mathbf{40}b + \mathbf{210} + \mathbf{315}
  \end{array}\right.
\end{equation}
in order to clearly distinguish all projectors.

While the first linear constraint \eqref{eqn:linconst1} is straightforward to solve for $\mathfrak{sl}(4)$, things now become more involved. First, note that the constraint acts trivially on the index $C$. We suppress this index and rewrite \eqref{eqn:linconst1}
\begin{equation}\label{eqn:C1sl5}
  C_1{}_{a_1 a_2 a_3, b_1 b_2 b_3, d_1 d_2, e_1 e_2} = \sigma_1 \Gamma_{a_1 a_2, a_3 b_1 b_2} \epsilon_{b_3 d_1 d_2 e_1 e_2} = 0
\end{equation}
in terms of the permutations
\begin{gather}
	\sigma_1 = (6\,5\,4\,3) + (3\,5\,2\,4\,1) - (6\,5\,4\,3\,2) - (3\,5\,6\,2\,4\,1) +	(6\,5\,4\,3\,2\,1) - (6\,10\,2\,7\,4\,8\,5\,9\,1) - \nonumber \\
	(6\,10\,5\,9\,4\,8\,2\,7\,1) + (6\,10\,2\,3\,7\,4\,8\,5\,9\,1) + (6\,10\,5\,9\,4\,8\,2\,3\,7\,1) + (3\,5\,1) (4\,6\,2) \nonumber \\- (3\,7\,1)(6\,10\,5\,9\,4\,8,2) - (3\,7\,4\,8\,5\,9\,1)(6\,10\,2)
\end{gather}
which act on the ten remaining indices. This form allows for the constraint to be solved by linear algebra techniques. To this end, we choose the explicit basis 
\begin{align}\label{eqn:basis1010b}
  (d_1 d_2),\, (e_1 e_2) \in V_{\mathbf{10}} &= \big\{ (d_1 d_2) \,|\, d_1,\,d_2 \in \{1\, \dots\, 5\} \wedge d_1 < d_2 \big\} \nonumber \\
  (a_1 a_2 a_3),\, (b_1 b_2 b_3) \in V_{\overline{\mathbf 10}} &= \big\{ (a_1 a_2 a_3) \,|\, a_1,\,a_2,\,a_3 \in \{1\, \dots\, 5\} \wedge a_1 < a_2 < a_3 \big\}
\end{align}
for the irreps $\mathbf{10}$ and $\overline{\mathbf 10}$ appearing in \eqref{eqn:C1sl5}. Keeping the properties of the totally anti-symmetric tensor in mind, we interpret $\sigma_1$ as a linear map from $\Gamma$ to $C_1$
\begin{equation}
  \sigma_1 : V_{\mathbf{10}} \times V_{\overline{\mathbf 10}} \rightarrow V_{\overline{\mathbf 10}} \times V_{\overline{\mathbf 10}} \times  V_{\mathbf{10}} \times V_{\mathbf{10}}\,.
\end{equation}
Solutions of the first linear constraint are in the kernel of this map and can be associated to the projectors of the irreps we have discussed above. An explicit calculation shows that
\begin{equation}
  \sigma_1 ( P_{\mathbf{1}} + P_{\mathbf{24}} ) = 0 \quad \text{but} \quad
  \sigma_1 P_{\mathbf{75}} \ne 0
\end{equation}
holds. Hence, the most general solution can be written in terms of the projector
\begin{equation}
  P_1 = P_{\mathbf{10}a} + P_{\mathbf{10}b} + P_{\mathbf{15}} + P_{\mathbf{40}a} + P_{\mathbf{175}a}\,.
\end{equation}

Next, we have to check which of these irreps survive the transition from the connection $\Gamma_{AB}{}^C$ to $X_{AB}{}^C$. In analogy to \eqref{eqn:C1sl5}, we express \eqref{eqn:XfromGamma} in terms of permutations
\begin{equation}\label{eqn:simgaX}
	\sigma_X = () -(3\,1)(4\,2)+(3\,5\,1)(4\,6\,2)-(3\,5\,1)(4\,6\,7\,2) + (3\,5\,7\,2\,4\,6\,1)
\end{equation}
through
\begin{equation}
  X_{a_1 a_2, b_1 b_2, c_1 c_2 c_3} = \sigma_X P_1 \Gamma_{a_1 a_2, b_1 b_2, c_1 c_2 c_3}\,.
\end{equation}
Note that the first linear constraint is already implemented in this equation by the projector $P_1$. Again, we apply the techniques presented in appendix \ref{app:SLnrepresentations} to decompose
\begin{equation}
  \sigma_X P_1 P_{\mathbf{10}\times\mathbf{10}\times\overline{\mathbf 10}} = \frac{12}5 P_{\mathbf{10}ab} + P_{\mathbf{10}c} + 4 P_{\mathbf{15}} + 3 P_{\mathbf{40}a}
\end{equation}
into orthogonal projectors on different $\mathfrak{sl}(5)$ irreps where $P_{\mathbf{10}ab}$ is defined as
\begin{equation}
  P_{\mathbf{10}ab} = \frac5{12} (P_{\mathbf{10}a}-P_{\mathbf{10}b}) \sigma_X (P_{\mathbf{10}a} + P_{\mathbf{10}b})\,.
\end{equation}
It embeds just another ten-dimensional irrep $\mathbf{10}ab$ into $\mathbf{10}a$ and $\mathbf{10}b$. In the following, we only focus on the $\mathbf{15}$ and $\mathbf{40}$. These are exactly the irreps which survive the linear constraint on the embedding tensor in seven-dimensional maximal gauged supergravities\footnote{In \cite{Samtleben:2005bp} a tree index tensor $Z^{ab,c}$ represents the $\overline{\mathbf{40}}$. Here we use the dual version. Both are connected by \eqref{eqn:40bto40} and capture the same data.} \cite{Samtleben:2005bp}. As presented in appendix A of \cite{Berman:2012uy}, the remaining two ten-dimensional irreps can be combined to one $\mathbf{10}$ capturing trombone gaugings. Since we are only interested in a proof of concept, we do not discuss trombone gaugings. They are considered in \cite{Bosque:2016fpi} which takes the embedding tensor irreps $\mathbf{10}+\mathbf{15}+\overline{\mathbf{40}}$ as starting point. We a priori did not restrict the allowed groups $G$. But trying to implement generalized diffeomorphisms on them exactly reproduces the correct irreps of the embedding tensor. Originally, they arise from supersymmetry conditions \cite{deWit:2002vt}. Here, we do not make any direct reference to supersymmetry. Hence, it is remarkable that we still reproduce this result.

Now, let us discuss the remaining linear constraint \eqref{eqn:linconst2}. We proceed in the same fashion as for the first one and write
\begin{equation}\label{eqn:C2sigma2}
  C_2{}_{a_1 a_2 a_3, b_1 b_2 b_3, c_1 c_2, d_1 d_2, e_1 e_2} = \sigma_2 X_{a_1 a_2, a_3 b_1, b_2 b_3 c_1} \epsilon_{c_2 d_1 d_2 e_1 e_2}
\end{equation}
in terms of a sum of permutations denoted as $\sigma_2$ which is of a similar form as \eqref{eqn:simgaX} but contains 54 different terms. Thus, we do not write it down explicitly. In the basis \eqref{eqn:basis1010b}, $\sigma_2$ gives rise to the linear map
\begin{equation}
  \sigma_2 : V_{\mathbf{10}} \times V_{\mathbf{10}} \times V_{\overline{\mathbf 10}} \rightarrow V_{\overline{\mathbf 10}} \times V_{\overline{\mathbf 10}} \times  V_{\mathbf{10}} \times V_{\mathbf{10}} \times V_{\mathbf{10}}
\end{equation}
whose kernel contains the $\mathbf{15}$, but not the $\mathbf{40}a$. However, from maximal gauged supergravities in seven dimensions \cite{Samtleben:2005bp}, we know that also gaugings in the dual $\overline{\mathbf{40}}$ are consistent. How do we resolve this contradiction? First, we implement the components of this irrep in terms of the tensor $Z^{ab,c}$ and connect it to the $\mathbf{40}a$, we discussed above, through
\begin{equation}\label{eqn:40bto40}
  (X_{\mathbf{40a}})_{a_1 a_2, b_1 b_2, c_1 c_2 c_3} = \epsilon_{a_1 a_2 d_1 d_2 [b_1} Z^{d_1 d_2, e_1} \epsilon_{b_2] c_1 c_2 c_3 e_1} 
\end{equation}
with the expected property
\begin{equation}
  P_{\mathbf{40a}} (X_{\mathbf{40a}})_{a_1 a_2, b_1 b_2, c_1 c_2 c_3} =
    (X_{\mathbf{40a}})_{a_1 a_2, b_1 b_2, c_1 c_2 c_3}\,.
\end{equation}
Following the argumentation in \cite{Samtleben:2005bp}, we interpret $Z^{ab,c}$ as a 10$\times$5 matrix and calculate its rank
\begin{equation}
  s = \mathrm{rank} ( Z^{ab,c} )\,.
\end{equation}%
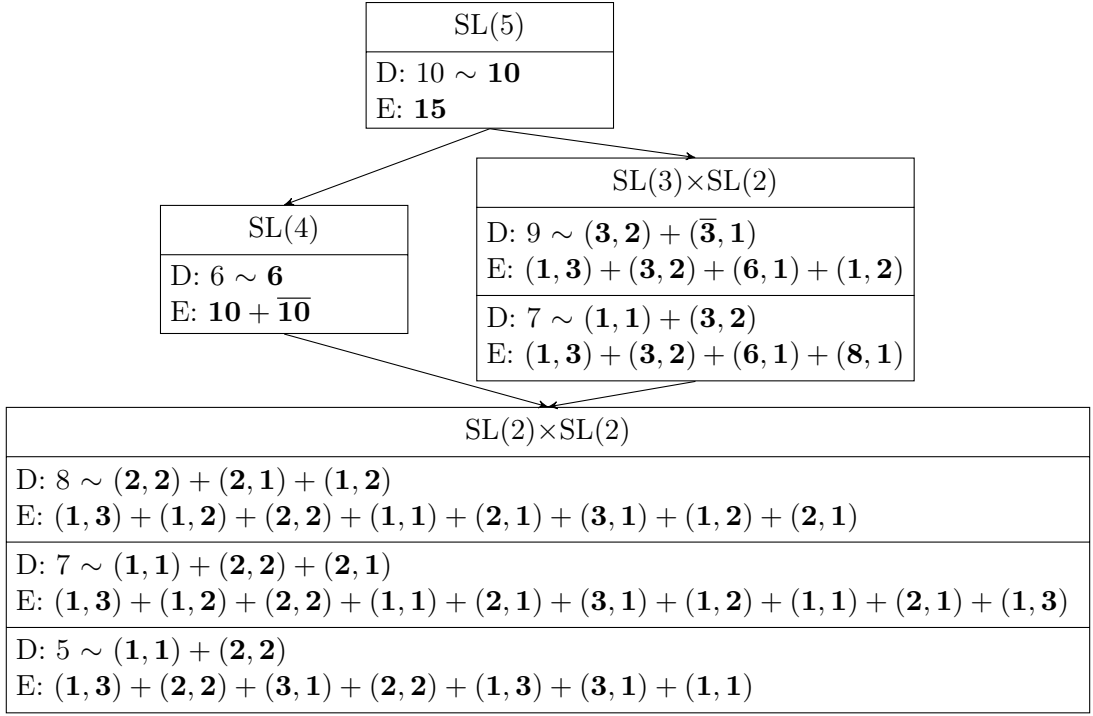
\begin{figure}
\centering\begin{tikzpicture}[every text node part/.style={align=center}, node distance=10em]
	\node[rectangle split, rectangle split parts=2, 
  	draw, minimum width=3cm, text width=3cm] (SL5)
		{ SL(5)
    	\nodepart{two}
     		D: 10 $\sim$  $\mathbf{10}$ \\
     		E: $\mathbf{15}$ };
	\node[rectangle split, rectangle split parts=2, 
  	draw, minimum width=3cm, text width=3cm, below left of=SL5] (SL4)
		{ SL(4)
    	\nodepart{two}
     		D: 6 $\sim$ $\mathbf{6}$ \\
     		E: $\mathbf{10} + \overline{\mathbf{10}}$ };
	\node[rectangle split, rectangle split parts=3, 
  	draw, minimum width=5.5cm, text width=5.5cm, below right of=SL5] (SL2SL3)
		{ SL(3)$\times$SL(2)
    	\nodepart{two}
     		D: 9 $\sim$ $(\mathbf{3}, \mathbf{2}) + (\overline{\mathbf 3}, \mathbf{1})$  \\
     		E: $(\mathbf{1},\mathbf{3}) + (\mathbf{3},\mathbf{2}) + (\mathbf{6},\mathbf{1}) + (\mathbf{1},\mathbf{2})$
    	\nodepart{three}
     		D: 7 $\sim$ $(\mathbf{1}, \mathbf{1}) + (\mathbf{3}, \mathbf{2})$ \\
     		E: $(\mathbf{1},\mathbf{3}) + (\mathbf{3},\mathbf{2}) + (\mathbf{6},\mathbf{1}) + (\mathbf{8},\mathbf{1})$ };
	\node[rectangle split, rectangle split parts=4, 
  	draw, minimum width=8.5cm, text width=14cm, below left of=SL2SL3, xshift=2em, yshift=-3em] (SL2SL2)
		{ SL(2)$\times$SL(2)
    	\nodepart{two}
        D: 8 $\sim$ $(\mathbf{2}, \mathbf{2}) + (\mathbf{2}, \mathbf{1}) + (\mathbf{1}, \mathbf{2})$  \\
     		E: $(\mathbf{1}, \mathbf{3}) + (\mathbf{1}, \mathbf{2}) + (\mathbf{2}, \mathbf{2}) + (\mathbf{1}, \mathbf{1}) + (\mathbf{2}, \mathbf{1}) + (\mathbf{3}, \mathbf{1}) + (\mathbf{1}, \mathbf{2}) + (\mathbf{2}, \mathbf{1})$
    	\nodepart{three}
        D: 7 $\sim$ $(\mathbf{1}, \mathbf{1}) + (\mathbf{2}, \mathbf{2}) + (\mathbf{2}, \mathbf{1})$  \\
     		E: $(\mathbf{1}, \mathbf{3}) + (\mathbf{1}, \mathbf{2}) + (\mathbf{2}, \mathbf{2}) + (\mathbf{1}, \mathbf{1}) + (\mathbf{2}, \mathbf{1}) + (\mathbf{3}, \mathbf{1}) + (\mathbf{1}, \mathbf{2}) + (\mathbf{1}, \mathbf{1}) + (\mathbf{2}, \mathbf{1}) + (\mathbf{1}, \mathbf{3})$ 
     	\nodepart{four}
        D: 5 $\sim$ $(\mathbf{1}, \mathbf{1}) + (\mathbf{2}, \mathbf{2})$  \\
     		E: $(\mathbf{1}, \mathbf{3}) + (\mathbf{2}, \mathbf{2}) + (\mathbf{3}, \mathbf{1}) + (\mathbf{2}, \mathbf{2}) + (\mathbf{1}, \mathbf{3}) + (\mathbf{3}, \mathbf{1}) + (\mathbf{1}, \mathbf{1})$ };
  \draw[->] (SL5.south) -- (SL4.north);
  \draw[->] (SL5.south) -- (SL2SL3.north);
  \draw[->] (SL4.south) -- (SL2SL2.north);
  \draw[->] (SL2SL3.south) -- (SL2SL2.north);
\end{tikzpicture}
\caption{Solutions of the linear constraints \eqref{eqn:linconst1} and \eqref{eqn:linconst2}. ``D:'' lists the dimension of the group manifold and the corresponding coordinate irreps. All components of the embedding tensor which are in the kernel of the linear constraints are denoted by ``E:''.}\label{fig:sollinconst}
\end{figure}%
The number of massless vector multiplets in the resulting seven-dimensional gauged supergravity is given by $10-s$. They contain the gauge bosons of the theory and transform in the adjoint representation of the gauge group $G$. Thus, we immediately deduce
\begin{equation}\label{eqn:dimG40}
  \mathrm{dim} \, G = 10 - s\,.
\end{equation}
In \DFTwzw{} the gauge group of the gauged supergravity which arises after a Scherk Schwarz compactification is in one-to-one correspondence with the group manifold we are considering \cite{Bosque:2015jda}. There is no reason why it should be different for gEFT. So, if we switch on gaugings in the $\overline{\mathbf 40}$, we automatically reduce the dimension of the group manifold representing the extended space. Possible ranks $s$ which are compatible with the quadratic constraint of the embedding tensor are $0 \le s\le 5$. For those cases we have to adapt the coordinates on the group manifold. To this end, we consider possible branching rules of SL(5) to its U-/T-duality subgroups given in table~\ref{tab:Udualitygroups}, e.g. SL(4), SL(3)$\times$SL(2) and SL(2)$\times$SL(2)
\begin{align}
  \mathbf{10} &\rightarrow \mathbf{4} + \mathbf{6} \label{eqn:10branching1}\\
  \mathbf{10} &\rightarrow (\mathbf{1}, \mathbf{1}) + (\mathbf{3}, \mathbf{2}) + (\overline{\mathbf 3}, \mathbf{1}) \label{eqn:10branching2}\\
  \mathbf{10} &\rightarrow (\mathbf{1}, \mathbf{1}) + (\mathbf{1}, \mathbf{1}) + (\mathbf{2}, \mathbf{1}) + (\mathbf{1}, \mathbf{2}) + (\mathbf{2}, \mathbf{2}) \label{eqn:10branching3}\,.
\end{align}
For the first one, we obtain a six-dimensional manifold whose coordinates are identified with the $\mathbf{6}$ of the branching rule \eqref{eqn:10branching1} after dropping the $\mathbf{4}$. In the adapted basis
\begin{align}\label{eqn:V4&V6}
  V_{\mathbf{4}}&= \{15,\,25,\,35,\,45\} & V_{\mathbf{6}} &= \{12,\,13,\,14,\,23,\,24,\,34\} \\
  V_{\overline{\mathbf 4}}&= \{234,\,134,\,124,\,123\} & V_{\overline{\mathbf 6}} &= \{345,\,245,\,235,\,145,\,135,\,125\}\,,
\end{align}
$\sigma_2$ is now restricted to
\begin{equation}\label{eqn:SL4sigma2}
  \sigma_2 : V_{\mathbf{6}} \times V_{\mathbf{6}} \times V_{\overline{\mathbf 6}} \rightarrow V_{\overline{\mathbf 6}} \times V_{\overline{\mathbf 6}} \times  V_{\mathbf{6}} \times V_{\mathbf{6}} \times V_{\mathbf{6}}\,,
\end{equation}
while the irreps $\mathbf{15}$ and $\mathbf{40}$ split into
\begin{align}
  \mathbf{15} &\rightarrow \xcancel{\mathbf{1}} + \xcancel{\mathbf{4}} + \mathbf{10} \\
  \mathbf{40} &\rightarrow \xcancel{\overline{\mathbf 4}} + \mathbf{6} + \overline{\mathbf 10} + \xcancel{\mathbf{20}}\,.
\end{align}
All crossed out irreps at least partially depend on $V_4$ or its dual which is not available as coordinate irrep anymore. Of course, the $\mathbf{10}$ from the $\mathbf{15}$ still satisfies all linear constraints. But now only the $\mathbf{6}$ is excluded by the second linear constraint \eqref{eqn:C2sigma2} with \eqref{eqn:SL4sigma2}, while the $\overline{\mathbf 10}$ is in its kernel. This result is in alignment with the SL(4) case we discuss in appendix~\ref{app:SLnrepresentations}. Hence, switching on specific gaugings in the $\mathbf{40}$ breaks indeed the U-duality group into a subgroup. An alternative approach \cite{Bosque:2016fpi} is to keep the full SL(5) covariance of the embedding tensors by not solving the linear constraints. However this technique obscures the interpretation of the extended space as a group manifold which is crucial for constructing the generalized frame $\mathcal{E}_A$ in the next section. Furthermore, breaking symmetries by non-trivial background expectation values for fluxes is a well-known paradigm. Only a tours with no fluxes has the maximal number of abelian isometries and should allow the full U-duality group. In case we restrict ourselves to a T-duality subgroup to solve the linear constrains, all \DFTwzw{} results are naturally embedded in the EFT formalism. For the remaining branchings \eqref{eqn:10branching2} and \eqref{eqn:10branching3}, we perform the same analysis in appendix~\ref{app:linconst2additional}. All results are summarized in figure~\ref{fig:sollinconst}.

\subsection{Quadratic Constraint}\label{sec:quadrconstr}
Finally, we come to the quadratic constraint \eqref{eqn:quadconstr} which simplifies drastically to
\begin{equation}
  X_{[BC]}{}^E X_{[ED]}{}^A + X_{[DB]}{}^E X_{[EC]}{}^{A} + X_{[CD]}{}^E X_{[EB]}{}^A = 0
\end{equation}
after solving the linear constraints which result in $Z^C{}_{AB} = 0$ for the remaining coordinates on the group manifold $G$. Now, it is identical to the Jacobi identity \eqref{eqn:jacobiident} which is automatically fulfilled for the Lie algebra $\mathfrak{g}$. Thus, the flat derivative satisfies the first Bianchi identity \eqref{eqn:BianchiFlatD}. For the covariant derivative \eqref{eqn:covderiv}, we compute the curvature and the torsion by evaluating the commutator
\begin{equation}
\left[ \nabla_A, \nabla_B \right] V_C = R_{ABC}{}^D V_D - T_{AB}{}^D \nabla_D V_C\,.
\end{equation}
Doing so, we obtain the curvature
\begin{equation}
R_{ABC}{}^D =  2 \Gamma_{[A|C}{}^E \, \Gamma_{|B]E}{}^D + X_{[AB]}{}^E \, \Gamma_{EC}{}^D\,,
\end{equation}
where we used that $\Gamma_{AB}{}^C$ is constant due to \eqref{eqn:XfromGamma} and the torsion
\begin{equation}
T_{AB}{}^C = - X_{[AB]}{}^C + 2 \Gamma_{[AB]}{}^C = Y^{CD}{}_{[A|E} \, \Gamma_{D|B]}{}^E\,,
\end{equation}
for which we used \eqref{eqn:XfromGamma} and \eqref{eqn:commutator}. In general both are non-vanishing. Using these equations, we can compute the first Bianchi identity
\begin{align}
  R_{[ABC]}{}^D + & \nabla_{[A} T_{BC]}{}^D - T_{[AB}{}^E T_{C]E}{}^D = \nonumber \\
  & 2 X_{[AB]}{}^E \, X_{[CE]}{}^D + 2 X_{[CA]}{}^E X_{[BE]}{}^D + 2 X_{[BC]}{}^E X_{[AE]}{}^D = 0
\end{align}
for $\nabla$. Again, it is fulfilled because of the Jacobi identity \eqref{eqn:jacobiident}. These results are in agreement with \DFTwzw{}. It is straightforward to check that all gaugings, given in table 3 of \cite{Samtleben:2005bp}, can be reproduced in the framework we presented in the first part of this paper. Explicit examples with with ten-dimensional groups CSO(1,0,4)/SO(5) and also a nine-dimensional group are discussed in section~\ref{sec:examples}.

\section{Section Condition Solutions}\label{sec:solSC}
So far, we implemented generalized diffeomorphisms on group manifolds $G$ which admit an embedding in one of the $U$-duality groups with $d\le 4$ in table~\ref{tab:Udualitygroups}. Still, they only close into a consistent gauge algebra if the SC \eqref{eqn:flatSC} is fulfilled. Hence, understanding the solutions of this constraint is very important and objective of this section. In the following, we adapt a technique for finding the most general SC solutions in \DFTwzw{} \cite{Hassler:2016srl} to our gEFT setup. It is based on a $H$-principle bundle over the physical subspace $M$=$G/H$. $H$ is a ($\dim G$-$\dim M$)-dimensional subgroup of $G$ with special properties which are explained in section~\ref{sec:HPrincipal}. As before, the construction follows closely the steps required in \DFTwzw{} and introduces generalizations whenever needed. We show in section~\ref{sec:gg} that each SC solution gives rise to a GG on $M$ which has two constituents: a twisted generalized Lie derivative and a generalized frame field. Both act on the generalized tangent bundle $T M \oplus \Lambda^2 T^* M$. Sometimes, the choice of the subgroup $H$ is not unique for a given $G$. Different subgroups result in dual backgrounds. Section~\ref{sec:dualbg} presents a way to systematically study these different possibilities. It works exactly as in \cite{Hassler:2016srl}, so we keep the discussion brief. Starting from the SC solution, we construct the generalized frame field $\mathcal{E}_A$ fulfilling \eqref{eqn:genlieongenviel} in section~\ref{sec:genframe}. We also introduce the additional constraint on the structure constants $X_{AB}{}^C$ which is required for this construction to work.

\subsection{Reformulation as \texorpdfstring{$H$}{H}-Principal Bundle}\label{sec:HPrincipal}
Following the discussion in \cite{Hassler:2016srl}, we first substitute the quadratic version \eqref{eqn:flatSC} of the SC by the equivalent linear constraint \cite{Berman:2012vc}
\begin{equation}\label{eqn:sclinear}
  v_a \, \epsilon^{aBC} D_B \, \cdot = 0
\end{equation}
which involves a vector field $v_a$ in the fundamental (SC irrep) of SL(5). This field can take different values on each point of $G$. In order to relate different points, remember that translations on $G$ are generated by the Lie algebra $\mathfrak{g}$. Especially, we are interested in the action of its generators in the representations
\begin{equation}\label{eqn:genfund&adj}
  \mathbf{5}: \quad (t_A)_b{}^c = X_{A,b}{}^c \quad \text{and} \quad \mathbf{10}: \quad (t_A)_B{}^C = X_{AB}{}^C = 2 X_{A, [b_1}{}^{[c_1} \delta_{b_2]}{}^{c_2]} = 2 (t_A)_{[b_1}{}^{[c_1} \delta_{b_2]}{}^{c_2]}\,.
\end{equation}
Both are captured by the embedding tensor. The corresponding group elements arise after applying the exponential map. Now, assume we have a set of fields $f_i$ with a coordinate dependence such that they solve the linear constraint \eqref{eqn:sclinear} for a specific choice of $v_a$. Then, there exists another set of fields $f_i'$ with a different coordinate dependence
\begin{equation}
  D_A f_i' = (\mathrm{Ad}_g)_A{}^B D_B f_i
    \quad \text{and} \quad
 (\mathrm{Ad}_g)_A{}^B t_B = g\, t_A g^{-1}
\end{equation}
which solve the linear constraint after transforming $v_a$ according to
\begin{equation}\label{eqn:shiftvag}
  v'_a = (g)_a{}^b v_b\,.
\end{equation}
Here, $(g)_a{}^b$ represents the left action of a group element $g$ on the vector $v_b$. This property of the linear constraint \eqref{eqn:sclinear} is due to the fact that a totally anti-symmetric tensor $\epsilon$ is SL$(5)$ invariant.

The situation is very similar to the one in \DFTwzw{}. Only the groups and their representations are different. A minor deviation from \cite{Hassler:2016srl} is the splitting of the $\mathbf{10}$ indices into two sets of subindices. In order to implement the section condition, we introduce a vector $v_a^0$ which gives rise to
\begin{equation}\label{eqn:sclinearv0}
  v_a^0 \epsilon^{a\beta C} t_\beta = 0 \quad \text{and} \quad
  v_a^0 \epsilon^{a\tilde\beta C} t_{\tilde\beta} \ne 0\,.
\end{equation}
It splits the generators $t_A$ of $\mathfrak{g}$
\begin{equation}\label{eqn:splittingtA}
  t_A = \begin{pmatrix} t_\alpha & t_{\tilde\alpha} \end{pmatrix} 
    \quad \text{and} \quad
  t_{\alpha} \in \mathfrak{m}\,,\quad t_{\tilde \alpha} \in \mathfrak{h}
\end{equation}
into a subalgebra $\mathfrak{h}$ and the complement $\mathfrak{m}$ with $\alpha$=1,\,\ldots,\,$\dim G/H$ and $\tilde\alpha$=$\tilde 1$,\,\dots,\,$\dim H$. We make this decomposition of $\mathfrak{g}$ manifest bysplitting the $\mathbf{10}$ index $A$ into two non-intersecting subindices $\alpha, \tilde{\alpha}$. The generators $t_{\tilde\alpha}$ generate the stabilizer subgroup $H \subset G$. Its elements leave $v^0_a$ invariant under the transformation \eqref{eqn:shiftvag}. This suggests to decompose each group element $g\in G$ into
\begin{equation}
  g = m h \quad \text{with} \quad h \in H
\end{equation}
while $m$ is a coset representative of the left coset $G/H$. Because the action of $h$ is free and transitive, we can interpret $G$ as a $H$-principal bundle
\begin{equation}\label{eqn:Hprincipal}
  \pi: G \rightarrow G/H = M
\end{equation}
over $M$, the physical manifold.

We now study this bundle in more detail. The discussion is closely related to the one in \cite{Hassler:2016srl}. So we keep it short, but still complete. A group element $g\in G$ is parameterized by the coordinates $X^I$. In order to implement the splitting \eqref{eqn:splittingtA}, we assign to the coset representative $m$ (generated by $t_\alpha$) the coordinates $x^i$ and to the elements $h\in H$ (generated by $t_{\tilde \alpha}$) the coordinates $x^{\tilde i}$. Doing so, results in
\begin{equation}\label{eqn:splittingcoord}
  X^I = \begin{pmatrix} x^i & x^{\tilde i} \end{pmatrix}
    \quad \text{with} \quad
  I = 1,\,\dots,\,\mathrm{dim}\,G\,, \quad
  i = 1,\,\dots,\,\mathrm{dim}\,G/H \quad \text{and} \quad
  \tilde i = \tilde 1,\,\dots,\,\mathrm{dim}\, H \,.
\end{equation}
In these adapted coordinates $\pi$ acts by removing the $x^{\tilde i}$ part of the $X^I$,
\begin{equation}\label{eqn:piX}
  \pi(X^I) = x^i\,.
\end{equation}
We also note that the corresponding differential map reads
\begin{equation}\label{eqn:pi*X}
  \pi_*( V^I \partial_I ) = V^i \partial_i\,.
\end{equation}
Each element of the Lie algebra $\mathfrak{g}$ generates a fundamental vector field on $G$. If we want to relate the two of them, we need to introduce the map
\begin{equation}\label{eqn:sharp}
  t_A^\sharp = E_A{}^I \partial_I
\end{equation}
which assigns a left-invariant vector field to each $t_A\in\mathfrak{g}$. It has the important property $\omega_L ( t_A^\sharp ) = t_A$ where
\begin{equation}\label{eqn:leftinvMCF}
  (\omega_L)_g = g^{-1} \partial_I g \, d X^I = t_A E^A{}_I d X^I
\end{equation}
is the left-invariant Maurer-Cartan form on $G$. Both \eqref{eqn:leftinvMCF} and \eqref{eqn:sharp} are completely fixed by the generalized background vielbein $E_A{}^I$ and its inverse transposed $E^A{}_I$. After taking into account the splitting of the generators \eqref{eqn:splittingtA} and the coordinates \eqref{eqn:splittingcoord}, they read
\begin{equation}\label{eqn:compbgvielbein}
  E^A{}_I = \begin{pmatrix} E^\alpha{}_i & 0 \\
    E^{\tilde\alpha}{}_i & E^{\tilde\alpha}{}_{\tilde i}
  \end{pmatrix} \quad \text{and} \quad
  E_A{}^I = \begin{pmatrix} E_\alpha{}^i & E_\alpha{}^{\tilde i} \\
    0 & E_{\tilde\alpha}{}^{\tilde i}
  \end{pmatrix}\,.
\end{equation}

We further equip the principal bundle with the $\mathfrak{h}$-valued connection one-form $\omega$. It splits the tangent bundle $T G$ into a horizontal/vertical bundle $H G$/$V G$. While the horizontal part
\begin{equation}
  H G = \{ X \in T G \,|\, \omega(X) = 0 \}
\end{equation}
follows directly from the connection one-form, the vertical one is defined as the kernel of the differential map $\pi_*$. We have to impose the two consistency conditions
\begin{equation}\label{eqn:constrconnection}
  \omega( t_{\tilde\alpha}^\sharp ) = t_{\tilde\alpha} \quad \text{and} \quad
  R_h^* \omega = Ad_{h^{-1}} \omega
\end{equation}
on $\omega$, where $R_g$ denotes right translations on $G$ by the group element $g\in G$. In analogy to \DFTwzw{} the connection one-form is chosen such that the bundle $H G$ solves the linear version \eqref{eqn:sclinear} of the SC. Following \cite{Hassler:2016srl}, we introduce the projector $P_m$ at each point $m$ of the coset space $G/H$ as a map
\begin{equation}
  P_m: \mathfrak{g} \rightarrow \mathfrak{h}, \quad P_m = t_{\tilde\alpha} (P_m)^{\tilde\alpha}{}_B \theta^B
\end{equation}
where we denote the dual one-form of the generator $t_A$ as $\theta^A$. $P_m$ is not completely arbitrary. It has to have the property
\begin{equation}\label{eqn:Pmprop}
  P_m t_{\tilde\alpha} = t_{\tilde\alpha}\, \quad \forall t_{\tilde\alpha} \in \mathfrak{h}\,.
\end{equation}
So far, this projector is only defined for coset representatives $m$ not for arbitrary group elements $g$. But, we can extend it to the full group manifold $G$ by
\begin{equation}\label{eqn:Pg}
  P_g = P_{m h} = \mathrm{Ad}_{h^{-1}} P_m \mathrm{Ad}_h\,.
\end{equation}
This allows us to derive the connection-one form
\begin{equation}\label{eqn:omegag}
  \omega_g = P_g \, (\omega_L)_g
\end{equation}
where $(\omega_L)_g$ is the left-invariant Maurer-Cartan from \eqref{eqn:leftinvMCF}.  As a result of \eqref{eqn:Pmprop}, it satisfies the constraints in \eqref{eqn:constrconnection}.

Finally, the $H$-principal bundle \eqref{eqn:Hprincipal} has sections $\sigma_i$ which are only defined in the patches $U_i \subset M$. They have the form
\begin{equation}\label{eqn:sigmai}
  \sigma_i (x^j) = \begin{pmatrix} \delta^j_k x^k & f_i^{\tilde j} \end{pmatrix}
\end{equation}
in the coordinates \eqref{eqn:splittingcoord} and are specified by the functions $f^{\tilde j}_i$. As for \DFTwzw, we choose those functions such that the pull back of the connection one-form $A_i = {\sigma_i}^* \omega$ vanishes in every patch $U_i$ \cite{Hassler:2016srl}. This is only possible if the corresponding field strength
\begin{equation}\label{eqn:F_i=0}
  F_i (X,Y) = d A_i (X,Y) + [A_i(X), A_i(Y)] = 0
\end{equation}
vanishes. In this case, $A_i$ is a pure gauge and can be locally ``gauged away''. It is very important to keep in mind that this field strength is not the one that describes the tangent bundle $T M$. Take for example the four sphere $S^4 \cong$SO(5)/SO(4). It is not parallelizable and thus its tangent bundle cannot be trivial. However, this has nothing to do with the field strength defined in \eqref{eqn:F_i=0}.

\subsection{Connection and Three-Form Potential}\label{sec:linkvomega}
For \DFTwzw{} the projector $P_m$ is related to the NS/NS two-form field $B_{ij}$. In the following we show that this result generalizes to the three-form $C_{ijk}$ for SL(5) gEFT. To this end, we study solutions to the linear version of the SC \eqref{eqn:sclinearv0} in more detail. By an appropriate SL(5) rotation, it is always possible to bring $v^0_a$ into the canonical form 
\begin{equation}\label{eqn:v0acanonical}
  v^0_a = \begin{pmatrix} 1 & 0 & 0 & 0 & 0 \end{pmatrix}\,.
\end{equation}
This allows us to fix an explicit basis 
\begin{equation}\label{eqn:decompSL510}
  \alpha = \{12,\,13,\,14,\,15\} \quad \text{and} \quad
  \tilde\alpha = \{23,\,24,\,25,\,34,\,35,\,45\}
\end{equation}
for the indices appearing in our construction. Furthermore, we introduce the tensor
\begin{equation}\label{eqn:etatensor}
  \eta^{\alpha\beta,\tilde\gamma} = \frac12 \epsilon^{1 \hat\alpha \hat\beta \tilde\gamma}
\end{equation}
where $\hat\beta$ labels the second fundamental index of the anti-symmetric pair (e.g. $\beta=13$ and $\hat\beta=3$). The lowered version of $\eta$ is defined in the same way
\begin{equation}
  \eta_{\alpha\beta,\tilde\gamma} = \epsilon_{1 \hat\alpha \hat\beta \tilde\gamma}
\end{equation}
and its normalization is chosen such that the relations
\begin{equation}\label{eqn:etasl5props}
  \eta^{\alpha\beta, \tilde\alpha} \eta_{\alpha\beta, \tilde\beta} = \delta^{\tilde\alpha}_{\tilde\beta}
    \quad \text{and} \quad
  \eta^{\alpha\beta, \tilde\alpha} \eta_{\gamma\delta, \tilde\alpha} = \delta^{[\alpha}_{[\gamma} \delta^{\beta]}_{\delta]}
\end{equation}
are satisfied. Using this tensor, we express the projector
\begin{equation}\label{eqn:PmC}
  (P_m)^{\tilde\alpha}{}_B = \begin{pmatrix} \eta^{\gamma\delta,\tilde\alpha} C_{\beta\gamma\delta} & \delta^{\tilde\alpha}_{\tilde\beta} \end{pmatrix}
\end{equation}
in terms of the totally anti-symmetric field $C_{\alpha\beta\gamma}$ on $M$. As we will see by considering the SC solution's GG in the next section, this field is related to the three-from flux
\begin{equation}\label{eqn:CfromC}
  C = \frac16 C_{\alpha\beta\gamma} E^\alpha{}_i E^\beta{}_j E^\gamma{}_k\, d x^i \wedge d x^j \wedge d x^k
\end{equation}
on the background. Remember that the projector \eqref{eqn:PmC} is chosen such that its kernel contains all the solutions of the linear SC \eqref{eqn:sclinear} for a fixed $v_a$. It is straightforward to identify
\begin{equation}\label{eqn:v->C}
  C_{\alpha\beta\gamma} = \frac{1}{v_1} \sum_{\delta} \epsilon_{1 \hat\alpha\hat\beta\hat\gamma\hat\delta} v_{\hat\delta}\,.
\end{equation}
However, this equation is only defined for $v_1 \ne 0$. Because \eqref{eqn:sclinear} is invariant under rescaling all values of $v_a$ specifying a distinct solution of the section condition are elements of $\mathbb{RP}^4$. This projective space has five patches $U_a=\{ v_a \in \mathbb{R}^5 | v_a = 1\}$ in homogeneous coordinates. From \eqref{eqn:v->C}, we see that the projector and therewith the connection only covers the subset $U_1$ for possible solutions of the section condition. As explained in the last subsection, a SC solution is characterized by the vanishing connection $A_i$. In this case, we can use \eqref{eqn:PmC} and \eqref{eqn:CfromC} to calculate the three-from flux
\begin{equation}\label{eqn:CfromE}
  C = - \frac16 \eta_{\alpha\beta,\tilde\gamma} E^\alpha{}_i E^\beta{}_j E^{\tilde\gamma}{}_k \, d x^i \wedge d x^j \wedge d x^k
\end{equation}
which appears in the GG of the theory.

Again, it is a convenient crosscheck to consider the symmetry breaking from SL(5) to SL(4) which we discussed in section~\ref{sec:linconstsl5}. Now, the index of $v_a$ runs only from $a=1,\dots,4$ and the linear constraint reads
\begin{equation}\label{eqn:SL4SC}
  v_a^0 \epsilon^{a\beta c} = 0
\end{equation}
with a four-dimensional totally anti-symmetric tensor $\epsilon$ and the explicit basis
\begin{equation}
  \alpha = \{12,\,13,\,14\}
    \quad \text{and} \quad
  \tilde\alpha = \{23,\,24,\,34\}\,,
\end{equation}
if we take $v^0_a = \begin{pmatrix} 1 & 0 & 0 & 0\end{pmatrix}$. At this point, we have to restrict $C$ from our previous discussion to the two-from
\begin{equation}
  C_{\alpha\beta4} = B_{\alpha\beta}
\end{equation}
in order the describe SC solutions with $v_5=0$. Applying this restriction to \eqref{eqn:PmC} and \eqref{eqn:v->C} gives rise to
\begin{equation}
  (P_m)^{\tilde \alpha}{}_B = \begin{pmatrix}\eta^{\gamma,\tilde\alpha} B_{\beta \gamma} & \delta^{\tilde\alpha}_{\tilde\beta} \end{pmatrix}
    \quad \text{and} \quad
  B_{\alpha\beta} = \frac{1}{v_1} \sum\limits_{\gamma} \epsilon_{1\hat\alpha\hat\beta\hat\gamma} v_{\hat\gamma}
\end{equation}
with
\begin{align}\label{eqn:etasl4}
  \eta^{\alpha,\tilde\beta} = \epsilon^{\alpha \tilde\beta}
    \quad &\text{and} \quad
  \eta_{\alpha,\tilde\beta} = \epsilon_{\alpha \tilde\beta}\,.
\intertext{Here the normalization for the $\eta$-tensor is chosen such that the analog relations}
  \eta^{\alpha,\tilde\alpha} \eta_{\alpha,\tilde\beta} = \delta^{\tilde\alpha}_{\tilde\beta}
    \quad &\text{and} \quad
  \eta^{\alpha,\tilde\alpha} \eta_{\beta, \tilde\alpha} = \delta^\alpha_\beta
\end{align}
to \eqref{eqn:etasl5props} hold. Furthermore, the same comments apply as above, but this time for $\mathbb{RP}^3$ instead of $\mathbb{RP}^4$. These results are in agreement with the ones for \DFTwzw{} in \cite{Hassler:2016srl}. Especially, the $\eta$-tensor gives rise the O(3,3) invariant metric
\begin{equation}
  \eta^{AB} = \epsilon^{AB} = \begin{pmatrix} 0 & \eta^{\alpha,\tilde\beta} \\
    \eta^{\beta,\tilde\alpha} & 0 \end{pmatrix}
\end{equation}
with indices $A$, $B$ in the coordinate irrep $\mathbf{6}$ of $\mathfrak{sl}(4)$. The only difference to \cite{Hassler:2016srl} is that we use a different basis for the Lie algebra in which the off-diagonal blocks $\eta^{\alpha,\tilde \beta}$ and $\eta^{\beta,\tilde \alpha}$ are not diagonal.

In general, it can get quite challenging to find the vanishing connection $A_i$=0 which is required to solve the SC. However, if $\mathfrak{m}$ and $\mathfrak{h}$ in the decomposition \eqref{eqn:splittingtA} form a symmetric pair with the defining property
\begin{equation}\label{eqn:symmetricpair}
  [\mathfrak{h},\mathfrak{h}] \subset \mathfrak{h} \,, \quad
  [\mathfrak{h},\mathfrak{m}] \subset \mathfrak{m} \quad \text{and} \quad
  [\mathfrak{m},\mathfrak{m}] \subset \mathfrak{h}\,,
\end{equation}
there is an explicit construction. It was worked out for \DFTwzw{} in \cite{Hassler:2016srl} and we adapt it to gEFT in the following. Starting point is the observation that the connection $A$ vanishes if 
\begin{equation}
  C_{ijk} = - \eta_{\alpha\beta,\tilde\gamma} E^\alpha{}_i E^\beta{}_j E^{\tilde\gamma}{}_k
\end{equation}
is totally anti-symmetric in the indices $i$, $j$, $k$. We rewrite this condition as
\begin{equation}\label{eqn:Ctotantisym}
  2 C_{ijk} - C_{kij} - C_{jki} = D_{ijk} = 0
\end{equation}
and study it further. To this end, it is convenient to introduce the notation 
\begin{equation}\label{eqn:def(...)}
  ( t_A , t_B , t_C ) = 2 \eta_{\alpha\beta,\tilde\gamma} - \eta_{\gamma\alpha,\tilde\beta} 
    - \eta_{\beta\gamma,\tilde\alpha}
\end{equation}
which allows us to express \eqref{eqn:Ctotantisym} as
\begin{equation}
  D_{ijk} = ( m^{-1} \partial_i m,\, m^{-1} \partial_j m,\, m^{-1} \partial_k m )
\end{equation}
after taking into account that $E^\alpha{}_i$ and $E^{\tilde\alpha}{}_i$ are certain components of the left-invariant Maurer-Cartan form \eqref{eqn:leftinvMCF} with a section where $h$ is the identity element of $H$. Following \cite{Hassler:2016srl}, we use the coset representative
\begin{equation}
  m = \exp\left(- f( x^i )\right)
\end{equation}
which gives rise to the expansion
\begin{equation}
  m^{-1} \partial_i m = \sum\limits_{n=0}^\infty \frac1{(n+1)!} [f, \partial_i f]_n
    \quad \text{with} \quad
  [f,t]_n = [\underbrace{f [\dots, [f}_{n\text{ times}},t] \dots ]]\,.
\end{equation}
Thus, we are left with checking that
\begin{equation}\label{eqn:hastobezero}
  D_{ijk} = \sum_{n_1=0}^\infty \, \sum_{n_2=0}^\infty \, \sum_{n_3=0}^\infty \frac{1}{
    (n_1 + 1)! (n_2 + 1)! (n_3 + 1)!} ([f,\partial_i f]_{n_1}, [f,\partial_j f]_{n_2}, 
    [f,\partial_k f]_{n_3} )
\end{equation}
is zero under the restriction \eqref{eqn:symmetricpair}. To do so, let us first simplify the notation by the abbreviation
\begin{equation}
  \langle n_1, n_2, n_3 \rangle_{ijk} := ([f,\partial_i f]_{n_1}, [f,\partial_j f]_{n_2}, [f,\partial_k f]_{n_3} )
\end{equation}
and rearrange the terms in \eqref{eqn:hastobezero} which results in
\begin{equation}
\label{eqn:Dijk}
  D_{ijk} = \sum\limits_{m=0}^\infty \, \sum\limits_{n_1+n_2+n_3=m} \,
    \frac{\langle n_1, n_2, n_3 \rangle_{ijk}}{(n_1 + 1)! (n_2 + 1)! (n_3 + 1)!}
    = \sum\limits_{m=0}^\infty S^m_{ijk}\,.
\end{equation}
This expression is zero if $S^m_{ijk}$ vanishes for all $m$. Therefore, it permits to do the calculation order by order. Let us start with
\begin{equation}
  S^0_{ijk} = \langle 0, 0, 0\rangle_{ijk} = 0\,.
\end{equation}
It vanishes because $(t_A, t_B, t_C)$ only gives a contribution if two of its arguments are in $\mathfrak{m}$ and one is in $\mathfrak{h}$ as it is obvious from the definition \eqref{eqn:def(...)}. Here all arguments are in $\mathfrak{m}$. The next order gives rise to
\begin{equation}
  S^m_{ijk} = \frac1{2!} \left( \langle 1, 0, 0 \rangle_{ijk} + \langle 0, 1, 0 \rangle_{ijk} + 
    \langle 0, 0, 1 \rangle_{ijk} \right) = 0
\end{equation}
and implements a linear constraint on the structure constants $X_{AB}{}^C$. It is equivalent to
\begin{equation}\label{eqn:linconstflatconn}
  ( [t, \mathfrak{m}], \mathfrak{m}, \mathfrak{m} ) + 
  ( \mathfrak{m}, [t, \mathfrak{m}], \mathfrak{m} ) + 
  ( \mathfrak{m}, \mathfrak{m}, [t, \mathfrak{m}] ) = 0
\end{equation}
where $t$ denotes a generator in the algebra $\mathfrak{sl}(5)$. Its components furnish the adjoint irrep $\mathbf{24}$. Note that the splitting of the flat coordinate indices $A$ into $\alpha$ and $\tilde\alpha$ singles out the direction $v_a^0$ in \eqref{eqn:v0acanonical}. Thus, it break SL(5) to SL(4) with the branching
\begin{equation}\label{eqn:sollinconstflatconn}
  \mathbf{24} \rightarrow \xcancel{\mathbf{1}} + \mathbf{4} + \overline{\mathbf{4}} + \mathbf{15}
\end{equation}
of the adjoint irrep. There is only one generator, corresponding to the crossed out irrep, which violates \eqref{eqn:linconstflatconn}. In quadratic order, we find
\begin{equation}
  S^2_{ijk} = \frac14 \left( \langle 1, 1, 0 \rangle_{ijk} + \langle 0, 1,1 \rangle_{ijk} 
    + \langle 1, 0, 1 \rangle_{ijk} \right) + \frac16 \left( \langle 2, 0, 0 \rangle_{ijk} + 
    \langle 0, 2, 0 \rangle_{ijk} + \langle 0, 0, 2 \rangle_{ijk} \right) = 0
\end{equation}
which represents a quadratic constraint on the structure constants. A solution is given by the symmetric pair \eqref{eqn:symmetricpair}. It implies that the first three terms are of the form $( \mathfrak{h}, \mathfrak{h}, \mathfrak{m})$ plus cyclic permutations, while the last three terms are covered by $(\mathfrak{m}, \mathfrak{m}, \mathfrak{m})$. As noticed before, all of them vanish independently. More generally, we now have
\begin{equation}\label{eqn:[f,df]_nevenodd}
  [f, \partial_i f]_n \subset \begin{cases} \mathfrak{h} & n \text{ odd} \\ \mathfrak{m} & n \text{ even}
  \end{cases}
\end{equation}
which implies 
\begin{equation}
  \langle n_1, n_2, n_3 \rangle_{ijk} = 0 \quad \text{if} \quad
  n_1 \,\mathrm{mod}\, 2 + n_2 \,\mathrm{mod}\, 2 + n_3 \,\mathrm{mod}\, 2 = 1\,.
\end{equation}
Take any contribution $\langle n_1, n_2, n_3 \rangle_{ijk}$ to $S^m{}_{ijk}$ in \eqref{eqn:Dijk} which is governed by $n_1 + n_2 + n_3 = m$. If $m$ is even then either two of the integers $n_1$, $n_2$, $n_3$ are odd while the third one is even, or they are all even. In both cases $\langle n_1, n_2, n_3 \rangle_{ijk}$ vanishes and so does the complete $S_m$ for even $m$. In combination with \eqref{eqn:[f,df]_nevenodd}, \eqref{eqn:linconstflatconn} becomes
\begin{equation}\label{eqn:shuffle<...>1}
  \langle n_1 + 1, n_2, n_3 \rangle_{ijk} + \langle n_1, n_2 + 1, n_3 \rangle_{ijk}  + 
    \langle n_1, n_2, n_3 + 1 \rangle_{ijk} = 0 \quad \text{for } n_1, n_2, n_3 \text{ even}\,.
\end{equation}
We use this identity to simplify the cubic contribution
\begin{equation}
  S^3_{ijk} = - \frac1{4!} \left( \langle 3, 0, 0\rangle_{ijk} + \langle 0, 3, 0\rangle_{ijk} +
    \langle 0, 0, 3\rangle_{ijk}\right) = 0
\end{equation}
which is equivalent to \eqref{eqn:shuffle<...>1} after substituting $1$ with $3$. Repeating this procedure again and again for $S^m_{ijk}$ with odd $m$, we finally obtain the conditions
\begin{equation}\label{eqn:shuffle<...>2l+1}
  \langle n_1 + 2 l + 1, n_2, n_3 \rangle_{ijk} + 
    \langle n_1 , n_2 + 2 l + 1, n_3 \rangle_{ijk} + \langle n_1, n_2, n_3 + 2 l + 1 \rangle_{ijk} = 0
    \quad \forall \,\, l \in \mathbb{N}
\end{equation}
(again with $n_1$, $n_2$, $n_3$ even) for the desired result \eqref{eqn:Ctotantisym} which proves $A_i$=0. Proving them, requires a generalization of \eqref{eqn:linconstflatconn} and exploits that the generator $t$ in this equation is an element of $\mathfrak{m}$. As a consequence, the commutator relations of the symmetric pair \eqref{eqn:symmetricpair} restrict $t$ to the $\mathbf{4}$ and $\overline{\mathbf{4}}$ in the decomposition \eqref{eqn:sollinconstflatconn}. So we see that \eqref{eqn:linconstflatconn} is not an independent constraint, but follows directly from having a symmetric pair. Denoting the two remaining, dual irreps as $x_i$ and $y^i$, where $i=1,\,\dots,\,4$\,, the relation
\begin{equation}
  [t, \partial_i t]_{2l + 1} = [t, \partial_i t]_1 \left( \frac{x_i y^i}4 \right)^l
\end{equation}
holds. It reduces \eqref{eqn:shuffle<...>2l+1} to \eqref{eqn:shuffle<...>1} and completes the prove. Finally, note that there is another case
\begin{equation}
  [ \mathfrak{m}, \mathfrak{m} ] \subset \mathfrak{m}
\end{equation}
for which one immediately has a flat connection. It implies that all terms in \eqref{eqn:Dijk} are of the form $(\mathfrak{m},\mathfrak{m},\mathfrak{m})$ and vanish.

\subsection{Generalized Geometry}\label{sec:gg}
All solutions of the SC which we discussed in the last two subsections are closely related to GG. In order to make this connection manifest, we have to introduce a map between $\mathfrak{h}$ and the vector space of two-forms $\Lambda^2 T_p^* M$ at each point $p\in M$. More specifically, we use the $\eta$-tensor \eqref{eqn:etatensor} to define the bijective map $\eta_p: \mathfrak{h} \rightarrow \Lambda^2 T_p^* M$ as
\begin{equation}\label{eqn:etamap}
  \eta_p( t_{\tilde\gamma} ) = \frac12 \left. \eta_{\alpha\beta,\tilde\gamma} E^{\alpha}{}_i E^{\beta}{}_j dx^i \wedge dx^j\right|_{\sigma(p)} \,.
\end{equation}
Its inverse
\begin{equation}
  \eta_p^{-1}( \nu ) = \left. \eta^{\alpha\beta, \tilde\gamma} t_{\tilde\gamma} \iota_{E_\alpha} \iota_{E_\beta} \nu \right|_{\sigma(p)}
\end{equation}
follows form the properties of the $\eta$-tensors and the vectors $E_a = E_a{}^i \partial_i$. With this map and $\pi_*$ \eqref{eqn:pi*X}, $\omega_g(X)$ \eqref{eqn:omegag} from section~\ref{sec:HPrincipal}, we are able to construct the generalized frame field \cite{Hull:2007zu,Coimbra:2011ky,Lee:2014mla}
\begin{equation}
  \hat E_A{}(p) = \pi_{*\,p} (t_A^\sharp) + \eta_p\,\omega_{\sigma(p)} ( t_A^\sharp)
\end{equation}
at each point $p$ of the physical space $M$. It is a map from a Lie algebra element $t_A$ to a vector in the generalized tangent space $T_p M \oplus \Lambda^2 T_p^* M$ of $M$ at $p$. Note that we suppress the index labeling the patch dependence of the section for the sake of brevity. However, the generalized frame field $\hat E_A$ depends explicitly on the section. For a non-trivial $H$-principal bundle, we find different frame fields in each patch and have to introduce transition functions accordingly.

Using the properties of the maps
\begin{equation}
  \pi_*(t_{\tilde\alpha}^\sharp) = 0 \,, \quad
  \omega \sigma_* = \sigma^* \omega = A = 0\,, \quad
  \pi_* \sigma_*  = \mathrm{id}_{T M} \quad \text{and} \quad
  \omega(t_{\tilde\alpha}^\sharp) = t_{\tilde\alpha}\,,
\end{equation}
we deduce the dual frame
\begin{equation}\label{eqn:vielbeinhatinv}
  \hat E^A{} (p, v, \tilde v) = \theta^A \Big( \eta_p^{-1} (\tilde v) + \iota_{\sigma_{*\,p}(v)} \,(\omega_L)_{\sigma(p)} \Big) \,.
\end{equation}
Here, we denote elements of the generalized tangent bundle as $V = v + \tilde v$ with $v \in T M$ and $\tilde v \in \Lambda^2 T^* M$. Finally, let us expand the generalized frame and its dual into components
\begin{equation}\label{eqn:genframe&dualform}
  \hat E_A = \begin{pmatrix}
    E_{\alpha}{}^i \partial_i + C_{\alpha\beta\gamma} E^{\beta}{}_i E^{\gamma}{}_j \, d x^i \wedge d x^j \\
     \eta_{\beta\gamma, \tilde\alpha} E^{\beta}{}_i E^{\gamma}{}_j \, d x^i\wedge d x^j \end{pmatrix}
    \quad \text{and} \quad
    \hat E^A (v,\tilde v) = \begin{pmatrix} E^{\alpha}{}_i v^i \\
      \eta^{\beta\gamma,\tilde\alpha} ( E_{\beta}{}^i E_{\gamma}{}^j \tilde v_{ij}
      - C_{\beta\gamma\delta} E^{\delta}{}_i v^i )
    \end{pmatrix}
\end{equation}
where the dependence on $p$ is understood and the indices labeling the patch are suppressed. In the calculation for the dual frame, one has to take into account
\begin{equation}\label{eqn:Cterm}
  \theta^{\tilde\alpha} \Big( \omega_L ( \sigma_* v ) \Big) = - C_{\beta\gamma\delta} \eta^{\gamma\delta,\tilde\alpha} E^{\beta}{}_i v^i
\end{equation}
which results from $\sigma^* \omega = 0$. This result makes perfect sense, because it reproduces the canonical vielbein of a SL(5) theory \cite{Hull:2007zu}
\begin{equation}
  \mathcal{V}_{\hat A}{}^{\hat I} = \begin{pmatrix} E_\alpha{}^i & E_\alpha{}^k C_{ijk} \\
    0 & E^\alpha{}_{[i} E^\beta{}_{j]}
  \end{pmatrix}
\end{equation}
and its inverse transposed. The $C_{ijk}$ in this expression is connected to the one we are using in \eqref{eqn:Cterm} by $C_{ijk} = C_{\alpha\beta\gamma} E^\alpha{}_i E^\beta{}_j E^\gamma{}_k$.

With the generalized frame and its inverse fixed, we are able to transport the generalized Lie derivative \eqref{eqn:genLieCov} to the generalized tangent bundle with the elements
\begin{equation}\label{eqn:VhatI}
  V^{\hat I} = \begin{pmatrix} v^i & \tilde v_{ij} \end{pmatrix} = V^A \hat E_A{}^{\hat I}
    \quad \text{and the dual} \quad
  V_{\hat I} = \begin{pmatrix} v_i & \tilde v^{ij} \end{pmatrix} V_A \hat E^A{}_{\hat I} \,.
\end{equation}
We distinguish the tangent bundle of the group manifold from the generalized tangent bundle by using hatted indices for the latter. In this index convention, \eqref{eqn:genframe&dualform} becomes
\begin{equation}\label{eqn:genframe&dual}
  \hat E_A{}^{\hat I} = \begin{pmatrix} E_\alpha{}^i & E_\alpha{}^k C_{kij} \\
    0 & \eta_{ij,\tilde\alpha} \end{pmatrix}
    \quad \text{and} \quad
   \hat E^A{}_{\hat I} = \begin{pmatrix} E^\alpha{}_i & 0 \\
    - C_{imn} \eta^{mn,\tilde\alpha} & \eta^{ij,\tilde\alpha} \end{pmatrix}
\end{equation}
with
\begin{equation}
  \eta^{ij,\tilde\alpha} = \eta^{\beta\gamma,\tilde\alpha} E_\beta{}^i E_\gamma{}^j
    \quad \text{and} \quad
  \eta_{ij,\tilde\alpha} = \eta_{\beta\gamma,\tilde\alpha} E^\beta{}_i E^\gamma{}_j\,.
\end{equation}
Employing the dual frame on the flat derivative, we obtain
\begin{equation}
  \partial_{\hat I} = \hat E^A{}_{\hat I} D_A = \begin{pmatrix} \partial_i & 0 \end{pmatrix}\,.
\end{equation}
For the infinitesimal parameter of a generalized diffeomorphism $\xi^{\hat J}$, we use the same convention as for $V^{\hat I}$ in \eqref{eqn:VhatI}. It is convenient to split the generalized Lie derivative into the two parts. First, we have
\begin{equation}\label{eqn:genlie0}
  \widehat{\mathcal{L}}_\xi V^{\hat I} = \xi^{\hat J} \partial_{\hat J} V^{\hat I} - V^{\hat J} \partial_{\hat J} \xi^{\hat I} + Y^{\hat I\hat J}{}_{\hat K\hat L} \partial_{\hat J} \xi^{\hat K} V^{\hat L}\,.
\end{equation}
Second, there is the curved version $\mathcal{F}_{\hat I\hat J}{}^{\hat K} = \mathcal{F}_{AB}{}^C \hat E^A{}_{\hat I} \hat E^B{}_{\hat J} \hat E_C{}^{\hat K}$ of
\begin{equation}\label{eqn:scFABC}
  \mathcal{F}_{AB}{}^C = X_{AB}{}^C - \widehat{\mathcal{L}}_{\hat E_A} \hat E_B{}^{\hat I} \hat E^C{}_{\hat I}\,.
\end{equation}
Together, they form the generalized Lie derivative
\begin{equation}\label{eqn:gengeometr}
  \mathcal{L}_\xi V^{\hat I} = \widehat{\mathcal{L}}_\xi V^{\hat I} + \mathcal{F}_{\hat J\hat K}{}^{\hat I} \xi^{\hat J} V^{\hat K}\,.
\end{equation}
In the following we show that $\widehat{\mathcal{L}}$ is the untwisted generalized Lie derivative of GG and $\mathcal{F}_{\hat I\hat J}{}^{\hat K}$ implements its twist with the non-vanishing form and vector components
\begin{align}
  \mathcal{F}^{ijkl}{}_{mn} &= X_{\tilde\alpha\tilde\beta}{}^{\tilde\gamma} \eta^{ij,\tilde\alpha}
    \eta^{kl,\tilde\beta} \eta_{mn,\tilde\gamma}
    \nonumber \\
  \mathcal{F}_i{}^{jkl} &= X_{\alpha\tilde\beta}{}^\gamma E^\alpha{}_i \eta^{jk,\tilde\beta}
    E_\gamma{}^l \nonumber \\
    \mathcal{F}^{ij}{}_k{}^l &= X_{\tilde\alpha\beta}{}^\gamma \eta^{ij,\tilde\alpha} E^\beta{}_k
    E_\gamma{}^l \nonumber \\
    \mathcal{F}_i{}^{jk}{}_{lm} &= \mathcal{F}_{\alpha\tilde\beta}{}^{\tilde\gamma} E^\alpha{}_i 
    \eta^{jk,\tilde\beta} \eta_{lm,\tilde\gamma} + \mathcal{F}_i{}^{jkn} C_{lmn} - \mathcal{F}^{nojk}{}_{lm}
    C_{ino} \nonumber \\
  \mathcal{F}^{ij}{}_{klm} &= \mathcal{F}_{\tilde\alpha\beta}{}^{\tilde\gamma} \eta^{ij,\tilde\alpha}
    E^\beta{}_k \eta_{lm,\tilde\gamma} + \mathcal{F}^{ij}{}_k{}^n C_{lmn} - \mathcal{F}^{ijno}{}_{lm}
    C_{kno} \nonumber \\
  \mathcal{F}_{ij}{}^k &= \mathcal{F}_{\alpha\beta}{}^\gamma E^\alpha{}_i E^\beta{}_j E_\gamma{}^k -
    \mathcal{F}_i{}^{lmk} C_{jlm} - \mathcal{F}^{lm}{}_j{}^k C_{ilm} \nonumber \\
  \mathcal{F}_{ijkl} &= \mathcal{F}_{\alpha\beta}{}^{\tilde\gamma} E^\alpha{}_i E^\beta{}_j
    \eta_{kl,\tilde\gamma} + \mathcal{F}_{ij}{}^m C_{klm} - \mathcal{F}_i{}^{mn}{}_{kl} C_{jmn} -
    \mathcal{F}^{mn}{}_{jkl} C_{imn} + \nonumber \\
    & \qquad
    \mathcal{F}_i{}^{mno} C_{jmn} C_{klo} + \mathcal{F}^{mn}{}_j{}^o C_{imn} C_{klo} -
    \mathcal{F}^{mnop}{}_{kl} C_{imn} C_{jop} \label{eqn:twistcontrib}
\end{align}
with
\begin{align}
  \mathcal{F}_{\alpha\beta}{}^\gamma &= X_{\alpha\beta}{}^\gamma - f_{\alpha\beta}{}^\gamma &
  \mathcal{F}_{\alpha\beta}{}^{\tilde\gamma} &= X_{\alpha\beta}{}^{\tilde\gamma} - G_{ijkl} E_\alpha{}^i 
    E_\beta{}^j \eta^{kl,\tilde\gamma}  \nonumber \\
  \mathcal{F}_{\alpha\tilde\beta}{}^{\tilde\gamma} &= X_{\alpha\tilde\beta}{}^{\tilde\gamma} 
    + 2 f_{\alpha\beta}{}^\gamma \eta_{\delta\gamma,\tilde\beta} \eta^{\delta\beta,\tilde\gamma} &
    \mathcal{F}_{\tilde\alpha\beta}{}^{\tilde\gamma} &= - \mathcal{F}_{\beta\tilde\alpha}{}^{\tilde\gamma} + f_{\alpha\gamma}{}^{\delta} \eta_{\beta\delta,\tilde\alpha} \eta^{\alpha\gamma,\tilde\gamma} \,. \label{eqn:genframetwist}
\end{align}
Here
\begin{equation}\label{eqn:fandGfluxEaI}
  f_{\alpha\beta}{}^\gamma = 2 E_{[\alpha}{}^i \partial_i E_{\beta]}{}^j E^\gamma{}_j
    \quad\text{and}\quad
  G = d C = \frac1{4!} G_{ijkl}\, d x^i\wedge d x^j \wedge d x^k \wedge d x^l
\end{equation}
are the geometric and four-form fluxes induced by generalized frame \eqref{eqn:genframe&dual}.

To explicitly check that $\widehat{\mathcal{L}}$ is equivalent to the familiar generalized Lie derivative \cite{Coimbra:2011ky,Berman:2011cg} of exceptional GG, we calculate its components. The evaluation of the first two terms in \eqref{eqn:gengeometr} is straightforward. However, the term containing the $Y$-tensor is more involved. Therefore, we proceed componentwise and start with
\begin{equation}
  Y^{AB}{}_{CD} \hat E_A{}^i \hat E_B{}^j = Y^{\alpha\beta}{}_{CD} \hat E_{\alpha}{}^i \hat E_{\beta}{}^j = 0\,.
\end{equation}
The last step takes into account that the indices $\alpha$ and $\beta$ are by design solutions of the SC. Thus, the vector components for the first two indices of the $Y$-tensor vanish. Furthermore, we know that the form part of the partial derivative $\partial_{\hat I}$ vanishes ($\partial^{ij} = 0$). Hence, the only contributing $Y$-tensor components are $Y_{ij}{}^{k}{}_{\hat{L}\hat{M}}$ which we evaluate now. To this end, we consider
\begin{equation}
  Y_{ij}{}^k{}_{\hat L\hat M} = -\delta_{[i}^k E_{j]}{}^{a5} \epsilon_{a B C} E^B{}_{\hat L} E^C{}_{\hat M}
\end{equation}
and use the dual generalized frame \eqref{eqn:genframe&dual} to obtain the non-vanishing component
\begin{equation}
  Y_{ij}{}^k{}_l{}^{mn} = -2 \delta_{[i}{}^k \delta_{j]}{}^{[m} \delta^{n]}{}_l\,.
\end{equation}
Due to the symmetry of the $Y$-tensor, we are now able to compute the third term in \eqref{eqn:genlie0} and obtain
\begin{equation}
  Y_{ij}{}^k{}_{\hat L\hat M} \partial_k \xi^{\hat L} V^{\hat M} = 
    \partial_i \xi^k \tilde v_{kj} + \partial_j \xi^k \tilde v_{ik} - \partial_i \tilde\xi_{jk} v^k - \partial_j \tilde\xi_{ki} v^k \,.
\end{equation}
Taking into account the first and the second term as well, we finally have
\begin{equation}\label{eqn:genLieGG}
  \widehat{\mathcal{L}}_\xi V^{\hat{I}} = \widehat{\mathcal{L}}_\xi \begin{pmatrix}
v^i \\ \tilde v_{ij}
\end{pmatrix} =  \begin{pmatrix}
 L_\xi v^i \\ L_\xi \tilde v_{ij} - 3v^k \partial_{[k} \tilde\xi_{ij]}
\end{pmatrix}\,,
\end{equation}
which is the generalized Lie derivative of exceptional GG \cite{Coimbra:2011ky,Berman:2011cg}.

As in subsection~\ref{sec:linkvomega}, we check our results by considering the restriction to the T-duality subgroup SL(4). In this case we have to modify the map $\eta_p: \mathfrak{h} \rightarrow T_p^* M$, which  is now defined as
\begin{equation}
  \eta_p(t_{\tilde\beta}) = \left. \eta_{\alpha, \tilde\beta} E^\alpha{}_i d x^i \right|_{\sigma(p)}\,,
\end{equation}
to take the different $\eta$-tensor \eqref{eqn:etasl4} for this duality group into account. Repeating all the steps from above, we find the generalized frame
\begin{equation}\label{eqn:genFrameDFT}
  \hat E_A = \begin{pmatrix}
    E_{\alpha}{}^i \partial_i + B_{\alpha\beta} E^{\beta}{}_i \, d x^i \\
     \eta_{\beta, \tilde\alpha} E^{\beta}{}_i \, d x^i \end{pmatrix}
    \quad \text{its dual} \quad
    \hat E^A (v,\tilde v) = \begin{pmatrix} E^{\alpha}{}_i v^i \\
      \eta^{\beta,\tilde\alpha} ( E_{\beta}{}^i \tilde v_i
      - B_{\beta\gamma} E^{\gamma}{}_i v^i )
    \end{pmatrix}
\end{equation}
 and the generalized Lie derivative of GG. It has the form \eqref{eqn:gengeometr} with
\begin{equation}
  \widehat{\mathcal{L}}_\xi V^{\hat I} =  \widehat{\mathcal{L}}_\xi \begin{pmatrix}
v^i \\ \tilde v_{i} \end{pmatrix} = \begin{pmatrix}
  L_\xi v^i \\ L_\xi \tilde v_i - 2 v^k \partial_{[k} \tilde \xi_{i]}
  \end{pmatrix}
\end{equation}
and the twist in \eqref{eqn:scFABC} which now has to be evaluated for the generalized frame in \eqref{eqn:genFrameDFT}. After an appropriate change of basis this expression matches the one derived in \cite{Hassler:2016srl}.

\subsection{Lie Algebra Cohomology and Dual Backgrounds}\label{sec:dualbg}
In general, the SC admits more than one solution. They arise from different choices of $v^0_a$ in \eqref{eqn:sclinearv0} and result in a distinguished splitting of the Lie algebra $\mathfrak{g}$ in the coset part $\mathfrak{m}$ and the subalgebra $\mathfrak{h}$. One can always restore the canonical form of $v^0_a$ \eqref{eqn:v0acanonical} by a SL(5) rotation. For this case the index assignment \eqref{eqn:decompSL510} remains valid and we only have to check whether the generators $t_{\tilde \alpha}$ form a Lie algebra $\mathfrak{h}$. This situation is very closely related to the \DFTwzw{} case discussed in \cite{Hassler:2016srl}. Thus, we also use Lie algebra cohomology to explore possible subgroups of the Lie group $\mathfrak{g}$.

Let us review the salient features of the construction. First, we only consider transformations in the coset SO(5)/SO(4)$\subset$ SL(5). All others, at most scale $v^0_a$ and thus leave the subalgebra $\mathfrak{h}$ invariant. A coset element 
\begin{equation}\label{eqn:deformrepr}
  \mathcal{T}_A{}^B = \exp ( \lambda \, t_A{}^B )
\end{equation}
is generated by applying the exponential map to a $\mathfrak{so}(5)$ generator $t$ acting on the coordinate irrep $\mathbf{10}$. It modifies the embedding tensor according to
\begin{equation}
  X'_{AB}{}^C = \mathcal{T}_A{}^D \mathcal{T}_B{}^E X_{DE}{}^F \mathcal{T}_F{}^C\,.
\end{equation}
We expand this expression in $\lambda$ to obtain
\begin{equation}
  X'_{AB}{}^C = X_{AB}{}^C + \lambda \delta X_{AB}{}^C + \lambda^2 \delta^2 X_{AB}{}^C + \dots
\end{equation}
and read off the $\mathfrak{g}$-valued two-forms
\begin{equation}
  c_n = t_C ( \delta^n X_{AB}{}^C ) \theta^A \wedge \theta^B\,.
\end{equation}
Only transformations with $\delta^n X_{\tilde\alpha \tilde\beta}{}^{\gamma} = 0$ are allowed. Otherwise $\mathfrak{h}$ fails to be a subalgebra. Finally, we have to check whether the restricted forms
\begin{equation}
  c_n = t_{\tilde\gamma} ( \delta^n X_{\tilde\alpha \tilde\beta}{}^{\tilde\gamma} ) \theta^{\tilde \alpha} \wedge \theta^{\tilde \beta} 
\end{equation}
are in the Lie algebra cohomology $H^2(\mathfrak{h},\mathfrak{h})$. If so, they give rise to a infinitesimal non-trivial deformation of $\mathfrak{h}$. Obstructions to the integrability of this deformation lie in $H^3(\mathfrak{h}, \mathfrak{h})$.

\subsection{Generalized Frame Field}\label{sec:genframe}
A significant application of the formalism presented in this paper is to construct the frame fields $\mathcal{E}_A{}^{\hat I}$ of generalized parallelizable manifolds $M$. In the following, we show that 
\begin{equation}\label{eqn:genparaframe}
  \mathcal{E}_A{}^{\hat I} = - M_A{}^B \hat E'_B{}^{\hat I}
\end{equation}
fulfills the defining equation \eqref{eqn:genLiegEFT} in the introduction if an additional linear constraint on the structure constants $X_{AB}{}^C$ holds. The derivation is done step by step starting with the frame $\hat E'_A{}^{\hat I}$. It differs from \eqref{eqn:genframe&dual} by using a three-from $\mathcal{C}$ instead of $C$ (see \eqref{eqn:genparaframeintro} in the introduction). So first, we calculate
\begin{equation}
  X'_{AB}{}^C = \widehat{\mathcal{L}}_{\hat E'_A} \hat E'_B{}^{\hat I} {E'}^C{}_{\hat I}
\end{equation}
which has the non-trivial components
\begin{align}
  X'_{\alpha\beta}{}^\gamma &= f_{\alpha\beta}{}^\gamma &
  X'_{\alpha\beta}{}^{\tilde\gamma} &= \mathcal{G}_{ijkl} E_\alpha{}^i E_\beta{}^j \eta^{kl,\tilde\gamma}
    \nonumber \\
  X'_{\alpha\tilde\beta}{}^{\tilde\gamma} &= 2 f_{\alpha\beta}{}^\gamma \eta_{\delta\gamma,\tilde\beta}
    \eta^{\delta\beta,\tilde\gamma} &
  X'_{\tilde\alpha\beta}{}^{\tilde\gamma} &= - X'_{\beta\tilde\alpha}{}^{\tilde\gamma} - f_{\alpha\gamma}{}^{\delta} \eta_{\beta\delta,\tilde\alpha} \eta^{\alpha\gamma,\tilde\gamma}\,.
\end{align}
As before $f_{\alpha\beta}{}^\gamma$ denotes the geometric flux \eqref{eqn:fandGfluxEaI} and
\begin{equation}
  \mathcal{G} = d \mathcal{C} = \frac{1}{4!} \mathcal{G}_{ijkl}\, d x^i \wedge d x^j \wedge d x^k \wedge d x^l
\end{equation}
is the field strength corresponding to $\mathcal{C}$. In \eqref{eqn:genparaframe}, $\hat E'_A{}^{\hat I}$ is twisted by the SL(5) rotation
\begin{equation}\label{eqn:SL5rot}
  M_B{}^A t_A = m^{-1} t_B m = (\mathrm{Ad}_{m^{-1}})_B{}^A t_A
\end{equation}
with the inverse transpose
\begin{equation}
  t_A M^A{}_B = m t_B m^{-1}\,.
\end{equation}
Next, we combine the two of them and evaluate
\begin{equation}
  X''_{AB}{}^C = \widehat{\mathcal{L}}_{M_A{}^D \hat E'_D} (M_B{}^E \hat E'_E{}^{\hat I})
    M^C{}_F \hat{E'}{}^F{}_{\hat I}\,.
\end{equation}
It is convenient to write the result as
\begin{equation}\label{eqn:X''ABC}
  X''_{AB}{}^C = X'''_{DE}{}^F M_A{}^D M_B{}^E M^C{}_F
    \quad \text{with} \quad
  X'''_{AB}{}^C = X'_{AB}{}^C + 2 T_{[AB]}{}^C + Y^{CD}{}_{EB} T_{DA}{}^E
\end{equation}
and
\begin{equation}
  T_{AB}{}^C = - \hat E'_A{}^{\hat I} \partial_{\hat I} M^D{}_B M_D{}^C\,.
\end{equation}
Taking into account the special form of $M_B{}^A$ in \eqref{eqn:SL5rot}, this tensor can be calculated:
\begin{equation}
  T_{AB}{}^C = - \hat E_A{}^i E^D{}_i X_{DB}{}^C = 
    \begin{pmatrix} - X_{\alpha B}{}^C +
    \eta^{\delta\epsilon,\tilde\delta} C_{\alpha\delta\epsilon} X_{\tilde\delta B}{}^C & 0\end{pmatrix}\,.
\end{equation}
In the second step, we remember that for a SC solution the connection $A$ vanishes. This allows us to identify $E_\alpha{}^i E^{\tilde\beta}{}_i = - \eta^{\gamma\delta,\tilde\beta} C_{\alpha\gamma\delta}$. 
By plugging the solution for $T_{AB}{}^C$ into \eqref{eqn:X''ABC}, we obtain the non-vanishing components
\begin{align}
  X'''_{\tilde\alpha\tilde\beta}{}^{\tilde\gamma} &= - X_{\tilde\alpha\tilde\beta}{}^{\tilde\gamma} &
  X'''_{\alpha\tilde\beta}{}^{\gamma} &= - X_{\alpha\tilde\beta}{}^{\gamma} &
  X'''_{\tilde\alpha\beta}{}^{\gamma} &= - X_{\tilde\alpha\beta}{}^{\gamma} &
    \nonumber \\
  X'''_{\alpha\tilde\beta}{}^{\tilde\gamma} &= -2 X'''_{\alpha\beta}{}^\gamma
    \eta_{\delta\gamma,\tilde\beta} \eta^{\delta\beta,\tilde\gamma} &
  X'''_{\tilde\alpha\beta}{}^{\tilde\gamma} &= - X'''_{\beta\tilde\alpha}{}^{\tilde\gamma}
    \nonumber \\
  X'''_{\alpha\beta}{}^\gamma &= - 2 X_{\alpha\beta}{}^\gamma + 
    2 X_{\tilde\alpha[\beta}{}^\gamma C_{\alpha]\delta\epsilon} \eta^{\delta\epsilon,\tilde\alpha}
    + f_{\alpha\beta}{}^\gamma
\end{align}
and
\begin{gather}
  X'''_{\alpha\beta}{}^{\tilde\gamma} = - 2 X_{\alpha\beta}{}^{\tilde\gamma} +
      2 X_{\gamma\alpha}{}^{\tilde\alpha} \eta_{\delta\beta,\tilde\alpha} \eta^{\gamma\delta,\tilde\gamma} -
      ( 2 X_{\gamma\beta}{}^{\delta} C_{\delta\epsilon\alpha} -
        4 X_{\gamma\alpha}{}^{\delta} C_{\delta\epsilon\beta} ) \eta^{\gamma\epsilon,\tilde\gamma} - 
          \nonumber \\
      2 X_{\alpha\beta}{}^\gamma C_{\delta\epsilon\gamma} \eta^{\delta\epsilon,\tilde\gamma}
      + \mathcal{G}_{ijkl} E_\alpha{}^i E_\beta{}^j \eta^{kl,\tilde\gamma} \label{eqn:X'''ABCcomp}
\end{gather}
after imposing the constraints
\begin{equation*}\label{eqn:linconst3}\tag{C3}
  X_{A\tilde\beta}{}^{\tilde\gamma} = -2 X_{A\beta}{}^\gamma
    \eta_{\delta\gamma,\tilde\beta} \eta^{\delta\beta,\tilde\gamma}
    \quad \text{and} \quad
  X_{\alpha\gamma}{}^\delta \eta_{\beta\delta,\tilde\alpha} \eta^{\alpha\gamma,\tilde\gamma} = 0\,.
\end{equation*}
At this point, \eqref{eqn:SL5rot} proves to be a good choice. Up to a sign, many components are already as we want them to be. This gets even better, if we take into account the explicit expression
\begin{equation}
  f_{\alpha\beta}{}^\gamma = X_{\alpha\beta}{}^\gamma - 2 X_{\tilde\alpha[\beta}{}^\gamma C_{\alpha]\delta\epsilon} \eta^{\delta\epsilon,\tilde\alpha}
\end{equation}
for the geometric flux which results in
\begin{equation}
  X'''_{\alpha\tilde\beta}{}^{\tilde\gamma} = - X_{\alpha\tilde\beta}{}^{\tilde\gamma}\,, \quad
  X'''_{\tilde\alpha\beta}{}^{\tilde\gamma} = - X_{\tilde\alpha\beta}{}^{\tilde\gamma}
    \quad \text{and} \quad
  X'''_{\alpha\beta}{}^{\gamma} = - X_{\alpha\beta}{}^\gamma
\end{equation}
after imposing the constraints \eqref{eqn:linconst3}. Finally, there is the last contribution \eqref{eqn:X'''ABCcomp} which should evaluate to $-X_{\alpha\beta}{}^{\tilde\gamma}$. It requires an appropriate choice for the four-form
\begin{equation}
  \mathcal{G}_{ijkl} = f(x^1, x^2, x^3, x^4) \epsilon_{ijkl}\,.
\end{equation}
Being the top-form on $M$, it only has one degree of freedom captured by the function $f$. With this ansatz, the last term in \eqref{eqn:X'''ABCcomp} becomes
\begin{equation}
  \mathcal{G}_{ijkl} E_\alpha{}^i E_\beta{}^j \eta^{lk,\tilde\gamma} = f \det (E_\rho{}^i) \epsilon_{1\hat\alpha\hat\beta\hat\gamma\hat\delta} \eta^{\gamma\delta,\tilde\gamma}\,.
\end{equation}
If we choose $f = \lambda \det (E^\rho{}_i)$ for an appropriate, constant $\lambda$, the miracle happens and we find $X'''_{\alpha\beta}{}^{\tilde\gamma} = -X_{\alpha\beta}{}^{\tilde\gamma}$. The key to this result is that the structure constants $X_{AB}{}^C$ are not arbitrary, but severely constrained by the linear constraints \eqref{eqn:linconst1}, \eqref{eqn:linconst2} and \eqref{eqn:linconst3}. The first two are solved in section~\ref{sec:linconstsl5} and we present the solutions to the remaining one at the end of this section. For the moment, we continue with
\begin{equation}\label{eqn:correctX'''}
  X'''_{AB}{}^C = - X_{AB}{}^C \quad \text{under \eqref{eqn:linconst1} - \eqref{eqn:linconst3}}\,.
\end{equation}
Structure constants of a Lie algebra are preserved under the adjoint action \eqref{eqn:SL5rot}. Thus, we immediately imply
\begin{equation}
  X''_{AB}{}^C = X'''_{AB}{}^C = - X_{AB}{}^C\,.
\end{equation}
Up to the minus sign, this is exactly the result we are looking for. In order to get rid of this wrong sign, we introduce an additional minus in the generalized frame field $\mathcal{E}_A{}^{\hat I}$ \eqref{eqn:genparaframe}. The result is equivalent to \eqref{eqn:genparaframeintro} in the introduction. As argued above, the three-from $\mathcal{C}$ it contains has to be chosen such that
\begin{equation}\label{eqn:G=lambdavol}
  \mathcal{G} = d \mathcal{C} = \lambda \det(E^\rho{}_i) d x^1 \wedge d x^2 \wedge d x^3 \wedge d x^4 = 
    \lambda \mathrm{vol}\,,
\end{equation}
where $\mathrm{vol}$ is the volume form on $M$ induces by the frame field $E^\alpha{}_i$.

Finally, we have to find the solutions of the linear constraint \eqref{eqn:linconst3}. Otherwise the construction above does not apply. In order to identify these solutions, we discuss embedding tensor components in the $\mathbf{15}$ \cite{Samtleben:2005bp}
\begin{equation}\label{eqn:emb1055b}
  X_{abc}{}^d = \delta_{[a}^d Y_{b]c}
\end{equation}
parameterized by the symmetric matrix $Y_{ab}$ and in the $\overline{\mathbf{40}}$ \cite{Samtleben:2005bp}
\begin{equation}\label{eqn:embeddingt40}
  X_{abc}{}^d = -2 \epsilon_{abcef} Z^{ef,d}
\end{equation}
given in terms of the tensor $Z^{ab,c}$ with $Z^{ab,c}$=$Z^{[ab],c}$ and $Z^{[ab,c]}$=0. The structure constants of the corresponding Lie algebra $\mathfrak{g}$ follow from the further embedding of them into $\mathbf{10}\times\mathbf{10}\times\overline{\mathbf{10}}$ \cite{Samtleben:2005bp}
\begin{equation}\label{eqn:emb101010b}
  X_{AB}{}^C = X_{a_1 a_2, b_1 b_2}{}^{c_1 c_2} = 2 X_{a_1 a_2[b_1}{}^{[c_1} \delta_{b_2]}^{c_2]}\,.
\end{equation}
If there are only contributions from the $\mathbf{15}$, this expression is identical to the structure constants because the corresponding group manifold is ten-dimensional.  We study this case first. Splitting the indices $A$, $B$, $C$, \dots{} into a coset component $\alpha$ and a subalgebra part $\tilde\alpha$ according to \eqref{eqn:decompSL510} singles out one direction in the fundamental irrep of SL(5). It is given by $v^0_a$ in \eqref{eqn:v0acanonical} and results in the branching
\begin{equation}\label{eqn:sollinconst3_15}
  \mathbf{15} \rightarrow \mathbf{1} + \xcancel{\mathbf{4}} + \mathbf{10} 
\end{equation}
from SL(5) to SL(4). The linear constraint \eqref{eqn:linconst3} is violated by the crossed out irreps. If we only take into account the remaining ones, all terms containing $C_{\alpha\beta\gamma}$ in $X'''_{\alpha\beta}{}^{\tilde\gamma}$ vanish. For \eqref{eqn:correctX'''} to hold, we further require that the relation
\begin{equation}
  X_{\alpha\beta}{}^{\tilde\gamma} - 2 X_{\gamma\alpha}{}^{\tilde\alpha} \eta_{\delta\beta,\tilde\alpha} \eta^{\gamma\delta,\tilde\gamma} = \lambda \epsilon_{1\hat\alpha\hat\beta\hat\gamma\hat\delta} \eta^{\gamma\delta,\tilde\gamma}
\end{equation}
is satisfied. This is the case, if we identify
\begin{equation}\label{eqn:lambdafromY11}
  \lambda = -\frac34 \, Y_{11}\,.
\end{equation}
For all the remaining gaugings in \eqref{eqn:sollinconst3_15}, one should in principal be able to construct a generalized parallelizable space $M$. However, this construction relies on finding a flat connection $A$ in order to solve the SC in the first place. In general deriving this connection can turn out to be complicated. However, as explained at the end of~\ref{sec:linkvomega}, if $M$ is a symmetric space there is a simple procedure to immediately solve this challenge. Luckily all remaining irreps in \eqref{eqn:sollinconst3_15} give rise to a symmetric pairs $\mathfrak{m}$ and $\mathfrak{h}$ so that one can immediately solve the SC. Furthermore, the solutions to the quadratic constraint \eqref{eqn:quadconstr} are known as well. The resulting group manifolds depend on the eigenvalues of the symmetric, real matrix $Y_{ab}$. If $p$ of them are positive, $q$ are negative and $r$ are zero, we find
\begin{equation}
  G = \text{CSO}(p,q,r) = SO(p,q) \ltimes \mathbb{R}^{(p+q) r}
    \quad \text{with} \quad
  p + q + r = 5\,.
\end{equation}
Our construction applies to all corresponding generalized frames $\mathcal{E}_A$. They where also constructed in \cite{Hohm:2014qga} by taking a clever ansatz in a distinguished coordinate system. Before this work, \cite{Lee:2014mla} presented the generalized frame for SO(5) ($p$=5, $q$=$r$=0), the four-sphere with $G$-flux.

With gaugings in the $\mathbf{40}$, group manifold with $\dim G$<10 are relevant. As discussed in section~\ref{sec:linconstsl5}, the irreps of the embedding tensor branch into the U-duality subgroups. Again, $v^0_a$ in \eqref{eqn:v0acanonical} distinguishes a direction and results in an additional branching. To see how this works, consider the SL(3)$\times$SL(2) solutions in figure~\ref{fig:sollinconst}. Starting with $\dim G$=9, the relevant embedding tensor components
\begin{equation}\label{eqn:sollinconst3_40_9}
  (\mathbf{1},\mathbf{3}) + (\mathbf{3},\mathbf{2}) + (\mathbf{6},\mathbf{1}) + (\mathbf{1},\mathbf{2}) \rightarrow (\mathbf{1},\mathbf{3}) + \xcancel{(\mathbf{1},\mathbf{2})} + (\mathbf{2},\mathbf{2}) + 
  (\mathbf{1},\mathbf{1}) + \xcancel{(\mathbf{2},\mathbf{1})} + (\mathbf{3},\mathbf{1}) + (\mathbf{1},\mathbf{2})
\end{equation}
branch from SL(3)$\times$SL(2) to SL(2)$\times$SL(2). All crossed out irreps decent from the $\mathbf{4}$ in \eqref{eqn:sollinconst3_15}. Only the last irrep $(\mathbf{1},\mathbf{2})$ originates from the $\mathbf{40}$. It does not admit a symmetric pair. Still, one is able to construct a generalized frame field for the four-tours with geometric flux in section~\ref{sec:T4Gflux} which is realized by a gauging in this irrep. For the scaling factor $\lambda$ in \eqref{eqn:G=lambdavol} the relation \eqref{eqn:lambdafromY11} still applies. One can go on and repeat this discussion for group manifolds with $\dim G$<9. We do not perform it here, because all the examples we present in the next section are covered by the cases above.

\section{Examples}\label{sec:examples}
It is instructive to study some explicit examples for the construction described in the previous sections. In the following, we present the four-torus with $G$-flux, its dual backgrounds and the four-sphere with $G$-flux. While the former is well-known from conventional EFT, it illustrates how dual backgrounds arise in our formalism. Furthermore, it allows to study group manifolds $G$ with $\dim G$<10, which arise from gaugings in the $\mathbf{40}$. In this case SL(5) is broken to SL(3)$\times$SL(2). A more sophisticated setup is the four-sphere with $G$ flux which corresponds to the group manifolds SO(5). It is was studied in \cite{Lee:2014mla,Hohm:2014qga} and so permits to compare the resulting generalized frame field $\mathcal{E}_A$ with the literature.

\subsection{Duality-Chain of the Four-Torus with \texorpdfstring{$G$}{G}-Flux}\label{sec:T4Gflux}
In string theory there is the well-known duality chain\cite{Shelton:2005cf}
\begin{equation}
 H_{ijk} \leftrightarrow f_{ij}{}^k \leftrightarrow Q_i{}^{jk} \leftrightarrow R^{ijk} \,,
\end{equation}
where each adjacent background is related by a single T-duality which maps IIA $\leftrightarrow$ IIB string theory. In this section, we show how parts of this chain result from different SC solutions on a ten- and a nine-dimensional group manifold. In order to uplift these examples to M-theory, we need to consider only IIA backgrounds and two T-dualities taking IIA $\leftrightarrow$ IIA string theory. Thus, the above duality chain splits into the two distinct duality chains
\begin{equation}
  H_{ijk} \leftrightarrow Q_i{}^{jk}
\end{equation}
and
\begin{equation}
 f_{ij}{}^k \leftrightarrow R^{ijk} \,.
\end{equation}
Similarly, when considering M-theory, we apply three U-duality transformations to ensure that we map M-theory to itself. One may think of this as taking a $T^3$ in the limit of vanishing volume. Indeed, if we had taken only a $S^1$ of vanishing radius, we would have obtained weakly-coupled IIA string theory. A $T^2$ of vanishing volume gives IIB string theory (think of taking repeated small radius limits of the two circles of $T^2$). In this case, we have weakly coupled IIA compactified on a small circle. Applying T-duality to this circle results in IIB in the decompactification limit. Thus, for every two-cycle of vanishing volume we see that we open up a new dual direction. Having a $T^3$ of vanishing volume means we loose three directions but open up three new ones (one for each of the three two-cycles in $T^3$). So we finally arrived at an eleven-dimensional background again. Another way to see this is to identify two of the directions of the U-duality with the two directions of the T-duality and the third one with the M-theory circle. This also ensures the correct dilaton transformation. From this arguments it becomes clear that the M-theory $T^4$ duality chain is also split and we have
\begin{equation}\label{eqn:chain1}
 G_{ijkl} \leftrightarrow Q_{i}{}^{jkl}
\end{equation}
and
\begin{equation}\label{eqn:chain2}
 f_{ij}{}^k \leftrightarrow R^{i,jklm} \,.
\end{equation}
As we will see only the former can be realized in our framework. This finding is in agreement with the \DFTwzw{} case, where the $R$-flux background does not posses a maximally isotropic subalgebra $\mathfrak{h}$ \cite{Hassler:2016srl}.

The splitting \eqref{eqn:chain1} and \eqref{eqn:chain2} of the duality chain is also manifest in the embedding tensor \cite{Blair:2014zba}. For $SL(5)$, it has two irreducible representations (not counting the trombone which we neglect in this paper). Each chain represents one of these irreps. Duality transformations are implemented by SL(5) rotations. These clearly do not mix different irreps.

\subsubsection{Gaugings in the \texorpdfstring{$\mathbf{15}$}{15}}\label{sec:T4_15}
Let us start with the first chain. It is fully contained in the irrep $\mathbf{15}$ \cite{Blair:2014zba} which we express in terms of the symmetric tensor $Y_{ab}$. The resulting embedding tensor is \eqref{eqn:emb1055b} and the corresponding structure constants arise from \eqref{eqn:emb101010b}. It is always possible to diagonalize the symmetric matrix $Y_{ab}$ by a SO(5) transformation. For gaugings in the $\mathbf{15}$ only, the quadratic constraint is fulfilled automatically. A four-torus with $\mathbf{g}$ units of $G$-flux is given by the explicit choice
\begin{equation}\label{eqn:gaugingT4Gflux}
  Y_{ab} = - 4 \mathbf{g} \, \diag( 1,\,0,\,0,\,0,\,0 )\,.
\end{equation}
This particular choice is compatible with the vector $v_0^a$ in \eqref{eqn:v0acanonical} and the decomposition \eqref{eqn:decompSL510} of the $\mathbf{10}$ index $A = (\alpha, \tilde\alpha)$. It gives rise to the group manifold $G =\text{CSO}(1,0,4)$ with an abelian subgroup $H$ which is generated by all infinitesimal translations in $\mathbb{R}^6$. We use the $21$-dimensional, faithful representation of $\mathfrak{g}$ derived in appendix~\ref{app:faithfulrepr} to obtain the matrix representation
\begin{align}
  m &= \exp( t_1 x^1 ) \exp( t_2 x^2 ) \exp( t_3 x^3 ) \exp{ t_4 x^4} \quad \text{and} \label{eqn:matrixexpm}\\
  h &= \exp( t_{\tilde 1} x^{\tilde 1}) \exp( t_{\tilde 2} x^{\tilde 2}) \exp( t_{\tilde 3} x^{\tilde 3})
  \exp(t_{\tilde 4} x^{\tilde 4}) \exp( t_{\tilde 5} x^{\tilde 5}) \exp(t_{ \tilde 6} x^{\tilde 6}) \label{eqn:matrixexph}
\end{align}
of the Lie group $G$. This group is not compact and therefore does not represent the background we are interested in (clearly a torus is compact). Thus, we have to quotient by the discrete subgroup CSO(1,0,4,$\mathbf{Z}$) from the left. Doing so is equivalent to impose the coordinate identifications \eqref{eqn:coordident1} and \eqref{eqn:coordident2} which are derived in appendix~\ref{app:faithfulrepr}.

For this setup, the connection $A = A^{\tilde \alpha} t_{\tilde \alpha}$ reads
\begin{align}
  A^{\tilde 1} &= \Big[ ( \mathbf{g} x^2 + C_{134} ) d x^1 + C_{234} \, d x^2 \Big] \,, &
  A^{\tilde 2} &= \Big[ ( \mathbf{g} x^3 - C_{124} ) d x^1 + C_{234} \, d x^3 \Big] \,,
    \nonumber \\
  A^{\tilde 3} &= \Big[ ( \mathbf{g} x^3 - C_{124} ) d x^2 - C_{134} \, d x^3 \Big] \,, &
  A^{\tilde 4} &= \Big[ ( \mathbf{g} x^4 + C_{123} ) d x^1 + C_{124} \, d x^4 \Big] \,,
    \nonumber \\
  A^{\tilde 5} &= \Big[ ( \mathbf{g} x^4 + C_{123} ) d x^2 - C_{134} \, d x^4 \Big] \,, &
  A^{\tilde 6} &= \Big[ ( \mathbf{g} x^4 + C_{123} ) d x^3 + C_{124} \, d x^4 \Big]
\end{align}
in the patch we are considering. For the three-form field
\begin{equation}\label{eqn:CT4Gflux}
  C = \frac{\mathbf{g}}2 ( x^1\,d x^2 \wedge d x^3 \wedge d x^4 - x^2\,d x^1 \wedge d x^3 d \wedge x^4 + x^3\,d x^1 \wedge d x^2 \wedge d x^4 - x^4\,d x^1\wedge x^2 \wedge x^3 )\,,
\end{equation}
with the flux contribution
\begin{equation}\label{eqn:GfieldhatE}
  G_{\hat E} = d C = 2 \mathbf{g}\, d x^1\wedge d x^2 \wedge d x^3 \wedge d x^4
\end{equation}
to the generalized frame field $\hat E_A$, the field strength $F = d A$ vanishes. In order to set $A=0$ in the current patch, we apply the transformation $g\rightarrow g \exp( t_{\tilde\alpha} \lambda^{\tilde\alpha} )$ to all group elements with
\begin{align}
  \lambda^{\tilde 1} &= - \frac{\mathbf{g}}2 x^1 x^2\,, &
  \lambda^{\tilde 2} &= - \frac{\mathbf{g}}2 x^1 x^3\,, &
  \lambda^{\tilde 3} &= - \frac{\mathbf{g}}2 x^2 x^3\,, \nonumber \\
  \lambda^{\tilde 4} &= - \frac{\mathbf{g}}2 x^1 x^4\,, &
  \lambda^{\tilde 5} &= - \frac{\mathbf{g}}2 x^2 x^4\,, &
  \lambda^{\tilde 6} &= - \frac{\mathbf{g}}2 x^3 x^4\,. \label{eqn:gaugetrafoGflux}
\end{align}
It results in the desired $A=0$ and the background generalized vielbein
\begin{equation}\label{eqn:EAiGflux}
  E^{\alpha}{}_i = \begin{pmatrix}
    1 & 0 & 0 & 0 \\ 0 & 1 & 0 & 0 \\ 0 & 0 & 1 & 0 \\ 0 & 0 & 0 & 1
  \end{pmatrix}\,, \quad
  E^{\tilde\alpha}{}_i = \frac{\mathbf{g}}2 \begin{pmatrix}
    x^2 & - x^1 & 0 & 0 \\
    x^3 & 0 & -x^1 & 0 \\
    0 & x^3 & -x^2 & 0 \\
    x^4 & 0 & 0 & -x^1 \\
    0 & x^4 & 0 & -x^2 \\
    0 & 0 & x^4 & -x^3
  \end{pmatrix}
    \quad \text{and} \quad
  E^{\tilde\alpha}{}_{\bar i} = \begin{pmatrix}
    1 & 0 & 0 & 0 & 0 & 0 \\
    0 & 1 & 0 & 0 & 0 & 0 \\
    0 & 0 & 1 & 0 & 0 & 0 \\
    0 & 0 & 0 & 1 & 0 & 0 \\
    0 & 0 & 0 & 0 & 1 & 0 \\
    0 & 0 & 0 & 0 & 0 & 1
  \end{pmatrix}\,.
\end{equation}
This gauging gives rise to a symmetric space. Thus, we also could have used the coset representative
\begin{equation}
  m = \exp ( t_1 x^1 + t_2 x^2 + t_3 x^3 + t_4 x^4 )
\end{equation}
instead of \eqref{eqn:matrixexpm} to find the same result. However, it is nice to demonstrate the full procedure at least once. With \eqref{eqn:EAiGflux} we calculate the generalized frame field $\hat E_A{}^{\hat I}$, its dual and finally the twist $\mathcal{F}_{\hat I\hat J\hat K}$ of the generalized Lie derivative \eqref{eqn:gengeometr}. It has contributions \eqref{eqn:twistcontrib} from the four-form
\begin{equation}
  G_{\mathcal{F}} = \frac{1}{4!} \mathcal{F}_{ijkl} \, d x^i\wedge d x^j\wedge d x^k\wedge d x^l =
    - \mathbf{g}\, d x^1 \wedge d x^2 \wedge d x^3 \wedge d x^4
\end{equation}
only. In total, we obtain the expected $\mathbf{g}$ units of $G$-flux
\begin{align}
  G = G_{\hat E} + G_{\mathcal{F}} &= \frac{1}{4!} X_{\alpha\beta}{}^{\tilde \gamma} E^\alpha{}_i E^\beta{}_j \eta_{\gamma\delta,\tilde \gamma} E^\gamma{}_k E^\delta{}_l \, d x^i \wedge d x^j \wedge d x^k \wedge d x^l \nonumber \\
  &=  \mathbf{g}\, d x^1 \wedge d x^2 \wedge d x^3 \wedge d x^4 \,.
\end{align}
on the background after combining this contribution with $G_{\hat E}$ from the generalized frame.

This result is very similar to the one obtained for the torus with $H$-flux in \cite{Hassler:2016srl}. Again the flux is split between the twist term and the frame field in a particular way. However, this splitting arises naturally from the principal bundle construction. To see how this works, we calculate the flux contribution from the frame field
\begin{equation}\label{eqn:GhatEfromC}
  G_{\hat E} = d C = - \frac16 E^\alpha{}_i E^\beta{}_j d ( \eta_{\alpha\beta,\tilde\gamma} E^{\tilde \gamma}{}_j d x^j) \wedge d x^i \wedge d x^j
\end{equation}
by using the relation \eqref{eqn:CfromE}. We identify
\begin{equation}
  \mathcal{A}_{\alpha\beta} = \eta_{\alpha\beta,\tilde \gamma} E^{\tilde\gamma}{}_i \, d x^i
\end{equation}
with the connection of a $T^6$ bundle over the tours. Thus, each independent $\mathcal{A}_{\alpha\beta}$, like e.g. $\mathcal{A}_{12}$, is the connection of a circle bundle. The first Chern class of these bundles are defined as
\begin{equation}
  c_{\alpha\beta} = d \mathcal{A}_{\alpha\beta}\,.
\end{equation}
and by plugging the result \eqref{eqn:EAiGflux} for $E^{\tilde\alpha}{}_i$ into this equation, we obtain the independent classes
\begin{align}
  c_{21} &= \mathbf{g}\, d x^3 \wedge d x^4 \,, &
  c_{13} &= \mathbf{g}\, d x^2 \wedge d x^4 \,, &
  c_{41} &= \mathbf{g}\, d x^2 \wedge d x^3 \,, \nonumber \\
  c_{32} &= \mathbf{g}\, d x^1 \wedge d x^4 \,, &
  c_{24} &= \mathbf{g}\, d x^1 \wedge d x^3 \,, &
  c_{43} &= \mathbf{g}\, d x^1 \wedge d x^2\,,
\end{align}
explicitly. Each of them represents a class in the integer valued cohomology $H^2(\mathcal{S}_{\alpha\beta}, M)=\mathbb{Z}$ of the circle bundle $\mathcal{S}_{\alpha\beta}$ over $M=T^4$. Furthermore, they are not trivial, which shows that the principal bundle we constructed is non-trivial, too. If we denote the cohomology class of a closed form $\omega$ by $[ \omega ]$, we can rewrite \eqref{eqn:GhatEfromC} as
\begin{equation}\label{eqn:[GhatE]fromChern}
  [ G_{\hat E} ] = \frac13 ( [c_{21}] + [c_{13}] + [c_{41}] + [c_{32}] + [c_{24}] + [c_{43}] )\,.
\end{equation}
Because $G_{\hat E}$ is a top form on the $T^4$ it lives in the integer valued de Rham cohomology $H^4_{\mathrm{dR}}(M)$ which is isomorph to $H^2(\mathcal{S}_{\alpha\beta}, M)$. Thus, there is no obstruction in comparing the Chern numbers with $[G_{\hat E}]$ and \eqref{eqn:[GhatE]fromChern} makes perfect sense. All different $S^1$ factors in the $H$-principal bundle are equal. So it is natural that they share the same Chern number, namely one. In this case \eqref{eqn:[GhatE]fromChern} forces
\begin{equation}
  [G_{\hat E}] = 2 \mathbf{g}
\end{equation}
which is compatible with our result \eqref{eqn:GfieldhatE}.

It is interesting to note that in this example the field strength $F = d A$ for the $H$-principal bundle vanishes everywhere on $M$. Still it is not possible to completely gauge away the connection $A$. Because the gauge transformation $\lambda^{\tilde a}$ in \eqref{eqn:gaugetrafoGflux} is not globally well-defined on $M$. This is clearly a result of the discrete subgroup which was modded out from the left to make $G$ compact. One could think that this effect is related to the topological non-trivial $G$-flux in this background. But it is not, as the four-sphere with $G$-flux in the next section proves. There, one can get rid of the connection everywhere. However, locally we can always solve the SC and construct the generalized frame field
\begin{equation}\label{eqn:genframeT4Gflux}
  \mathcal{E}_\alpha = - E_\alpha{}^i \partial_i + \iota_{E_\alpha}  \mathcal{C}' \,, \quad
  \mathcal{E}_{\tilde\alpha} = - \frac{1}{2} \eta_{ij,\tilde\alpha} \, d x^i \wedge d x^j
\end{equation}
where we take $E_\alpha{}^i$ as the inverse transpose from the frame in \eqref{eqn:EAiGflux} and
\begin{equation}
  \mathcal{C}' = \mathbf{g} \left(
    2 x^4\, d x^1 \wedge d x^2 \wedge d x^3 
    + x^3\, d x^1 \wedge d x^2 \wedge d x^4
    - x^2\, d x^1 \wedge d x^3 \wedge d x^4
    + x^1\, d x^2 \wedge d x^3 \wedge d x^4 \right)\,.
\end{equation}
with
\begin{equation}
  G = d \mathcal{C}' = \mathbf{g} \, d x^1 \wedge d x^2 \wedge d x^3 \wedge d x^4\,.
\end{equation}
The gauging \eqref{eqn:gaugingT4Gflux} represents the irrep $\mathbf{1}$ in the solution \eqref{eqn:sollinconst3_15} of the third linear constraint. Thus, this frame results from the construction in section~\ref{sec:genframe} with $\lambda = 3 \mathbf{g}$ and 
\begin{equation}
  \mathcal{C} = -3 \mathbf{g} \, x^4 \, d x^1 \wedge d x^2 \wedge d x^3 \,,
\end{equation}
resulting in the required
\begin{equation}
  \mathcal{G} = d \mathcal{C} = 3 \mathbf{g} \, d x^1 \wedge d x^2 \wedge d x^3 \wedge d x^4\,.
\end{equation}

Now, we perform a deformation of this solution by applying a $\mathcal{T}_A{}^B$ which generates the SO(5) rotation
\begin{equation} \label{eqn:rotGtoQflux}
  \mathcal{T}_a{}^b = \begin{pmatrix}
    0  & 1 & 0 & 0 & 0 \\
    -1 & 0 & 0 & 0 & 0 \\
    0  & 0 & 1 & 0 & 0 \\
    0  & 0 & 0 & 1 & 0 \\
    0  & 0 & 0 & 0 & 1
  \end{pmatrix}
    \quad \text{as} \quad
  \mathcal{T}_{a_1 a_2}{}^{b_1 b_2} =  2 \delta_{[a_1}{}^{[b_1} \mathcal{T}_{a_2]}{}^{b_2]}\,.
\end{equation}
After this rotation the subalgebra $\mathfrak{h}$ becomes non-abelian and is governed by the non-vanishing commutator relations
\begin{equation}
  [t_{\tilde 1}, t_{\tilde 2} ] = \mathbf{g}\, t_{\tilde 3}\,, \quad
  [t_{\tilde 1}, t_{\tilde 4} ] = \mathbf{g}\, t_{\tilde 5} \quad \text{and} \quad
  [t_{\tilde 2}, t_{\tilde 4} ] = \mathbf{g}\, t_{\tilde 6}\,.
\end{equation}
In this frame, solving the SC is easier than for the one we considered above. This is because the field strength $A$ vanishes automatically for $C=0$. As a result, the vielbein is trivial with $E^{\tilde a}{}_i$ vanishing whereas the remaining components $E^\alpha{}_i$ and $E^{\tilde \alpha}{}_{\tilde i}$ are equivalent to the previous results in \eqref{eqn:EAiGflux}. The generalized frame field $\hat E_A$ does not contribute to the fluxes of the background. Thus, the only contribution comes from the twist \eqref{eqn:twistcontrib}
\begin{equation}\label{eqn:QformX}
  Q_i{}^{jkl} = \mathcal{F}_i{}^{jkl} - \mathcal{F}^{jk}{}_i{}^l = 2 X_{\alpha\tilde\beta}{}^{\gamma} E^\alpha{}_i \eta^{jk,\tilde\beta} E_\gamma{}^l
\end{equation}
which is totally anti-symmetric in the indices $i$, $j$, $l$. It is convenient to recast this quantity as
\begin{equation}
  Q_{ij} = \frac1{3!} Q_i{}^{klm} \epsilon_{klmj} =  -\mathbf{g} \diag( 1,\, 0,\,  0,\, 0 )
\end{equation}
where $\epsilon_{klmj}$ is the totally anti-symmetric tensor in four dimensions. So we conclude that this background has $\mathbf{g}$ units of $Q$-flux. As it arises by a SO(5) transformation from the previous one with $\mathbf{g}$ units of $G$-flux, we found a direct realization of the duality chain \eqref{eqn:chain1}.

This gauging is in the $\mathbf{10}$ of \eqref{eqn:sollinconst3_15}. So we are able to construct the generalized frame
\begin{equation}
  \mathcal{E}_\alpha = - E_\alpha{}^i \partial_i \,, \quad
  \mathcal{E}_{\tilde\alpha} = \eta_{ij,\tilde\alpha} \beta^{ijk} \partial_k - \frac12 \eta_{ij,\tilde\alpha}\, d x^i \wedge d x^j
\end{equation}
with $\mathcal{C}=0$ and the totally anti-symmetric $\beta^{ijk}$ whose non-vanishing components read
\begin{equation}
  \beta^{234} = - \frac{\mathbf{g}}2 x^1\,.
\end{equation}
It sources the $Q$-flux
\begin{equation}
  Q_i{}^{jkl} = - 2 \partial_i \beta^{jkl}
\end{equation}
in \eqref{eqn:QformX}. An alternative way to obtain a generalized frame with the same properties is to rotate the generalized frame field in the previous duality frame \eqref{eqn:genframeT4Gflux} by $\mathcal{T}_A{}^B$ in \eqref{eqn:rotGtoQflux}.

\subsubsection{Gaugings in the \texorpdfstring{$\mathbf{40}$}{40}}\label{sec:T4_40}
In order to realize the twisted four-torus from which the second chain \eqref{eqn:chain2} starts, we consider the embedding tensor solution \eqref{eqn:embeddingt40} with the non-vanishing components \cite{Blair:2014zba}
\begin{equation}\label{eqn:ZfFlux}
  Z^{23,3} = - Z^{32,3} = \frac{\mathbf{f}}2
\end{equation}
to obtain $\mathbf{f}$ units of geometric flux. As before the structure constants of the Lie algebra $\mathfrak{g}$ arise from \eqref{eqn:emb101010b}. However, this algebra is not ten-dimensional anymore. As discussed in section~\ref{sec:linconstsl5} gaugings in the $\mathbf{40}$ reduce the dimension of the group manifold according to \eqref{eqn:dimG40}. Thus, the $G$ we consider here is nine-dimensional and admits a SL(3)$\times$SL(2) structure as shown in figure~\ref{fig:sollinconst}. Its coordinates decompose into the two irreps
\begin{equation}
  (\mathbf{3},\mathbf{2}): \{1,\,2,\,\tilde 1,\,\tilde 2,\,\tilde 3,\,\tilde 4\}
    \quad \text{and} \quad
  (\overline{\mathbf{3}}, \mathbf{1}): \{3,\,4,\,\tilde 5\}
\end{equation}
with the adapted version
\begin{equation}
  \alpha = \{12,\,13,\,14,\,15\}
    \quad \text{and} \quad
  \tilde\alpha = \{24,\,25,\,34,\,35,\,45\}
\end{equation}
of the basis \eqref{eqn:decompSL510} for the components of the $\mathbf{10}$ indices $\alpha$ and $\tilde\alpha$. In this basis, the non-vanishing commutator relations, defining $\mathfrak{g}$, read
\begin{equation}\label{eqn:defgexample40}
  [t_{\tilde 5}, t_3] = \mathbf{f}\, t_{\tilde 2}\,,\quad
  [t_{\tilde 5}, t_4] = \mathbf{f}\, t_{\tilde 4} \quad \text{and} \quad
  [t_3, t_4] = \mathbf{f}\, t_2\,.
\end{equation}
Together, the six generators appearing in these three relations form the algebra $\mathfrak{cso}(1,0,3)$ with the center $\{t_2,\,t_{\tilde 2},\,t_{\tilde 4}\}$. While the remaining $t_1$, $t_{\tilde 1}$ and $t_{\tilde 3}$ give rise to a three-dimensional abelian factor. There is a 16-dimensional faithful representation for $\mathfrak{g}$ which is presented in appendix~\ref{app:faithfulrepr}. We use it to calculate the coset elements $m$ according to \eqref{eqn:matrixexpm}, while elements of the subgroup $H$ are given by
\begin{equation}
  h = \exp( t_{\tilde 1} x^{\tilde 1}) \exp( t_{\tilde 2} x^{\tilde 2}) \exp( t_{\tilde 3} x^{\tilde 3})
  \exp(t_{\tilde 4} x^{\tilde 4}) \exp( t_{\tilde 5} x^{\tilde 5})\,.
\end{equation}
As in the duality chain in the last subsection, the identifications \eqref{eqn:coordident3} and \eqref{eqn:coordident4} on the coordinates of the group manifold are required here, too. Otherwise, we would not obtain a compact background. It is a fibration
\begin{equation}
  T^2=F \hookrightarrow M \rightarrow B=T^2
\end{equation}
where a point on the fiber $F$ is labeled by the coordinates $x^1$, $x^2$, while the base $B$ is parameterized by the remaining coordinates $x^3$ and $x^4$. The fiber is contained in the coordinate irrep $(\mathbf{2},\mathbf{3})$ and the base is part of $(\mathbf{1},\overline{\mathbf{3}})$. Again, the gauge potential $A$ vanishes for $C=0$ automatically. Thus, there is a solution of the SC with the generalized background vielbein
\begin{equation}\label{eqn:EAifflux}
  E^{\alpha}{}_i = \begin{pmatrix}
    1 & 0 & 0 & 0 \\ 0 & 1 & f x^4 & 0 \\ 0 & 0 & 1 & 0 \\ 0 & 0 & 0 & 1
  \end{pmatrix}\,, \quad
  E^{\tilde\alpha}{}_i = - \mathbf{f} \begin{pmatrix}
    0 & 0 & 0 & 0 \\
    0 & 0 & x^{\tilde 5} & 0 \\
    0 & 0 & 0 & 0 \\
    0 & 0 & 0 & x^{\tilde 5} \\
    0 & 0 & 0 & 0
  \end{pmatrix}
    \quad \text{and} \quad
  E^{\tilde\alpha}{}_{\bar i} = \begin{pmatrix}
    1 & 0 & 0 & 0 & 0 \\
    0 & 1 & 0 & 0 & 0 \\
    0 & 0 & 1 & 0 & 0 \\
    0 & 0 & 0 & 1 & 0 \\
    0 & 0 & 0 & 0 & 1 \\
  \end{pmatrix}\,.
\end{equation}
It comes with the non-vanishing geometric flux
\begin{equation}
  f^2 = \partial_{[i} E^2{}_{j]} d x^i \wedge d x^j = - \mathbf{f}\, d x^3 \wedge d x^4
\end{equation}
along the same lines as the \DFTwzw{} example three-torus with $f$-flux in \cite{Hassler:2016srl}. As for \DFTwzw{}, the twist term in the generalized Lie derivative \eqref{eqn:gengeometr} vanishes for this background.

It is instructive to take a closer look at the GG of this setup. Because the group manifold does not have the full ten dimensions things are more subtle.  Remember that in general the SC of SL(3)$\times$SL(2) EFT admits two different solutions. First, there are those reproducing eleven-dimensional supergravity with three internal directions and second there are solutions resulting in ten-dimensional type IIB (only two internal directions)\cite{Hohm:2015xna}. This fact is manifest from the SL(5) perspective we take. Each solution of \eqref{eqn:SL4SC} is labeled by a distinct $v^0_a$ in the $\mathbf{5}$ of SL(5) which branches as
\begin{equation}
  \mathbf{5}\rightarrow (\mathbf{1},\mathbf{2}) + (\mathbf{3}, \mathbf{1})
\end{equation}
to SL(3)$\times$SL(2). The first irrep in this direct sum captures SC solutions with a eleven-dimensional SUGRA description and the second one corresponds to type IIB. The latter case is implemented on the two-dimensional fiber $F$. Furthermore, the splitting of $M$ into base $B$ and fiber $F$ allows to distinguish between three different kinds of two-forms on $\Lambda^2 T^* M$. Those with all legs on the base or the fiber and further the ones with one leg on the base and the other leg on the fiber. Over each point $p$ of $M$, $\Lambda^2 T_p^* M$ is a six dimensional vector space. Nevertheless $\mathfrak{h}$, is only five-dimensional. Hence, the map $\eta_p$ in \eqref{eqn:etamap} is not bijective anymore. This is a problem because this property is essential to our construction in section~\ref{sec:gg}. However, we restore it by removing two-forms whose legs are completely on the base from the codomain of $\eta_p$. They are not part of the resulting GG. Apart from that \eqref{eqn:genLieGG} is still valid. Especially, we are able to construct the generalized frame field $\mathcal{E}_A$ because the gauging for this example is the surviving $(\mathbf{1},\mathbf{2})$ of \eqref{eqn:sollinconst3_40_9}. For the commutator relations in \eqref{eqn:defgexample40}, one sees that the resulting physical manifold $M$ is not a symmetric space because both $[\mathfrak{h},\mathfrak{m}]\subset\mathfrak{m}$ and $[\mathfrak{m},\mathfrak{m}]\subset\mathfrak{h}$ are violated. Still one is able find a SC solution (as we did) because $\mathfrak{m}$ is a subalgebra of $\mathfrak{g}$ with $[\mathfrak{m},\mathfrak{m}]\subset \mathfrak{m}$.  The corresponding generalized frame field is
\begin{equation}
  \mathcal{E}_\alpha = E'_\alpha{}^i \partial_i\,, \quad
  \mathcal{E}_{\tilde\alpha} = \frac12 \eta_{\beta\gamma,\tilde\alpha} 
    {E'}^\beta{}_i {E'}^\gamma{}_i \, d x^i \wedge d x^j
\end{equation}
with the frame
\begin{equation}
  E'_\alpha{}^i = \begin{pmatrix} -1 & 0 & 0 & 0 \\
    0 & -1 & 0 & 0 \\
    0 & 0 & -1 & 0 \\
    0 & x^3 \mathbf{f} & 0 & - 1
  \end{pmatrix}
    \quad \text{and the dual} \quad
  {E'}^\alpha{}^i = \begin{pmatrix} 
    -1 &  0 &  0 &  0 \\
     0 & -1 &  0 & - x^3 \mathbf{f} \\
     0 &  0 & -1 &  0 \\
     0 &  0 &  0 & -1
  \end{pmatrix}\,.
\end{equation}
However, this step is redundant because the twist $\mathcal{F}_{\hat I\hat J}{}^{\hat K}$ already vanished for $\hat E_A$.\enlargethispage{1em}

Let us finally come to the dual background with $R$-flux in \eqref{eqn:chain2}. For our specific choice of $v^0_a$ in \eqref{eqn:v0acanonical}, it is completely fixed by the four independent components $Z^{a1,1}$ ($a$=1, \dots, 4) of the $\mathbf{40}$ in the embedding tensor \eqref{eqn:embeddingt40} \cite{Blair:2014zba}. Clearly the SO(5) transformation
\begin{equation}\label{eqn:trafof->R}
  \mathcal{T}_a{}^b = \begin{pmatrix}
    0  & 0 & 1 & 0 & 0 \\
    0  & 1 & 0 & 0 & 0 \\
    -1 & 0 & 0 & 0 & 0 \\
    0  & 0 & 0 & 1 & 0 \\
    0  & 0 & 0 & 0 & 1
  \end{pmatrix}
\end{equation}
brings \eqref{eqn:ZfFlux} into this form. However, there are two problems with the resulting setup. First, the generators $t_{\tilde\alpha}$ do not yield a subalgebra $\mathfrak{h}$ after applying $\mathcal{T}$. In \DFTwzw{}, we face the same situation. It is in agreement with the completely non-geometric nature of the $R$-flux. If we would find a SC solution with our technique for the $R$-flux, there would be a geometric interpretation in terms of a manifold $M$ equipped with a GG. This is not the case. But there is also another subtlety which is absent in \DFTwzw{}. Remember that SL(5) gets broken to SL(3)$\times$SL(2) for the torus with geometric flux because the corresponding structure constants originate from the $\mathbf{40}$. But the transformation \eqref{eqn:trafof->R} is not an element of this reduced symmetry group. Hence, the second background in the chain \eqref{eqn:chain2} does not admit the most general SC solutions we discuss in this paper. There are of course still solutions, where the fluctuations are constant.

\subsection{Four-Sphere with \texorpdfstring{$G$}{G}-Flux}
In order to obtain a four-sphere with radius $R$ as the physical manifold $M$ after solving the SC, we have to consider the group manifold SO(5). It arises from a embedding tensor solution in the $\mathbf{15}$ with
\begin{equation}
  Y_{ab} = -\frac4R \, \diag(1,\,1,\,1,\,1,\,1)\,.
\end{equation}
In comparison to the previous examples in section~\ref{sec:T4Gflux} it is much simpler to obtain a faithful representation of the corresponding Lie algebra $\mathfrak{g}$=$\mathfrak{so}$(5). A canonical choice are the anti-symmetric matrices
\begin{equation}
  (t_A)_b{}^c = -\frac12 X_{Ab}{}^c
\end{equation}
which arise directly from the embedding tensor \eqref{eqn:emb1055b} and act on the fundamental irrep of $\mathfrak{g}$. In contrast to \eqref{eqn:matrixexpm}, we now parameterize coset representatives by
\begin{align}\label{eqn:mS4}
  m = \exp\Big[ R \, \phi^1  &\left(\cos(\phi^2) t_1 + \sin(\phi^2)\cos(\phi^3) t_2 + \right.\nonumber \\ 
  &\left. \sin(\phi^2) \sin(\phi^3) \cos(\phi^4) t_3 + \sin(\phi^2) \sin(\phi^3) \sin(\phi^4) t_4\right) \Big] \,,
\end{align}
where the angels represent spherical coordinates with
\begin{equation}
  \phi^1\,,\, \phi^2\,,\,  \phi^3 \in [0, \pi] \quad \text{and} \quad
  \phi^4 \in [0, 2\pi)\,,
\end{equation}
while the elements of the subgroup still are calculated by \eqref{eqn:matrixexpm}. Together $\mathfrak{m}$ and $\mathfrak{h}$ form a symmetric part. As shown at the end of section~\ref{sec:linkvomega}, the particular choice \eqref{eqn:mS4} for $m$ has the advantage that the gauge potential $A$ automatically vanishes for
\begin{equation}
  C = R^3 \tan\left(\frac{\phi^1}2\right) \sin^3(\phi^1) \sin^2(\phi^2) \sin(\phi^3) \, d \phi^2 \wedge d \phi^3 \wedge d \phi^4 \,.
\end{equation}
The corresponding field strength
\begin{equation}
  G_{\hat E} = d C = 4 R^3 \cos\left(\frac{\phi^1}2\right)\sin^3\left(\frac{\phi^1}2\right)\left((1+3\cos(\phi^1)\right) \sin^2 (\phi^2) \sin (\phi^3 ) \, d \phi^1 \wedge d \phi^2 \wedge d \phi^3 \wedge d \phi^4
\end{equation}
is in the trivial cohomology class of $H^4_{\mathrm{dR}(S^4)}$ because the integral
\begin{equation}
  \int_{S^4} G_{\hat E} = 0
\end{equation}
is zero. At the same time $C$ and therewith the connection $A^{\tilde\alpha}$ are globally well-defined. In contrast to the four-torus with $G$-flux in the last section, we can gauge away the connection globally although the background has $G$-flux in a non-trivial cohomology class, too. Another interesting quantity one can compute is the first Pontryagin class for the connection
\begin{equation}
  \mathcal{A}^{\tilde\alpha} = E^{\tilde\alpha}{}_i \, d x^i\,.
\end{equation}
This quantity is analogous to the Chern classes, we computed for the $T^6$-bundle in the $T^4$ with $G$-flux background. It vanishes completely.

To write down the generalized frame field \eqref{eqn:genframe&dual}, we furthermore need the vielbein
\begin{equation}
  E^\alpha{}_i = R \begin{pmatrix}
    c_2 & - s_1 s_2 & 0 & 0 \\
    c_3 s_2 & c_2 c_3 s_1 & - s_1 s_2 s_3 & 0 \\
    c_4 s_2 s_3 & c_2 c_4 s_1 s_3 & c_3 c_4 s_1 s_2 & - s_1 s_2 s_3 s_4 \\
    s_2 s_3 s_4 & c_2 s_1 s_3 s_4 & c_3 s_1 s_2 s_4 & c_4 s_1 s_2 s_3
  \end{pmatrix}
\end{equation}
with $c_i = \cos(\phi^i)$ and $s_i = \sin(\phi^i)$ which is part of the left-invariant Maurer-Cartan form $E^A{}_I$ in \eqref{eqn:compbgvielbein}. It gives rise to the metric
\begin{equation}
  d s^2 = E^\alpha{}_i \delta_{\alpha\beta} E^\beta{}_j \, d \phi^i d \phi^j = R^2 \left( (d \phi^1)^2 +
    s_1^2  (d \phi^2)^2 + s_1^2 s_2^2 (d \phi^3)^2 + s_1^2 s_2^2 s_3^2 (d \phi^3)^2 \right)
\end{equation}
on a round sphere with radius $R$. Equipped with a solution of the SC for $G$=SO(5), we are able to apply the construction in section~\ref{sec:genframe} and obtain the generalized frame field $\mathcal{E}_A$ with $\mathcal{C}$ such that
\begin{equation}\label{eqn:mcG_S^4}
  \mathcal{G} = d \mathcal{C} = 3 R^3 \sin^3(\phi^1) \sin^2(\phi^2) \sin(\phi^3) \, 
  d \phi^1 \wedge d \phi^2 \wedge d \phi^3 \wedge d \phi^4 = \frac3R \mathrm{vol}\,.
\end{equation}
Because the complete result is not very compact, we do not present it here. Instead, we present an alternative parameterization of the group elements $m$ in terms of Cartesian coordinates
\begin{align}
  y^1 &= R \cos(\phi^1)  &
  y^2 &= R \sin(\phi^1) \cos(\phi^2) \nonumber \\
  y^3 &= R \sin(\phi^1) \sin(\phi^2) \cos(\phi^3) &
  y^4 &= R \sin(\phi^1) \sin(\phi^2) \sin(\phi^3) \cos(\phi^4) \nonumber \\
  y^5 &= R \sin(\phi^1) \sin(\phi^2) \sin(\phi^3) \sin(\phi^4)\,.
\end{align}
They have the advantage that they yield a very simple coset representative
\begin{equation}
  m = \frac{1}{R} \begin{pmatrix}
    y^1 & -y^2 & -y^3 & -y^4 & -y^5 \\
    y^2 & y^{22} & y^{23} & y^{24} & y^{25} \\
    y^3 & y^{23} & y^{33} & y^{34} & y^{35} \\
    y^4 & y^{24} & y^{34} & y^{44} & y^{45} \\
    y^5 & y^{25} & y^{35} & y^{45} & y^{55}
  \end{pmatrix}
    \quad \text{with} \quad
  y^{ij} = R \delta^{ij} - \frac{y^i y^j}{R + y^1}
\end{equation}
and allow a direct comparison of our results with \cite{Lee:2014mla}. On the order hand, we have to implement the additional constraint
\begin{equation}
  \sum_{i=1}^5 (y^i)^2 = R
\end{equation}
in all equations that follow. As before, we calculate
\begin{equation}
  E^\alpha{}_i = \frac1R \begin{pmatrix}
    -y^2 & y^{22} & y^{23} & y^{24} & y^{25} \\
    -y^3 & y^{23} & y^{33} & y^{34} & y^{35} \\
    -y^4 & y^{24} & y^{34} & y^{44} & y^{45} \\
    -y^5 & y^{25} & y^{35} & y^{45} & y^{55}
  \end{pmatrix} = E_\alpha{}^i 
\end{equation}
for the components of the left-invariant Maurer-Cartan form. Finding the vectors $E_\alpha{}^i$ is a bit more challenging here than before because $E^\alpha{}_i$ is not a square matrix and therefore not invertible. However, it is completely fixed by $E_\alpha{}^i E^\beta{}_i = \delta_\alpha^\beta$ and furthermore requiring that all vectors $E_\alpha{}^i$ are perpendicular to the radial direction $\vec r$=$(y^1\,y^2\,y^3\,y^4\,y^5)^T$. Now, we calculate the vector part $\mathcal{E}_A{}^i$ of the generalized frame which we denote as $V_A{}^i$ in order to permit a direct comparison of our results with the ones in \cite{Lee:2014mla}. Its components are
\begin{equation}
  V_A{}^i = \frac1R \left( \delta_{a1}^i y^{a2}  - \delta_{a2}^i y^{a1} \right)
\end{equation}
where we split the $\mathbf{10}$ index $A$ into the two fundamental indices $a_1$ and $a_2$. One can check that they generate the algebra $\mathfrak{so}(5)$ under the Lie derivative $L$, namely
\begin{equation}\label{eqn:so5vectors}
  L_{V_A} V_B = X_{AB}{}^C V_C\,.
\end{equation}
Furthermore, it is convenient to study the two-forms
\begin{equation}
  \sigma_A = \frac12 \mathcal{E}_A{}_{ij} \, d y^i \wedge d y^j
\end{equation}
which evaluate to
\begin{equation}
  \sigma_A = - \frac1R \epsilon_{a_1 a_2 ij} d y^i \wedge d y^j\,.
\end{equation}
Analogous to \eqref{eqn:so5vectors}, they generate the Lie algebra $\mathfrak{g}$ under the Lie derivative
\begin{equation}
  L_{V_A} \sigma_B = X_{AB}{}^C \sigma_C\,.
\end{equation}
Finally, we need the volume form
\begin{align}
  \mathrm{vol} &= \frac{1}{4!} \epsilon_{1\hat\alpha\hat\beta\hat\gamma\hat\delta} 
    E^\alpha{}_i E^\beta{}_j E^\gamma{}_k E^\delta{}_l \, d y^i \wedge d y^j \wedge d y^k \wedge d y^l
     \nonumber \\
    & = \frac1{4! R} \epsilon_{ijklm} y^i d y^j \wedge d y^k \wedge d y^l \wedge d y^m
\end{align}
which fulfills the relation\cite{Lee:2014mla}\footnote{In comparison to \cite{Lee:2014mla}, we use structure coefficients $X_{AB}{}^C$ with the opposite sign. For example, we have $X_{\tilde 1\tilde 2}{}^{\tilde 3}=X_{23,24}{}^{34}$=$R^{-1}$ while from (2.5) in \cite{Lee:2014mla} one gets $X_{23,24}{}^{34}$=$-R^{-1}$. So the vectors $V_A$ and the forms $\sigma_A$, which we calculate, also have a flipped sign compared to their results. However, \eqref{eqn:connectionVAsigmaA} is the same.}
\begin{equation}\label{eqn:connectionVAsigmaA}
  \iota_{V_A} \mathrm{vol} = \frac{R}3 d \sigma_A\,.
\end{equation}
Hence, we reproduce all ingredients which were discussed in \cite{Lee:2014mla} to show that the $S^4$ with four-form flux is parallelizable. Following this paper, we plug the generalized frame field
\begin{equation}
  \mathcal{E}_A = V_A + \sigma_A + \iota_{V_A} \mathcal{C} 
\end{equation}
into the generalized Lie derivative
\begin{equation}
  \widehat{\mathcal{L}}_{\mathcal{E}_A} \mathcal{E}_B = L_{V_A} V_B + L_{V_A} \sigma_B + \iota_{[V_A,V_B]} \mathcal{C} - \iota_{V_B} ( d \sigma_A  - \iota_{V_A} d \mathcal{C} )
\end{equation}
where the last term vanishes for a $\mathcal{C}$ governed by \eqref{eqn:mcG_S^4}. In principal, we could scale $\sigma_A$ and $\mathcal{C}$ by the same constant factor to obtain another generalized frame field which still fulfills \eqref{eqn:genlieongenviel}. In \cite{Lee:2014mla} it was fixed by imposing appropriate equations of motion. It is interesting to have a closer look at these equations. They originate from eleven-dimensional SUGRA with the action
\begin{equation}\label{eqn:S11dSUGRA}
  S = \frac{1}{2 \kappa_{11}^2} \int d^{11} x \sqrt{-G} \left( \mathcal{R} - \frac{1}2 | d \mathcal{C} |^2 \right)
\end{equation}
for the bosonic sector. $G$ is the metric in eleven dimensions, $\mathcal{R}$ denotes the corresponding curvature scalar and $\mathcal{C}$ is a three-form gauge field. Using the Freund-Rubin ansatz \cite{Freund:1980xh} to solve the equations of motion for this action on the spacetime AdS$_7 \times$S$^4$, we find
\begin{equation}
  \mathcal{R}_{S^4} = \frac{12}{R^2} = \frac{4}{3} | d \mathcal{C} |^2
    \quad \text{or} \quad
  |\mathcal{G}|^2 = \frac9{R^2}\,.
\end{equation}
Furthermore after applying the relations
\begin{equation}
  \mathcal{G} \wedge \star \,\mathcal{G} = | F |^2 \, \mathrm{vol} \quad \text{and} \quad \star \mathrm{vol} = 1
\end{equation}
where $\star$ is the Hodge star operator on the $S^4$, one finds that this result is in agreement with \eqref{eqn:mcG_S^4}. Clearly, this results depends on the relative factors between the two terms in the action \eqref{eqn:S11dSUGRA} which are fixed by supersymmetry. In SL(5) EFT this particular reltaion between the gravity sector and the form-field originates from the requirement that the generalized frame field is an SL(5) element. Naively, $\mathcal{E}_A{}^{\hat I}$ has 100 independent components. They can be organized according to
\begin{equation}
  \mathbf{10} \times \overline{\mathbf{10}} = \xcancel{\mathbf{1}} + \mathbf{24} + \xcancel{\mathbf{75}}\,,
\end{equation}
but only the ones in the adjoint irrep are non-vanishing. This property is automatically implemented in our approach as can be seen from \eqref{eqn:genparaframe}. The generalized frame field $\hat E'_B{}^{\hat I}$ has the frame field $E_\beta{}^i$ and the three-form $\mathcal{C}$ as constituents. They furnish the irreps $\mathbf{1}+\mathbf{15}$ and $\mathbf{4}$ of SL(4) which arise from the branching
\begin{equation}
  \mathbf{24} \rightarrow \mathbf{1} + \mathbf{4} + \overline{\mathbf{4}} + \mathbf{15}
\end{equation}
of SL(5). Thus, $\hat E'_B{}^{\hat I}$ is an SL(5) element. By construction, $M_A{}^B$ shares this property. So it is no surprise that $\mathcal{E}_A{}^{\hat I}$, which results from the multiplication of the two, is an SL(5) element, too. As a consequence, we find the correct scaling factor for the four-form flux. 

\section{Conclusion}\label{sec:conclusion}
In this work, we present a technique to explicitly construct the generalized frame fields for generalized parallelizable coset spaces $M$=$G/H$ in four dimensions. It is based on the idea of making the Lie group $G$ in the extended space of EFT manifest. This can be done in the framework of gEFT \cite{Bosque:2016fpi} and is closely related to the concept underlying \DFTwzw{} \cite{Blumenhagen:2014gva,Blumenhagen:2015zma,Bosque:2015jda,Hassler:2016srl}. As we discuss in the first part of this paper, there are several restrictions on $G$. They are closely related to the embedding tensor of the U-duality group SL(5) in four dimensions. We only use the extended space as a technical tool. In the end, one has to get rid of all the unphysical directions in this space by solving the SC. It deviates in gEFT from the SC known in the conventional formulation \cite{Bosque:2016fpi}. Collecting clues from \DFTwzw{} \cite{Hassler:2016srl}, we are able to solve it by choosing a particular embedding of the physical subspace $M$ in $G$. Each SC solution comes with a canonical generalized frame field $\hat E_A$ and a GG governed by a twisted generalized Lie derivative. But for a generalized parallelizable space, the frame field is defined with respect to the untwisted generalized Lie derivative. As a consequence, in a last step we need to modify $\hat E_A$ such that it adsorbs the twisted part and becomes $\mathcal{E}_A$ for which the defining relation of a generalized parallelization \eqref{eqn:genlieongenviel} holds. There are three linear constraints which are required for the steps outlined above to go through. After solving them, we find among other things that our construction applies to all gaugings which are purely in the $\mathbf{15}$. The corresponding generalized frames are already known from \cite{Lee:2014mla,Hohm:2014qga}. However, we are also able to treat groups $G$ which originate from the $\mathbf{40}$. Their dimension is smaller than ten and the U-duality group SL(5) has to be broken. Nevertheless, our results still apply and we provide an example in section~\ref{sec:T4_40}.

Let us finally mention that this paper marks an important success in the \DFTwzw{}/gEFT program the authors have initiated by approaching a long standing question in the DFT/EFT and GG community: How to systematically construct generalized frame fields satisfying all required consistency conditions? Of course, we did not completely answer this question. But we presented all necessary tools for SL(5) EFT. Especially, the treatment in the sections~\ref{sec:dftwzwmotiv}-\ref{sec:genLie} apply to other U-duality groups, too. Hence, there does not seem to be an obstruction to extend the results in this work to other dimensions appearing in table~\ref{tab:Udualitygroups}. Studying the required linear constraints, one should be able to find a large class of generalized parallelizable spaces $M$ with $\dim M$ $\ne$ 4. Because of the very close connection between these spaces, maximal gauged supergravities and the embedding tensor formalism, one might even hope to eventually obtain a full classification of them. The significance of such a classification for the understanding of consistent coset reductions was already emphasised in the introduction of this paper.

\acknowledgments
We gratefully acknowledge that Emanuel Malek was involved in the initial stages of this project. We like to thank him for many important discussions during the preparation of this paper. We also would like to thank
David Berman,
Martin Cederwall,
Olaf Hohm,
Arnau Pons Domenech,
and
Daniel Waldram
for helpful discussions. Furthermore, FH thanks the ITS at the City University of New York and the theory group at Columbia University for their hospitality during parts of this project. The work of PB and DL is supported by the ERC Advanced Grant ``Strings and Gravity" (Grant No. 320045). FH is supported by the NSF CAREER grant PHY-1452037 and also acknowledges support from the Bahnson Fund at UNC Chapel Hill as well as the R. J. Reynolds Industries, Inc. Junior Faculty Development Award from the Office of the Executive Vice Chancellor and Provost at UNC Chapel Hill and from the NSF grant PHY-1620311.

\appendix
\section{SL(\texorpdfstring{$n$}{n}) Representation Theory}\label{app:SLnrepresentations}
In the first part of this appendix, we review how projectors on $\mathfrak{sl}(N)$ irreps can be constructed from Young symmetrizers. Further, we show how these projectors are used to explicitly decompose tensor products of irreps into direct sums. As first application of this concept, the linear constraints encountered in section~\ref{sec:genLie} are solved for the T-duality group SL(4) in the second part.

\subsection*{Theory: Young Tableaux and Projectors on Irreps}
Let us fix some convention first: A Young diagram is a set of $n$ boxes which are arranged in rows and columns starting from the left. The number of boxes in each row may not increase while going from the top of the diagram to the bottom. An example for $n=6$ is
\begin{equation}\label{eqn:ex321}
  \ydiagram{3,2,1}\,.
\end{equation}
It is in one-to-one correspondence to the partition $(3,2,1)$ of $6$. A diagram becomes a Young tableau, if we write the numbers from one to $n$ into the boxes. In general, there are $n!$ different ways to do so. If the numbers in a tableau are increasing in every row and column at the same time, it is called a standard tableau. The number of standard tableaux for a given diagram can be calculated by the hook length formula: For each box in a diagram $\lambda$ one counts the number of boxes in the same row $i$ to its right and boxes in the same column $j$ below it. For the box itself, one has to add one to the result to obtain the hook length $h_\lambda(i,j)$. From this data the number of standard tableaux is calculated as
\begin{equation}
  d_{\mathrm{std}} = \frac{n!}{\prod h_\lambda(i,j) }\,.
\end{equation}
Take the example~\ref{eqn:ex321}, here we obtain
\begin{equation*}
  \ytableaushort{531,31,1} \quad\text{for each box and }
  d_{\mathrm{std}} = \frac{6!}{5\cdot3^2} = 16\,.
\end{equation*}
Starting from a Young tableau $t$, one combines all permutations from the symmetric group $S_n$ which only shuffle elements within each row of $t$ into the row group $R_t$. Similarly, all permutations which only shuffle elements in columns are assigned to the column group $C_t$. Together $R_t$ and $C_t$ give rise to the Young symmetrizer
\begin{equation}
  e_t = \sum_{\pi\in R_t,\,\sigma\in C_t} \mathrm{sign}(\sigma) \sigma\circ\pi\,.
\end{equation}
An instructive example is
\begin{equation}\label{eqn:ytabexample}
  t = \ytableaushort{12,3} \quad\text{and}\quad
  e_t = \big( () - (1\,3) \big)\big( () + (1\,2) \big) = () + (1\,2) - (1\,3) - (3\,2\,1)\,,
\end{equation}
where we use cycle notation for elements of $S_3$. We are interested in applying $e_t$ to tensors, such as $X_{a_1 \dots a_n}$, where the permutations act on the indices. For instance, with the tableau $t$ from \eqref{eqn:ytabexample} we find
\begin{equation}\label{eqn:youngsym21}
  e_t X_{a_1 a_2 a_3} = X_{a_1 a_2 a_3} + X_{a_2 a_1 a_3} - X_{a_3 a_2 a_1} - X_{a_2 a_3 a_1}\,.
\end{equation}
It is straightforward to check that the resulting tensor is anti-symmetric with respect to the first two indices $a_1$, $a_2$ and moreover the total antisymmetrization $X_{[a_1 a_2 a_3]}$ vanishes. If the indices $a_i=1,\cdots,N$ are in the fundamental of $\mathfrak{sl}(N)$, the resulting tensor $e_t X_{a_1 a_2 a_3}$ is an irrep of the Lie algebra. Thus, the Young symmetrizer $e_t$ is proportional to the projector from a tensor product ($X_{a_1 a_2 a_3}$ is nothing else) to this irrep. This works for all other Young tableaux as well. In order to calculate the dimension of the irrep we project onto from the tableaux $t$, we first have to assign the number $N$ to the top left corner of the diagram $\lambda$ corresponding to $t$. In each column to the right we increase the number and in each row towards the bottom we decrease it. Taking again the diagram~\ref{eqn:ex321} as an instructive example, we have
\begin{equation*}
  \ytableausetup{mathmode, boxsize=2em}
  \begin{ytableau}
    \scriptstyle N & \scriptstyle N+1 & \scriptstyle N+2 \cr
    \scriptstyle N-1 & \scriptstyle N  \cr
    \scriptstyle N-2 \cr
  \end{ytableau}\,.
  \ytableausetup{boxsize=1em,aligntableaux=center}
\end{equation*}
These numbers are denoted in analogy to the hook length by $f_\lambda(i,j)$. Finally, the dimension of the irrep associated to $t$ is
\begin{equation}
  d_{\mathrm{irrep}} = \frac{\prod f_\lambda(i,j)}{\prod h_\lambda(i,j)}\,,
\end{equation}
which gives rise to the dimension $N(N^2-1)/3$ for the Young symmetrizer \eqref{eqn:youngsym21}. For $N=5$ this yields $\overline{\mathbf{40}}$, exactly one of the two irreps in the embedding tensor.

As already mentioned $e_t$ is only proportional to a projector and fulfills
\begin{equation}
  e_t e_t = k_t e_t\,,
\end{equation}
where $k_t$ is a constant depending on the tableau $t$. We use this to define the projector onto $t$ as
\begin{equation}
  P_t = \frac{1}{k_t} e_t \quad \text{with} \quad P_t^2 = P_t\,.
\end{equation}
Such projectors come with the following properties:
\begin{itemize}
  \item Projectors of tableaux corresponding to different diagrams are orthogonal.
  \item Projectors of standard tableaux are linear independent. They can be combined to a system of orthogonal projects $P_{\lambda, i}$. Here $\lambda$ labels the diagram they decent from.
  \item The sum of all these projectors for all diagrams with $n$ boxes is the identity of $S_n$.
\end{itemize}
Now, assume that we have a projector $P$ to a reducible representation and want to decompose it into a sum of orthogonal projectors $P_{\lambda, i}$ onto irreps as
\begin{equation}\label{eqn:decompP}
  P = \sum_\lambda \sum_i P_{\lambda, i}\,.
\end{equation}
As mentioned above, these orthogonal projectors arise from a sum
\begin{equation}
  P_{\lambda, i} = \sum\limits_t (c_{\lambda, i})_t e_t \circ P
\end{equation}
over different projectors originating from standard tableaux for a specific diagram $\lambda$. However, the coefficients $(c_{\lambda, i})_t$ in this expansion still have to be fixed. This can be done by requiring that the commutator
\begin{equation}
  [P, P_{\lambda, i}] = \sum_t (c_{\lambda, i})_t [e_t \circ P, P] = 0
\end{equation}
of $P$ with each of the $P_{\lambda, i}$ vanishes. For the resulting nullspace an orthonormal basis is chosen:
\begin{equation}
  P_{\lambda, i} \circ P_{\lambda, j} = \begin{cases} P_{\lambda, i} & i = j\\ 0 & i \ne j\,. \end{cases}
\end{equation}

\subsection*{Application: Linear Constrains for SL(4)}
In order to solve the linear constraints from section~\ref{sec:genLie}, we first decompose the constraint quantity $\Gamma_{AB}{}^C$ into irreps. Here, we work with the Lie algebra $\mathfrak{sl}(4)$, thus indices denoted by capital letters are in the irrep $\mathbf{6}$ and small letter indices label the fundamental representation $\mathbf{4}$. As explained in the first part of this appendix, Young symmetrizer act on the latter representation. Hence, we first translate
\begin{equation}
  \Gamma_{AB}{}^C \rightarrow \Gamma_{[a_1 a_2], [b_1 b_2]}{}^{[c_1 c_2]}\,.
\end{equation}
We further have to distinguish between raised and lowered indices. While the former live in the $\mathbf{6}$, the latter are in dual $\overline{\mathbf 6}$\footnote{Note that for $\mathfrak{sl}(4)$ the six-dimensional representation is real, e.g. $\mathbf{6} = \overline{\mathbf 6}$. Thus, in general we do not need to distinguish between the two of them. However, it is still a good bookkeeping device.}. Changing from an irrep to its dual is done by contraction with the totally anti-symmetric tensor
\begin{equation}\label{eqn:lowerdouleind}
  \Gamma_{a_1 a_2, b_1 b_2, c_1 c_2} = \Gamma_{a_1 a_2, b_1 b_2}{}^{d_1 d_2} \epsilon_{d_1 d_2 c_1 c_2}\,.
\end{equation}
In total, the connection has 216 independent components which are organized through the following irreps
\begin{equation}\label{eqn:666}
  \mathbf{6}\times\mathbf{6}\times\overline{\mathbf 6} = 3 (\mathbf{6}) + \mathbf{10} + \overline{\mathbf 10} + \mathbf{50} + 2 (\mathbf{64})\,.
\end{equation}
They are in one-to-one correspondence with their Young diagrams
\begin{equation}
  \ydiagram{1,1}\times\left(\,\ydiagram{1,1}\times\ydiagram{1,1}\,\right) = 3\, \ydiagram{2,2,1,1} + \ydiagram{3,1,1,1} + \ydiagram{2,2,2} + \ydiagram{3,3} + 2\, \ydiagram{3,2,1}\,.
\end{equation}
On the right hand side of this equation we identify the projector
\begin{equation}\label{eqn:P666}
  P_{\mathbf{6}\times\mathbf{6}\times\overline{\mathbf 6}} = \frac18 \big( () - (1\,2) \big) \big( () - (3\,4) \big) \big( () - (5\,6) \big)
\end{equation}
on a reducible representation. In order to correctly decompose it into a sum \eqref{eqn:decompP} of projectors onto irreps, we further have to take the diagrams $(1,1,1,1,1,1)$ and $(2,1,1,1,1)$ into account, even if they clearly vanish for $\mathfrak{sl}(4)$. Still, they contribute to the decomposition into irreps of the symmetric group $S_6$. While the first gives rise to one projector, the second leads to two orthogonal projectors. As in \eqref{eqn:P666}, we suppress their contribution in the following. When a diagram appears more than once in a decomposition, there are different ways to organize the corresponding projectors. Here we use the following scheme:
\begin{equation}\label{eqn:P666names}
  \mathbf{6} \times ( \mathbf{6}\times \overline{\mathbf 6} ) = \mathbf{6} \times (\mathbf{1} + \mathbf{15} + \mathbf{20}') = \left\{ \begin{array}{ll}
    \mathbf{6} \times \mathbf{1} &= \mathbf{6}a \\
    \mathbf{6} \times \mathbf{15} &= \mathbf{6}b + \mathbf{10} + \overline{\mathbf 10} + \mathbf{64}a \\
    \mathbf{6} \times \mathbf{20}' &= \mathbf{6}c + \mathbf{50} + \mathbf{64}b
  \end{array}\right. \,.
\end{equation}
In this convention, we can finally write down the resulting decomposition
\begin{equation}\label{eqn:P6662}
  P_{\mathbf{6}\times\mathbf{6}\times\overline{\mathbf 6}} = P_{\mathbf{6}a} + P_{\mathbf{6}b} + P_{\mathbf{6}c} + P_{\mathbf{10}} + P_{\overline{\mathbf 10}} + P_{\mathbf{50}} + P_{\mathbf{64}a} + P_{\mathbf{64}b}\,.
\end{equation}
Now we are ready to discuss the first linear constraint \eqref{eqn:linconst1}. In fundamental indices, it reads
\begin{align}
  C_{a_1 a_2, b_1 b_2, c_1 c_2, d_1 d_2, e_1 e_2} =&\, \epsilon_{a_1 a_2 b_1 b_2} ( - \Gamma_{c_1 c_2, d_1 d_2, e_1 e_2} - \Gamma_{c_1 c_2, e_1 e_2, d_1 d_2 } ) +  \nonumber \\
  &\, \epsilon_{d_1 d_2 e_1 e_2} ( \Gamma_{c_1 c_2, b_1 b_2, a_1 a_2} - \Gamma_{c_1 c_2, a_1 a_2, b_1 b_2 } )
\end{align}
after substituting the $Y$-tensor
\begin{equation}
  Y^{a_1 a_2, b_1 b_2}{}_{c_1 c_2, d_1 d_2} = \frac{1}{4}\epsilon^{a_1 a_2 b_1 b_2} \epsilon_{c_1 c_2 d_1 d_2}
\end{equation}
and lowering all indices as described in \eqref{eqn:lowerdouleind}. Clearly, this expression vanishes if the terms in the brackets next to the totally anti-symmetric tensors vanish. They are not independent. Thus, we are left with the constraint
\begin{equation}
  \Gamma_{a_1 a_2, b_1 b_2, c_1 c_2} + \Gamma_{a_1 a_2, c_1 c_2, b_1 b_2} = 0\,,
\end{equation}
which we recast in terms of a projector
\begin{equation}
  2 P_1 \Gamma_{a_1 a_2, b_1 b_2, c_1 c_2} = 0 \quad \text{with} \quad
  P_1 = \frac{1}{2} \big( () + (3\,5) (4\,6) \big)\,.
\end{equation}
All irreps in the decomposition \eqref{eqn:P6662} that are not in the nullspace of this projector and thus violate \eqref{eqn:linconst1} have to vanish. To this end, we replace \eqref{eqn:P6662} by
\begin{equation}
  (1 - P_1) P_{\mathbf{6}\times\mathbf{6}\times\overline{\mathbf 6}} = 
    P_{\mathbf{6}b} + P_{\mathbf{10}} + P_{\overline{\mathbf 10}} +    P_{\mathbf{64}a}\,.
\end{equation}
Intriguingly, this equation gives us exactly the components of the embedding tensor for half-maximal, electrically gauged supergravities in seven dimensions. However, not all of these irreps survive the linear constraint applied to them \cite{Samtleben:2005bp}. Let us also check whether this is the case for our setup. Therefore, we calculate $X_{AB}{}^C$ according to \eqref{eqn:XfromGamma}. In components, this equation gives rise to
\begin{equation}
  X_{a_1 a_2, b_1 b_2, c_1 c_2} = \Gamma_{a_1 a_2, b_1 b_2, c_1 c_2} - \Gamma_{b_1 b_2, a_1 a_2, c_1 c_2} + \Gamma_{c_1 c_2,a_1 a_2, b_1 b_2}
\end{equation}
or written in terms of permutations
\begin{equation}
  \sigma_X = () - (1\,3)(2\,4) + (1\,3\,5)(2\,4\,6)
		\quad \text{as} \quad
	X_{a_1 a_2, b_1 b_2, c_1 c_2} = 
	\sigma_X (1 - P_1) \Gamma_{a_1 a_2, b_1 b_2, c_1 c_2}\,.
\end{equation}
Again, we are able to rewrite $\sigma_X$ in terms of orthogonal irrep projectors. Doing so, we finally obtain
\begin{equation}\label{eqn:sollinSL4}
  \sigma_X (1 - P_1) P_{\mathbf{6}\times\mathbf{6}\times\overline{\mathbf 6}} = 3 P_{\mathbf{10}} + 3 P_{\overline{\mathbf 10}}\,.
\end{equation}
These two irreps combine to the 20 independent components of the totally anti-symmetric tensor $F_{ABC}$. 

From the considerations in the last section, we already know that in this case all remaining linear constraints are solved. Furthermore, this decomposition gives us exactly the right factor of 3 between $\Gamma_{AB}{}^C$ and $X_{AB}{}^C$.

\section{Additional Solutions of the Linear Constraint}\label{app:linconst2additional}
In this appendix, we give the remaining solutions for the group manifolds presented in table~\ref{fig:sollinconst}. First, we continue with the SL(3)$\times$SL(2) case. The coordinates are represented by the branching \eqref{eqn:10branching2}
\begin{equation}
  \mathbf{10} \rightarrow (\mathbf{1}, \mathbf{1}) + (\mathbf{3}, \mathbf{2}) + \xcancel{(\overline{\mathbf 3}, \mathbf{1})}
\end{equation}
after removing the last term. Counting the dimensions of the remaining irreps, we see that there are seven independent directions on the manifold. Again, we choose a basis for the vector space
\begin{align}
  V_{(\mathbf{1},\mathbf{1})} &= \{12\} &
  V_{(\mathbf{3},\mathbf{2})} &= \{13,\,14,\,15,\,23,\,24,\,25\} &
  V_{(\overline{\mathbf 3}, \mathbf{1})} &= \{34,\,35,\,45\} \\
  V_{\overline{(\mathbf{1},\mathbf{1})}} &= \{345\} &
  V_{\overline{(\mathbf{3},\mathbf{2})}} &= \{245,\,235,\,234,\,145,\,135,\,134\} &
  V_{\overline{(\overline{\mathbf 3}, \mathbf{1})}} &= \{125,\,124,\,123\}
\end{align}
and check the implications on the representations of the embedding tensor
\begin{align}\label{eqn:branching15}
  \mathbf{15} &\rightarrow (\mathbf{1}, \mathbf{3}) + (\mathbf{3}, \mathbf{2}) + (\mathbf{6},\mathbf{1}) \\
  \mathbf{40} &\rightarrow \xcancel{(\mathbf{1}, \mathbf{2})} + \xcancel{(\overline{\mathbf 3}, \mathbf{1})} + \xcancel{(\mathbf{3},\mathbf{2})} + \xcancel{(\overline{\mathbf 3}, \mathbf{3})} + \xcancel{(\overline{\mathbf 6}, 2)} + (\mathbf{8}, \mathbf{1})\,.\label{eqn:81survive}
\end{align}
While there are no restriction on the irreps resulting from the branching of the $\mathbf{15}$, the second linear constraint \eqref{eqn:linconst2} only allows the $(\mathbf{8},\mathbf{1})$ contribution from the $\mathbf{40}$. These are exactly the gaugings one would expect from the gauged supergravity point of view \cite{Samtleben:2005bp}. Another possible decomposition of the coordinates reads
\begin{equation}
  \mathbf{10} \rightarrow \xcancel{(\mathbf{1}, \mathbf{1})} + (\mathbf{3}, \mathbf{2}) + (\overline{\mathbf 3}, \mathbf{1})\,.
\end{equation}
It gives rise to a nine-dimensional group manifold. Again, all irreps in \eqref{eqn:branching15} are allowed and the
\begin{equation}\label{eqn:sl3sl2case2}
  \mathbf{40} \rightarrow (\mathbf{1}, \mathbf{2}) + \xcancel{(\overline{\mathbf 3}, \mathbf{1})} + \xcancel{(\mathbf{3},\mathbf{2})} + \xcancel{(\overline{\mathbf 3}, \mathbf{3})} + \xcancel{(\overline{\mathbf 6}, 2)} + \xcancel{(\mathbf{8}, \mathbf{1})}
\end{equation}
is restricted to the $(\mathbf{1},\mathbf{2})$ components. Of course, we could also consider the branching \eqref{eqn:10branching2} with both $(\mathbf{1},\mathbf{1})$ and $(\overline{\mathbf 3}, \mathbf{1})$ removed. This choice would result in a six-dimensional group manifold. However, doing the explicit calculation, we see that no irreps survives in this case.

Subsequently, we continue with this procedure for the T-duality subgroup SL(2)$\times$SL(2). First, we choose a basis for the vector space
\begin{align}
  V_{(\mathbf{1},\mathbf{1})} &= \{12\} &
  V_{(\mathbf{1},\mathbf{2})} &= \{15,\,25\} &
  V_{(\mathbf{2},\mathbf{2})} &= \{13,\,14,\,23,\,24\} \nonumber \\
  V_{(\mathbf{1}, \mathbf{1})} &= \{34\} &
  V_{(\mathbf{2},\mathbf{1})} &= \{35,\,45\}\\
  V_{\overline{(\mathbf{1},\mathbf{1})}} &= \{345\} &
  V_{\overline{(\mathbf{1},\mathbf{2})}} &= \{234,\,134\} &
  V_{\overline{(\mathbf{2},\mathbf{2})}} &= \{245,\,235,\,145,\,135\} \nonumber \\
  V_{\overline{(\mathbf{1}, \mathbf{1})}} &= \{125\} &
  V_{\overline{(\mathbf{2},\mathbf{1})}} &= \{124,\,123\}
\end{align}
which is adapted to the branching of the coordinates
\begin{equation}
  \mathbf{10} \rightarrow (\mathbf{1}, \mathbf{1}) + (\mathbf{1}, \mathbf{2}) + (\mathbf{2}, \mathbf{2}) + (\mathbf{1}, \mathbf{1}) + (\mathbf{2}, \mathbf{1})\,.
\end{equation}
By removing the irreps
\begin{equation}\label{eqn:SL410}
  \mathbf{10} \rightarrow (\mathbf{1}, \mathbf{1}) + \xcancel{(\mathbf{1}, \mathbf{2})} + (\mathbf{2}, \mathbf{2}) + \xcancel{(\mathbf{1}, \mathbf{1})} + (\mathbf{2}, \mathbf{1})
\end{equation}
from this decomposition, we obtain a seven-dimensional group manifold with the possible gaugings
\begin{align}
  \mathbf{15} \rightarrow &\;\;\;\;\,(\mathbf{1}, \mathbf{3}) + (\mathbf{1}, \mathbf{2}) + (\mathbf{2}, \mathbf{2}) + (\mathbf{1}, \mathbf{1}) + (\mathbf{2}, \mathbf{1}) + (\mathbf{3}, \mathbf{1})\,, \\
  \mathbf{40} \rightarrow &\;\;\;\;\,(\mathbf{1}, \mathbf{2}) + \xcancel{(\mathbf{1}, \mathbf{2})} + \xcancel{(\mathbf{2}, \mathbf{2})} +  (\mathbf{1}, \mathbf{1}) + (\mathbf{2}, \mathbf{1}) + (\mathbf{1}, \mathbf{3}) + \xcancel{(\mathbf{2}, \mathbf{3})} \nonumber \\ &+ \xcancel{(\mathbf{1}, \mathbf{2})} + \xcancel{(\mathbf{2}, \mathbf{2})} + \xcancel{(\mathbf{3}, \mathbf{2})} + \xcancel{(\mathbf{1}, \mathbf{1})} + \xcancel{(\mathbf{2}, \mathbf{1})} + \xcancel{(\mathbf{2}, \mathbf{1})} + \xcancel{(\mathbf{3}, \mathbf{1})} \,.\label{eqn:SL2SL2case1}
\end{align}
In order to see which irreps have to be removed, first note that the linear constraint for the $\mathbf{40}$ has an eight-dimensional solution space. In contrast to the previous cases, it is not possible to identify the crossed out irreps by their dimension alone. However, we can compare the linear constraint solutions to the ones obtained for the SL(4) case and see that they share 3 independent directions. For the branching
\begin{equation}
\label{eqn:SL410B}
  \overline{\mathbf{10}} \rightarrow (\mathbf{2},\mathbf{2}) + (\mathbf{3},\mathbf{1}) + ( \mathbf{1},\mathbf{3})
\end{equation}
of SL(4) to SL(2)$\times$SL(2), we see that these could furnish the irreps $(\mathbf{3},\mathbf{1})$ or $(\mathbf{1},\mathbf{3})$. Furthermore, this solution does not overlap with the ($\mathbf{8},\mathbf{1}$) from \eqref{eqn:81survive} which branches as
\begin{equation}
\label{eqn:SL3SL281}
  (\mathbf{8}, \mathbf{1}) \rightarrow (\mathbf{1}, \mathbf{1}) + 2 (\mathbf{2}, \mathbf{1}) + (\mathbf{3}, \mathbf{1})\,.
\end{equation}
Thus, $(\mathbf{1},\mathbf{3})$ is the only possible choice. A similar argumentation follows after taking into account the  $(\mathbf{1},\mathbf{2})$ of the SL(3)$\times$SL(2) case in \eqref{eqn:sl3sl2case2}. It shares two common directions with the solution of the linear constraint. The branching from SL(3)$\times$SL(2) to SL(2)$\times$SL(2) of this irrep is trivial
\begin{equation}
\label{eqn:SL3SL221}
(\mathbf{1},\mathbf{2}) \rightarrow (\mathbf{1},\mathbf{2})\,.
\end{equation}
Now, only three unidentified direction are left. They can be fixed just by their dimension. Doing so, we obtain the branching \eqref{eqn:SL2SL2case1}. This gauging is expected from the gauged supergravity point of view as well \cite{Samtleben:2005bp}.

Moreover, we find two more interesting cases which do not lie completely in one of the previous cases for SL(3)$\times$SL(2) and SL(4). The first one gives rise the eight-dimensional group manifolds with the coordinate irreps
\begin{equation}
  \mathbf{10} \rightarrow \xcancel{(\mathbf{1}, \mathbf{1})} + (\mathbf{1}, \mathbf{2}) + (\mathbf{2}, \mathbf{2}) + \xcancel{(\mathbf{1}, \mathbf{1})} + (\mathbf{2}, \mathbf{1})\,.
\end{equation}
Here, the solution space for the $\mathbf{40}$ part of the linear constraints has four independent directions. They are partially contained in the $(\mathbf{1}, \mathbf{2})$ and $(\mathbf{8},\mathbf{1})$ of SL(3)$\times$SL(2). With both the solution shares two directions each. According to \eqref{eqn:SL3SL281} and \eqref{eqn:SL3SL221}, we identify them with the irreps $(\mathbf{1}, \mathbf{2})$ and $(\mathbf{2}, \mathbf{1})$. These are the only irreps which can be switched on. There are no restrictions for the $\mathbf{15}$ part by the linear constraints. Thus, we obtain
\begin{align}
  \mathbf{15} \rightarrow &\;\;\;\;\,(\mathbf{1}, \mathbf{3}) + (\mathbf{1}, \mathbf{2}) + (\mathbf{2}, \mathbf{2}) + (\mathbf{1}, \mathbf{1}) + (\mathbf{2}, \mathbf{1}) + (\mathbf{3}, \mathbf{1}) \,, \\
  \mathbf{40} \rightarrow &\;\;\;\;\,(\mathbf{1}, \mathbf{2}) + \xcancel{(\mathbf{1}, \mathbf{2})} + \xcancel{(\mathbf{2}, \mathbf{2})} +  \xcancel{(\mathbf{1}, \mathbf{1})} + \xcancel{(\mathbf{2}, \mathbf{1})} + \xcancel{(\mathbf{1}, \mathbf{3})} + \xcancel{(\mathbf{2}, \mathbf{3})} \nonumber \\ &+ \xcancel{(\mathbf{1}, \mathbf{2})} + \xcancel{(\mathbf{2}, \mathbf{2})} + \xcancel{(\mathbf{3}, \mathbf{2})} + \xcancel{(\mathbf{1}, \mathbf{1})} + \xcancel{(\mathbf{2}, \mathbf{1})} + (\mathbf{2}, \mathbf{1}) + \xcancel{(\mathbf{3}, \mathbf{1})} \,.
\end{align}

Finally, there are five-dimensional group manifolds with the coordinate irreps
\begin{equation}
  \mathbf{10} \rightarrow (\mathbf{1}, \mathbf{1}) + \xcancel{(\mathbf{1}, \mathbf{2})} + (\mathbf{2}, \mathbf{2}) + \xcancel{(\mathbf{1}, \mathbf{1})} + \xcancel{(\mathbf{2}, \mathbf{1})}\,.
\end{equation}
In total, the solution space of the $\mathbf{40}$ part possesses $11$ independent directions. They are partially contained\footnote{There are four directions in the $(\mathbf{8},\mathbf{1})$ of SL(3)$\times$SL(2), but only one of them is not contained in the $\overline{\mathbf{10}}$ of SL(4).} in the $(\mathbf{8},\mathbf{1})$ of SL(3)$\times$SL(2) and sit completely in the $\overline{\mathbf{10}}$ of SL(4). Thus, we only obtain a new $(\mathbf{1},\mathbf{1})$ from \eqref{eqn:SL3SL281} and the right hand side of \eqref{eqn:SL410B}.  In contrast to the previous cases, only ten directions of the linear constraints' $\mathbf{15}$ part can be switched on. The solution for the $\mathbf{15}$ lies entirely in the $\mathbf{10}$ of SL(4). Taking into account the branching rule \eqref{eqn:SL410}, we find
\begin{align}
  \mathbf{15} \rightarrow &\;\;\;\;\,(\mathbf{1}, \mathbf{3}) + \xcancel{(\mathbf{1}, \mathbf{2})} + (\mathbf{2}, \mathbf{2}) + \xcancel{(\mathbf{1}, \mathbf{1})} + \xcancel{(\mathbf{2}, \mathbf{1})} + (\mathbf{3}, \mathbf{1}) \,, \\
  \mathbf{40} \rightarrow &\;\;\;\;\,\xcancel{(\mathbf{1}, \mathbf{2})} + \xcancel{(\mathbf{1}, \mathbf{2})} + (\mathbf{2}, \mathbf{2}) +  \xcancel{(\mathbf{1}, \mathbf{1})} + \xcancel{(\mathbf{2}, \mathbf{1})} + (\mathbf{1}, \mathbf{3}) + \xcancel{(\mathbf{2}, \mathbf{3})} \nonumber \\ &+ \xcancel{(\mathbf{1}, \mathbf{2})} + \xcancel{(\mathbf{2}, \mathbf{2})} + \xcancel{(\mathbf{3}, \mathbf{2})} + (\mathbf{1}, \mathbf{1}) + \xcancel{(\mathbf{2}, \mathbf{1})} + \xcancel{(\mathbf{2}, \mathbf{1})} + (\mathbf{3}, \mathbf{1}) \,.
\end{align}

All other solutions the linear constraints are completely contained in one of the previously discussed SL(4) or SL(3)$\times$SL(2) cases.

\section{Faithful Representations and Identifications}\label{app:faithfulrepr}
We first consider the Lie algebra of CSO(1,0,4) which is given in terms of the non-vanishing commutators
\begin{equation}
\label{eqn:appendixcliealgebra}
[t_{\hat\alpha}, t_{\hat\beta}] = \mathbf{g} \, t_{\hat\alpha\hat\beta}\,,
\end{equation}
where we assigned the generators
\begin{equation}
  t_A = \Big( t_1,\, t_2,\, t_3,\, t_4,\,
      t_{\tilde 1},\, t_{\tilde 2},\, t_{\tilde 3},\, t_{\tilde 4},\, t_{\tilde 5},\, t_{\tilde 6} \Big)\,.
\end{equation}
This algebra is relevant for the first duality chain \eqref{eqn:chain1} in section~\ref{sec:T4Gflux} and has the lower central series
\begin{equation}
  L_0= \{ t_1,\, t_2,\, t_3,\, t_4,\, t_{\tilde 1},\, t_{\tilde 2},\, t_{\tilde 3},\, t_{\tilde 4},\, t_{\tilde 5},\, t_{\tilde 6}\} \supset \{ t_{\tilde 1},\, t_{\tilde 2},\, t_{\tilde 3},\, t_{\tilde 4},\, t_{\tilde 5},\, t_{\tilde 6} \} \supset \{0\}\,.
\end{equation}
Following the procedure outlined in \cite{Bosque:2015jda}, we construct the $N=21$-dimensional subspace
\begin{align}
  V^2 = \{&t_1^2,\, t_1 t_2,\, t_1 t_3,\, t_1 t_4,\, t_2^2,\, t_2 t_3,\, t_2 t_4,\, t_3^2,\, t_3 t_4,\, t_4^2,\, t_{\tilde 1},\, t_{\tilde 2},\, t_{\tilde 3},\, t_{\tilde 4},\, t_{\tilde 5},\, t_{\tilde 6},  &&\ord \cdot = 2 \nonumber \\
    &t_1,\, t_2,\, t_3,\, t_4, &&\ord \cdot = 1 \nonumber \\
    &1 \} &&\ord \cdot = 0
\end{align}
of the universal enveloping algebra. The center of this algebra is given by $\{$ $t_{\tilde 1}$, $t_{\tilde 2}$, $t_{\tilde 3}$, $t_{\tilde 4}$, $t_{\tilde 5}$, $t_{\tilde 6}\}$. These six generators form an abelian subalgebra $\mathfrak{h}$. With this data, we are able to obtain the matrix representation for the generators $t_A$ by expanding the linear maps $\phi_{t_A}$ in the basis $V^2$. Finally the exponential maps \eqref{eqn:matrixexpm} and \eqref{eqn:matrixexph} give rise to the group elements
\begin{equation}
g = m h = \left(
\begin{array}{*{21}c} 
 1 & 0 & 0 & 0 & 0 & 0 & 0 & 0 & 0 & 0 & 0 & 0 & 0 & 0 & 0 & 0 & 0 & 0 & 0 & 0 & x^1 \\
 0 & 1 & 0 & 0 & 0 & 0 & 0 & 0 & 0 & 0 & 0 & 0 & 0 & 0 & 0 & 0 & 0 & 0 & 0 & 0 & x^2 \\
 0 & 0 & 1 & 0 & 0 & 0 & 0 & 0 & 0 & 0 & 0 & 0 & 0 & 0 & 0 & 0 & 0 & 0 & 0 & 0 & x^3 \\
 0 & 0 & 0 & 1 & 0 & 0 & 0 & 0 & 0 & 0 & 0 & 0 & 0 & 0 & 0 & 0 & 0 & 0 & 0 & 0 & x^4 \\
 x^1 & 0 & 0 & 0 & 1 & 0 & 0 & 0 & 0 & 0 & 0 & 0 & 0 & 0 & 0 & 0 & 0 & 0 & 0 & 0 & (x^1)^2/2 \\
 x^2 & x^1 & 0 & 0 & 0 & 1 & 0 & 0 & 0 & 0 & 0 & 0 & 0 & 0 & 0 & 0 & 0 & 0 & 0 & 0 & x^1 x^2 \\
 x^3 & 0 & x^1 & 0 & 0 & 0 & 1 & 0 & 0 & 0 & 0 & 0 & 0 & 0 & 0 & 0 & 0 & 0 & 0 & 0 & x^1 x^3 \\
 x^4 & 0 & 0 & x^1 & 0 & 0 & 0 & 1 & 0 & 0 & 0 & 0 & 0 & 0 & 0 & 0 & 0 & 0 & 0 & 0 & x^1 x^4 \\
 0 & x^2 & 0 & 0 & 0 & 0 & 0 & 0 & 1 & 0 & 0 & 0 & 0 & 0 & 0 & 0 & 0 & 0 & 0 & 0 & (x^2)^2/2 \\
 0 & x^3 & x^2 & 0 & 0 & 0 & 0 & 0 & 0 & 1 & 0 & 0 & 0 & 0 & 0 & 0 & 0 & 0 & 0 & 0 & x^2 x^3 \\
 0 & x^4 & 0 & x^2 & 0 & 0 & 0 & 0 & 0 & 0 & 1 & 0 & 0 & 0 & 0 & 0 & 0 & 0 & 0 & 0 & x^2 x^4 \\
 0 & 0 & x^3 & 0 & 0 & 0 & 0 & 0 & 0 & 0 & 0 & 1 & 0 & 0 & 0 & 0 & 0 & 0 & 0 & 0 & (x^3)^2 \\
0 & 0 & x^4 & x^3 & 0 & 0 & 0 & 0 & 0 & 0 & 0 & 0 & 1 & 0 & 0 & 0 & 0 & 0 & 0 & 0 & x^3 x^4 \\
0 & 0 & 0 & x^4 & 0 & 0 & 0 & 0 & 0 & 0 & 0 & 0 & 0 & 1 & 0 & 0 & 0 & 0 & 0 & 0 & (x^4)^2/2 \\
- \mathbf{g} \, x^2 & 0 & 0 & 0 & 0 & 0 & 0 & 0 & 0 & 0 & 0 & 0 & 0 & 0 & 1 & 0 & 0 & 0 & 0 & 0 & x^{\tilde 1} \\
- \mathbf{g} \, x^3 & 0 & 0 & 0 & 0 & 0 & 0 & 0 & 0 & 0 & 0 & 0 & 0 & 0 & 0 & 1 & 0 & 0 & 0 & 0 & x^{\tilde 2} \\
- \mathbf{g} \, x^4 & 0 & 0 & 0 & 0 & 0 & 0 & 0 & 0 & 0 & 0 & 0 & 0 & 0 & 0 & 0 & 1 & 0 & 0 & 0 & x^{\tilde 3} \\
0 & - \mathbf{g} \, x^3 & 0 & 0 & 0 & 0 & 0 & 0 & 0 & 0 & 0 & 0 & 0 & 0 & 0 & 0 & 0 & 1 & 0 & 0 & x^{\tilde 4} \\
0 & - \mathbf{g} \, x^4 & 0 & 0 & 0 & 0 & 0 & 0 & 0 & 0 & 0 & 0 & 0 & 0 & 0 & 0 & 0 & 0 & 1 & 0 & x^{\tilde 5} \\
0 & 0 & - \mathbf{g} \, x^4 & 0 & 0 & 0 & 0 & 0 & 0 & 0 & 0 & 0 & 0 & 0 & 0 & 0 & 0 & 0 & 0 & 1 & x^{\tilde 6} \\
0 & 0 & 0 & 0 & 0 & 0 & 0 & 0 & 0 & 0 & 0 & 0 & 0 & 0 & 0 & 0 & 0 & 0 & 0 & 0 & 1 \\
\end{array}
\right)
\end{equation}
with $\mathbf{g}$ representing the number of $G$-flux units the background carries. Working with such large matrices is cumbersome. So we represent $g$ instead by the ten tuple
$($ $x^1$, $x^2$, $x^3$, $x^4$, $x^{\tilde 1}$, $x^{\tilde 2}$, $x^{\tilde 3}$, $x^{\tilde 4}$, $x^{\tilde 5}$, $x^{\tilde 6}$ $)$. In this case, the group multiplication is given by
\begin{align}\label{eqn:groupmult}
  (x^1\,,x^2\,,x^3\,, x^4\,,x^{\tilde 1}\,,x^{\tilde 2}\,, x^{\tilde 3}\,, x^{\tilde 4}\,, &x^{\tilde 5}\,, 
    x^{\tilde 6}) (y^1\,,y^2\,,y^3\,, y^4\,,y^{\tilde 1}\,,y^{\tilde 2}\,, y^{\tilde 3}\,, y^{\tilde 4}\,,
    y^{\tilde 5}\,, y^{\tilde 6}) = \\ 
  (x^1+y^1\,,x^2+y^2\,,x^3+y^3\,,x^4+y^4\,,&- \mathbf{g} \,  x^2 y^1 + x^{\tilde 1} + y^{\tilde 1}\,, 
    - \mathbf{g} \, x^3 y^1 + x^{\tilde 2} + y^{\tilde 2} \nonumber \\
  - \mathbf{g} \, x^3 y^2 + x^{\tilde 3} + y^{\tilde 3}\,, - \mathbf{g} \,  &x^4 y^1 + x^{\tilde 4} + y^{\tilde 4}\,,
    - \mathbf{g} \, x^4 y^2 + x^{\tilde 5} + y^{\tilde 5}\,,- \mathbf{g} \,  x^4 y^3 + x^{\tilde 6} + y^{\tilde 6} )\,. \nonumber 
\end{align}
Let us check that this indeed gives rise to a group. The identity element is $e$=$($ $0$, $0$, $0$, $0$, $0$, $0$, $0$, $0$, $0$, $0$ $)$ and fulfills
\begin{equation}
  g e = e g = g\,.
\end{equation}
Furthermore, there is the inverse element
\begin{align}
  g^{-1}=(-x^1\,,-x^2\,,-x^3\,,-x^4\,, &- \mathbf{g} \, x^1 x^2 -  x^{\tilde 1}\,, - \mathbf{g} \, x^1 x^3 - x^{\tilde 2}\,, \\ 
 &- \mathbf{g} \, x^2 x^3 - x^{\tilde 3}\,, - \mathbf{g} \, x^1 x^4 - x^{\tilde 4}\,, - \mathbf{g} \,  x^2 x^4 -  x^{\tilde 5}\,, - \mathbf{g} \, x^3 x^4 - x^{\tilde 6}) \nonumber
\end{align}
fulfilling
\begin{equation}
  g^{-1} g = g g^{-1} = e\,.
\end{equation}
Because $\mathbf{g}$ is an integer, the group multiplication \eqref{eqn:groupmult} does not only close over the real numbers, but also for $x^i$ and $x^{\tilde i}$ being integers. Thus, CSO(1,0,4,$\mathbb{Z}$) is a subgroup of CSO(1,0,4) and we can mod it out by considering the right coset CSO(1,0,4,$\mathbb{Z}$)\textbackslash CSO(1,0,4) which gives rise to the equivalence relation
\begin{equation}
  g_1 \sim g_2 \quad \text{if and only if} \quad g_1 = k g_2 \quad \text{with} \quad
  g_1\,, g_2 \in \mathrm{CSO}(1,0,4) \quad \text{and} \quad
  k \in \mathrm{CSO}(1,0,4,\mathbb{Z})\,.
\end{equation}
After substituting $k=(n^1\,,n^2\,,n^3\,,n^4\,,n^{\tilde 1}\,,n^{\tilde 2}\,,n^{\tilde 3}\,,n^{\tilde 4}\,,n^{\tilde 5}\,,n^{\tilde 6})$ with $n^i$, $n^{\tilde i} \in\mathbb{Z}$, we obtain the identifications
\begin{align}
  (x^1\,,x^2\,,x^3\,,x^4\,,& x^{\tilde 1}\,,x^{\tilde 2}\,, x^{\tilde 3}\,, x^{\tilde 4}\,, 
    x^{\tilde 5}\,, x^{\tilde 6}) \sim \\ 
  (x^1+n^1\,,x^2+n^2\,,x^3+n^3\,,x^4+n^4\,,&- \mathbf{g} \, x^1 n^2 + x^{\tilde 1} + n^{\tilde 1}\,, 
    - \mathbf{g} \, x^1 n^3 + x^{\tilde 2} + n^{\tilde 2} \nonumber \\
  - \mathbf{g} \, x^2 n^3 + x^{\tilde 3} + n^{\tilde 3}\,,  - \mathbf{g} \, & x^1 n^4 + x^{\tilde 4} + 
    n^{\tilde 4}\,, - \mathbf{g} \, x^2 n^4 + x^{\tilde 5} + n^{\tilde 5}\,,- \mathbf{g} \,
    x^3 n^4 + x^{\tilde 6} + n^{\tilde 6} ) \nonumber
\end{align}
from \eqref{eqn:groupmult}. Especially, we have
\begin{align}
\label{eqn:coordident1}
  (x^1\,,x^2\,,x^3\,, x^4\,,& x^{\tilde 1}\,,x^{\tilde 2}\,, x^{\tilde 3}\,, x^{\tilde 4}\,, x^{\tilde 5}\,, x^{\tilde 6})  \sim (x^1 + 1\,,x^2\,,x^3\,, x^4\,,x^{\tilde 1}\,,x^{\tilde 2}\,, x^{\tilde 3}\,, x^{\tilde 4}\,, x^{\tilde 5}\,, x^{\tilde 6})  \\
  & \sim (x^1\,,x^2 + 1\,,x^3\,, x^4\,,x^{\tilde 1} - \mathbf{g} \, x^1 \,,x^{\tilde 2}\,, x^{\tilde 3}\,, x^{\tilde 4}\,, x^{\tilde 5}\,, x^{\tilde 6}) \nonumber  \\
  & \sim (x^1\,,x^2\,,x^3 + 1\,, x^4\,,x^{\tilde 1}\,,x^{\tilde 2} - \mathbf{g} \, x^1\,, x^{\tilde 3} - \mathbf{g} \, x^2\,, x^{\tilde 4}\,, x^{\tilde 5}\,, x^{\tilde 6})\nonumber  \\
  & \sim (x^1\,,x^2\,,x^3\,, x^4 + 1\,,x^{\tilde 1}\,,x^{\tilde 2}\,, x^{\tilde 3}\,, x^{\tilde 4} - \mathbf{g} \, x^1\,, x^{\tilde 5} - \mathbf{g} \, x^2 \,, x^{\tilde 6} - \mathbf{g} \, x^3) \nonumber  
\end{align}
for the physical coordinates and
\begin{align}
\label{eqn:coordident2}
  (x^1\,,x^2\,,x^3\,, x^4\,,x^{\tilde 1}\,,x^{\tilde 2}\,, x^{\tilde 3}\,, x^{\tilde 4}\,, x^{\tilde 5}\,, x^{\tilde 6})  & \sim (x^1\,,x^2\,,x^3\,, x^4\,,x^{\tilde 1} + 1\,,x^{\tilde 2}\,, x^{\tilde 3}\,, x^{\tilde 4}\,, x^{\tilde 5}\,, x^{\tilde 6})   \\
  & \sim (x^1\,,x^2\,,x^3\,, x^4\,,x^{\tilde 1}\,,x^{\tilde 2} + 1\,, x^{\tilde 3}\,, x^{\tilde 4}\,, x^{\tilde 5}\,, x^{\tilde 6})  \nonumber \\
  & \sim (x^1\,,x^2\,,x^3\,, x^4\,,x^{\tilde 1}\,,x^{\tilde 2}\,, x^{\tilde 3} + 1\,, x^{\tilde 4}\,, x^{\tilde 5}\,, x^{\tilde 6})  \nonumber \\
  & \sim (x^1\,,x^2\,,x^3\,, x^4\,,x^{\tilde 1}\,,x^{\tilde 2}\,, x^{\tilde 3}\,, x^{\tilde 4} + 1\,, x^{\tilde 5}\,, x^{\tilde 6})  \nonumber \\
  & \sim (x^1\,,x^2\,,x^3\,, x^4\,,x^{\tilde 1}\,,x^{\tilde 2}\,, x^{\tilde 3}\,, x^{\tilde 4}\,, x^{\tilde 5} + 1\,, x^{\tilde 6})  \nonumber \\
  & \sim (x^1\,,x^2\,,x^3\,, x^4\,,x^{\tilde 1}\,,x^{\tilde 2}\,, x^{\tilde 3}\,, x^{\tilde 4}\,, x^{\tilde 5}\,, x^{\tilde 6} + 1)  \nonumber  
\end{align}
for the remaining ones. Taking into account these identifications, the left invariant Maurer-Cartan form
\begin{equation}
\label{eqn:appendixvielbein}
E^A{}_I = \begin{pmatrix}
1 & 0 & 0 & 0 & 0 & 0 & 0 & 0 & 0 & 0 \\
0 & 1 & 0 & 0 & 0 & 0 & 0 & 0 & 0 & 0 \\
0 & 0 & 1 & 0 & 0 & 0 & 0 & 0 & 0 & 0 \\
0 & 0 & 0 & 1 & 0 & 0 & 0 & 0 & 0 & 0 \\
\mathbf{g} \, x_2 & 0 & 0 & 0 & 1 & 0 & 0 & 0 & 0 & 0 \\
\mathbf{g} \, x_3 & 0 & 0 & 0 & 0 & 1 & 0 & 0 & 0 & 0 \\
0 & \mathbf{g} \, x_3 & 0 & 0 & 0 & 0 & 1 & 0 & 0 & 0 \\
\mathbf{g} \, x_4 & 0 & 0 & 0 & 0 & 0 & 0 & 1 & 0 & 0 \\
0 & \mathbf{g} \, x_4 & 0 & 0 & 0 & 0 & 0 & 0 & 1 & 0 \\
0 & 0 & \mathbf{g} \, x_4 & 0 & 0 & 0 & 0 & 0 & 0 & 1
\end{pmatrix}\,,
\end{equation}
is globally well defined, namely
\begin{align}
 E_1 &= d x^1  \\
 E_2 &= d x^2 \nonumber \\
 E_3 &= d x^3 \nonumber \\
 E_4 &= d x^4 \nonumber \\
  E^1 &= d x^{\tilde 1} + \mathbf{g} \, x^2 d x^1 = d (x^{\tilde 1}- \mathbf{g} \, x^1) + (x^2 + 1) \, \mathbf{g} \,  d x^1 \nonumber \\
  E^2 &= d x^{\tilde 2} + \mathbf{g} \, x^3 d x^1 = d (x^{\tilde 2} - \mathbf{g} \, x^1) + (x^3 + 1) \, \mathbf{g} \, d x^1 \nonumber \\
  E^3 &= d x^{\tilde 3} + \mathbf{g} \, x^3 d x^2 = d (x^{\tilde 3} - \mathbf{g} \, x^2) + (x^3 + 1) \, \mathbf{g} \, d x^2 \nonumber \\
  E^4 &= d x^{\tilde 4} + \mathbf{g} \, x^4 d x^1 = d (x^{\tilde 4} - \mathbf{g} \, x^1) + (x^4 + 1) \, \mathbf{g} \, d x^1 \nonumber \\
  E^5 &= d x^{\tilde 5} + \mathbf{g} \, x^4 d x^2 = d (x^{\tilde 5} - \mathbf{g} \, x^2) + (x^4 + 1) \, \mathbf{g} \, d x^2 \nonumber \\
  E^6 &= d x^{\tilde 6} + \mathbf{g} \, x^4 d x^3 = d (x^{\tilde 6} - \mathbf{g} \, x^3) + (x^4 + 1) \, \mathbf{g} \, d x^3\,. \nonumber 
\end{align}
\newpage
For the second duality chain \eqref{eqn:chain2}, the nine-dimensional Lie algebra $\mathfrak{g}$ in \eqref{eqn:defgexample40} is relevant. We perform the exponential maps \eqref{eqn:matrixexpm} as well as \eqref{eqn:matrixexph}, to obtain the group element
\begin{equation}
g = m h = \left(
\begin{array}{cccccccccccccccc}
 1 & 0 & 0 & 0 & 0 & 0 & 0 & 0 & 0 & 0 & 0 & 0 & 0 & 0 & 0 & x^{\tilde 5} \\
 0 & 1 & 0 & 0 & 0 & 0 & 0 & 0 & 0 & 0 & 0 & 0 & 0 & 0 & 0 & x^3 \\
 0 & 0 & 1 & 0 & 0 & 0 & 0 & 0 & 0 & 0 & 0 & 0 & 0 & 0 & 0 & x^4 \\
 0 & 0 & 0 & 1 & 0 & 0 & 0 & 0 & 0 & 0 & 0 & 0 & 0 & 0 & 0 & x^1 \\
 0 & 0 & 0 & 0 & 1 & 0 & 0 & 0 & 0 & 0 & 0 & 0 & 0 & 0 & 0 & x^{\tilde 1} \\
 0 & 0 & 0 & 0 & 0 & 1 & 0 & 0 & 0 & 0 & 0 & 0 & 0 & 0 & 0 & x^{\tilde 3} \\
 x^{\tilde 5} & 0 & 0 & 0 & 0 & 0 & 1 & 0 & 0 & 0 & 0 & 0 & 0 & 0 & 0 & ({x^{\tilde 5}})^2 /2 \\
 x^3 & {x^{\tilde 5}} & 0 & 0 & 0 & 0 & 0 & 1 & 0 & 0 & 0 & 0 & 0 & 0 & 0 & x^3 {x^{\tilde 5}} \\
 x^4 & 0 & {x^{\tilde 5}} & 0 & 0 & 0 & 0 & 0 & 1 & 0 & 0 & 0 & 0 & 0 & 0 & x^4 {x^{\tilde 5}} \\
 0 & x^3 & 0 & 0 & 0 & 0 & 0 & 0 & 0 & 1 & 0 & 0 & 0 & 0 & 0 & (x^3)^2 / 2 \\
 0 & x^4 & x^3 & 0 & 0 & 0 & 0 & 0 & 0 & 0 & 1 & 0 & 0 & 0 & 0 & x^3 x^4 \\
 0 & 0 & x^4 & 0 & 0 & 0 & 0 & 0 & 0 & 0 & 0 & 1 & 0 & 0 & 0 & (x^4)^2 / 2 \\
 0 & -\mathbf{f} x^4 & 0 & 0 & 0 & 0 & 0 & 0 & 0 & 0 & 0 & 0 & 1 & 0 & 0 & x^2 \\
 -\mathbf{f} x^4 & 0 & 0 & 0 & 0 & 0 & 0 & 0 & 0 & 0 & 0 & 0 & 0 & 1 & 0 &
   x^{\tilde 4} -\mathbf{f} x^4 {x^{\tilde 5}} \\
 -\mathbf{f} x^3 & 0 & 0 & 0 & 0 & 0 & 0 & 0 & 0 & 0 & 0 & 0 & 0 & 0 & 1 &
   x^{\tilde 2}-\mathbf{f} x^3 {x^{\tilde 5}} \\
 0 & 0 & 0 & 0 & 0 & 0 & 0 & 0 & 0 & 0 & 0 & 0 & 0 & 0 & 0 & 1 \\
\end{array}
\right)
\end{equation}
with $\mathbf{f}$ representing the number of $F$-flux units the background carries. Again we represent $g$ in terms of the nine tuple $($ $x^1$, $x^2$, $x^3$, $x^4$, $x^{\tilde 1}$, $x^{\tilde 2}$, $x^{\tilde 3}$, $x^{\tilde 4}$, $x^{\tilde 5}$ $)$ instead of working with this big matrix. In this case, the group multiplication is given by
\begin{align}
\label{eqn:groupmult2}
  (x^1\,,x^2\,,x^3\,, x^4\,,x^{\tilde 1}\,,x^{\tilde 2}\,, x^{\tilde 3}\,, x^{\tilde 4}\,, x^{\tilde 5}) (&y^1\,,y^2\,,y^3\,, y^4\,,y^{\tilde 1}\,,y^{\tilde 2}\,, y^{\tilde 3}\,, y^{\tilde 4}\,, y^{\tilde 5}) = \\ 
  (x^1+y^1\,,- \mathbf{f} \, x^4 y^3 + x^2+y^2 \,,x^3+y^3\,,x^4+y^4\,,& \, x^{\tilde 1} + y^{\tilde 1}\,, \mathbf{f} \, x^{\tilde 5} y^3 + x^{\tilde 2} + y^{\tilde 2} \nonumber \\ x^{\tilde 3} + y^{\tilde 3}\,, & \, \mathbf{f} \,  x^{\tilde 5} y^4 + x^{\tilde 4} + y^{\tilde 4}\,, x^{\tilde 5} + y^{\tilde 5} )\,. \nonumber 
\end{align}
To verify that $g$ is a group, first consider the identity element $e$=$($ $0$, $0$, $0$, $0$, $0$, $0$, $0$, $0$, $0$ $)$, It satisfies
\begin{equation}
  g e = e g = g\,.
\end{equation}
Moreover, the inverse element is given by
\begin{align}
  g^{-1}=(-x^1\,,- \mathbf{f} \, x^3 x^4 -x^2\,,-x^3\,,-x^4\,, &-  x^{\tilde 1}\,, \mathbf{f} \, x^3 x^{\tilde 5} - x^{\tilde 2}\,, - x^{\tilde 3}\,, \mathbf{f} \, x^4 x^{\tilde 5} - x^{\tilde 4}\,, -  x^{\tilde 5}) 
\end{align}
fulfilling
\begin{equation}
  g^{-1} g = g g^{-1} = e\,.
\end{equation}
Since $\mathbf{f}$ is an integer, the group multiplication \eqref{eqn:groupmult2} does not only close over the real numbers, but also for $x^i$ and $x^{\tilde i}$ being integers. Hence, we can mod out the discrete subgroup $G_{\mathbb{Z}}$ formed by restricting all coordinates to integers from the left. This results in the equivalence relation
\begin{equation}
  g_1 \sim g_2 \quad \text{if and only if} \quad g_1 = k g_2 \quad \text{with} \quad
  g_1\,, g_2 \in G \quad \text{and} \quad
  k \in G_{\mathbb{Z}}\,.
\end{equation}
Finally, we substitute $k=(n^1\,,n^2\,,n^3\,,n^4\,,n^{\tilde 1}\,,n^{\tilde 2}\,,n^{\tilde 3}\,,n^{\tilde 4}\,,n^{\tilde 5})$ with $n^i$, $n^{\tilde i} \in\mathbb{Z}$ and find the identifications
\begin{align}
  (x^1\,,x^2\,,x^3\,, x^4\,,x^{\tilde 1}\,,x^{\tilde 2}\,, x^{\tilde 3}\,, x^{\tilde 4}\,, &x^{\tilde 5}) \sim \\ 
  (x^1+n^1\,,-\mathbf{f} \, x^3 n^4 + x^2+n^2\,,x^3+n^3\,,x^4+n^4\,,&x^{\tilde 1} + n^{\tilde 1}\,, \mathbf{f} \, x^3 n^{\tilde 5} + x^{\tilde 2} + n^{\tilde 2} \nonumber \\ x^{\tilde 3} + n^{\tilde 3}\,,  &\mathbf{f} \, x^4 n^{\tilde 5} + x^{\tilde 4} + n^{\tilde 4}\,, x^{\tilde 5} + n^{\tilde 5}) \nonumber
\end{align}
from \eqref{eqn:groupmult2}. Particularly, for the physical coordinates
\begin{align}
\label{eqn:coordident3}
  (x^1\,,x^2\,,x^3\,, x^4\,, x^{\tilde 1}\,,x^{\tilde 2}\,, x^{\tilde 3}\,, x^{\tilde 4}\,, x^{\tilde 5})  &\sim (x^1 + 1\,,x^2\,,x^3\,, x^4\,,x^{\tilde 1}\,,x^{\tilde 2}\,, x^{\tilde 3}\,, x^{\tilde 4}\,, x^{\tilde 5})  \\
  & \sim (x^1\,,x^2 + 1\,,x^3\,, x^4\,,x^{\tilde 1}\,,x^{\tilde 2}\,, x^{\tilde 3}\,, x^{\tilde 4}\,, x^{\tilde 5}) \nonumber  \\
  & \sim (x^1\,,x^2\,,x^3 + 1\,, x^4\,,x^{\tilde 1}\,,x^{\tilde 2}\,, x^{\tilde 3}\,, x^{\tilde 4}\,, x^{\tilde 5})\nonumber  \\
  & \sim (x^1\,,x^2 - \mathbf{f} x^3 \,,x^3\,, x^4 + 1\,,x^{\tilde 1}\,,x^{\tilde 2}\,, x^{\tilde 3}\,, x^{\tilde 4}\,, x^{\tilde 5})  \nonumber
\end{align}
and for the remaining coordinates
\begin{align}\label{eqn:coordident4}
  (x^1\,,x^2\,,x^3\,, x^4\,,x^{\tilde 1}\,,x^{\tilde 2}\,, x^{\tilde 3}\,, x^{\tilde 4}\,, x^{\tilde 5})  & \sim (x^1\,,x^2\,,x^3\,, x^4\,,x^{\tilde 1} + 1\,,x^{\tilde 2}\,, x^{\tilde 3}\,, x^{\tilde 4}\,, x^{\tilde 5}) \\
  & \sim (x^1\,,x^2\,,x^3\,, x^4\,,x^{\tilde 1}\,,x^{\tilde 2} + 1\,, x^{\tilde 3}\,, x^{\tilde 4}\,, x^{\tilde 5})  \nonumber \\
  & \sim (x^1\,,x^2\,,x^3\,, x^4\,,x^{\tilde 1}\,,x^{\tilde 2}\,, x^{\tilde 3} + 1\,, x^{\tilde 4}\,, x^{\tilde 5})  \nonumber \\
  & \sim (x^1\,,x^2\,,x^3\,, x^4\,,x^{\tilde 1}\,,x^{\tilde 2}\,, x^{\tilde 3}\,, x^{\tilde 4} + 1\,, x^{\tilde 5})  \nonumber \\
  & \sim (x^1\,,x^2\,,x^3\,, x^4\,,x^{\tilde 1}\,,x^{\tilde 2} + \mathbf{f} \, x^3 \,, x^{\tilde 3}\,, x^{\tilde 4} + \mathbf{f} \, x^4\,, x^{\tilde 5} + 1) \nonumber \,. 
\end{align}
After taking these identifications into account, we compute the left-invariant Maurer-Cartan form
\begin{equation}
\label{eqn:appendixvielbein2}
E^A{}_I = \begin{pmatrix}
1 & 0 & 0 & 0 & 0 & 0 & 0 & 0 & 0 \\
0 & 1 & \mathbf{f} \, x^4 & 0 & 0 & 0 & 0 & 0 & 0 \\
0 & 0 & 1 & 0 & 0 & 0 & 0 & 0 & 0 \\
0 & 0 & 0 & 1 & 0 & 0 & 0 & 0 & 0  \\
0 & 0 & 0 & 0 & 1 & 0 & 0 & 0 & 0 \\
0 & 0 & - \mathbf{f} \, x^{\tilde 5} & 0 & 0 & 1 & 0 & 0 & 0 \\
0 & 0 & 0 & 0 & 0 & 0 & 1 & 0 & 0 \\
0 & 0 & 0 & - \mathbf{f} \, x^{\tilde 5} & 0 & 0 & 0 & 1 & 0 \\
0 & 0 & 0 & 0 & 0 & 0 & 0 & 0 & 1
\end{pmatrix}\,.
\end{equation}
Taking into account the identifications \eqref{eqn:coordident3} and \eqref{eqn:coordident4}, it is straightforward to check that this $E^A{}_I$ is globally well defined:
\begin{align}
 E_1 &= d x^1 \\
 E_2 &= d x^2 + \mathbf{f} \, x^4 dx^3 = d( x^2 - \mathbf{f} \, x^3 ) + ( x^4 + 1 ) \, \mathbf{f} \, dx^3 \nonumber \\
 E_3 &= d x^3 \nonumber \\
 E_4 &= d x^4 \nonumber \\
  E^1 &= d x^{\tilde 1} \nonumber \\
  E^2 &= d x^{\tilde 2} - \mathbf{f} \, x^{\tilde 5} dx^3 = d( x^{\tilde 2} + \mathbf{f} \, x^3 ) - ( x^{\tilde 5} + 1 ) \, \mathbf{f} \, dx^3 \nonumber \\
  E^3 &= d x^{\tilde 3} \nonumber \\
  E^4 &= d x^{\tilde 4} - \mathbf{f} \, x^{\tilde 5} dx^4 = d( x^{\tilde 4} + \mathbf{f} \, x^4 ) - ( x^{\tilde 5} + 1 ) \, \mathbf{f} \, dx^4 \nonumber \\
  E^5 &= d x^{\tilde 5}\,. \nonumber 
\end{align}

\bibliography{literatur}
\bibliographystyle{JHEP}
\end{document}